\documentclass[11pt,preprint]{aastex}

\input epsf.tex

\newcommand\iso[2]{$^{\rm #1}$#2}

\def\teff{\mbox{T$_{\rm eff}$}}
\def\logg{\mbox{log $g$}}
\def\vt{\mbox{v$_{\rm t}$}}
\def\BmV0{\mbox{$(B-V)^{\rm 0}$}}
\def\VmK0{\mbox{$(V-K)^{\rm 0}$}}
\def\MV0{\mbox{$M_{\rm V}^{\rm 0}$}}
\def\mM{\mbox{$(m-M)^{\rm 0}$}}
\def\Msun{\mbox{$M_{\odot}$}}
\def\carbiso{\mbox{${\rm ^{12}C/^{13}C}$}}
\def\etal{\mbox{et al.}}
\def\eg{\mbox{e.g.}}
\def\ie{\mbox{i.e.}}

\begin{document}

\title{
The Chemical Composition Contrast between M3 and M13 Revisited: New
Abundances for 28 Giant Stars in M3\footnote{
Based on data obtained with) the Keck~I Telescope of the W. M. Keck Observatory,
which is operated by the California Association for Research In Astronomy
(CARA, Inc.) on behalf of the University of California, the California
Institute of Technology and the National Aeronautics and Space 
Administration (NASA)}.
}

\author{
Christopher Sneden\altaffilmark{2},
Robert P. Kraft\altaffilmark{3}, 
Puragra Guhathakurta\altaffilmark{3}, \\
Ruth C. Peterson\altaffilmark{3}, 
and
Jon P. Fulbright\altaffilmark{4}
}

\altaffiltext{2}{Department of Astronomy and McDonald Observatory,
University of Texas, Austin, TX 78712; chris@verdi.as.utexas.edu}

\altaffiltext{3}{UCO/Lick Observatory, Dept of Astronomy and Astrophysics, 
University of California, Santa Cruz, CA 95064; kraft@ucolick.org, 
raja@ucolick.org, peterson@ucolick.org}

\altaffiltext{4}{Carnegie Observatories, 813 Santa Barbara St, Pasadena, 
CA 91101; jfulb@ociw.edu}

\begin{abstract}

We report new chemical abundances of 23 bright red giant members of
the globular cluster M3, based on high-resolution (R~$\sim$~45000) spectra 
obtained with the Keck~I telescope. 
The observations, which involve the use of multislits in the HIRES Keck~I 
spectrograph, are described in detail.

Combining these data with a previously-reported small sample of M3 giants 
obtained with the Lick 3m telescope, we compare metallicities 
and [X/Fe] ratios for 28 M3 giants with a 35-star sample in the 
similar-metallicity cluster M13, and with Galactic halo field stars 
having [Fe/H]~$<$~--1. 
For elements having atomic number A~$\geq$~A(Si), we derive little
difference in [X/Fe] ratios in the M3, M13 or halo field samples. 

All three groups exhibit C depletion with advancing evolutionary 
state beginning at the level of the red giant branch ``bump'', 
but the overall depletion of about 0.7 to 0.9~dex seen in the clusters
is larger than that associated with the field stars. 
The behaviors of O, Na, Mg and Al are distinctively different among 
the three stellar samples. 
Field halo giants and subdwarfs have a positive correlation of Na with Mg, 
as predicted from explosive or hydrostatic carbon burning in 
Type~II supernova sites. 
Both M3 and M13 show evidence of high-temperature proton capture 
synthesis from the ON, NeNa, and MgAl cycles, while there is no evidence 
for such synthesis among halo field stars.
But the degree of such extreme proton-capture synthesis in M3 is smaller 
than it is in M13: the M3 giants exhibit only modest deficiencies of O 
and corresponding enhancements of Na, less extreme overabundances of Al, 
fewer stars with low Mg and correspondingly high Na, and no indication that 
O depletions are a function of advancing evolutionary state as has been 
claimed for M13. 

We have also considered NGC~6752, for which Mg isotopic abundances
have been reported by Yong \etal\ (2003).
Giants in NGC~6752 and M13 satisfy the same
anticorrelation of O abundances with the ratio 
(\iso{25}{Mg}+\iso{26}{Mg})/\iso{24}{Mg}, which measures the relative 
contribution of rare to abundant isotopes of Mg. 
This points to a scenario in which these abundance
ratios arose in the ejected material of 3-6 solar mass cluster stars,
material that was then used to form the atmospheres of the presently
evolving low-mass cluster stars. 
It also suggests that the low oxygen abundance seen among the most 
evolved M13 giants arose in hot-bottom O to N processing in these same 
intermediate-mass cluster stars. 
Thus mixing is required by the dependence of some abundance ratios
on luminosity, but an earlier nucleosynthesis process in a hotter environment
than giants or main-sequence stars is required by the variations previously 
seen in stars near the main sequence. 
The nature and the site of the earlier process is constrained but not 
pinpointed by the observed Mg isotopic ratio. 

\end{abstract}

\keywords{stars: abundances --- stars: Population II --- Galaxy:halo}

\section{INTRODUCTION}

Thirty years have elapsed since the initial discovery
(Osborn 1971\nocite{osb71}) of CN variations among giants in an otherwise 
mono-metallic globular cluster. 
A flood of evidence has now accumulated indicating that several light 
elements, not only carbon and nitrogen, but also oxygen, sodium and 
aluminum have large and often correlated abundance variations among the 
giants of individual clusters (see reviews by Kraft 1994\nocite{kra94}, 
Briley \etal\ 1994\nocite{bri94}, Wallerstein 
\etal\ 1997\nocite{wal97}, DaCosta 1998\nocite{dac98}). 
The range of variation in some of these elements often exceeds an order of 
magnitude, far more than expected from conventional ``first dredge-up'' 
predictions (see Iben \& Renzini 1984\nocite{ibe84}). 
It is clear that the surface chemical compositions of most 
cluster giants are reflections of the products of multiple proton-capture
chains (Denissenkov \& Denissenkova 1990\nocite{den90}; Cavallo \etal\ 
1996\nocite{cav96}; Langer, Hoffman \& Zaidins 1997\nocite{lan97}; 
Denissenkov \etal\ 1998\nocite{den98}) which involve conversion
not only of C to N, but also O to N, Ne to Na and Mg to Al. 

Less well understood is the site (or indeed sites) in which 
proton-capture syntheses actually take place. 
Two possible scenarios have been invoked.
In the first, the ``primordial'' picture, the variations are a result of 
pre-existing syntheses which made contributions to the gas out of which the 
presently observable low-mass stars were assembled. 
This approach has two subcases: {\it (1)} the stars formed from gas already 
possessing these variations at the epoch of cluster formation, a truly 
``primordial'' picture, or {\it (2)} the presently observed low-mass stars 
were ``polluted'' by the accretion of processed material ejected by 
intermediate-mass (3-6~\Msun) stars in their late evolutionary stages 
(Cottrell \& DaCosta 198l\nocite{cot81}, Cannon \etal\ 1998\nocite{can98}, 
Ventura \etal\ 2001\nocite{ven01}, Yong \etal\ 2003\nocite{yon03}). 
In either subcase, variations in C, N, O, Na and Al abundances should 
exist among main-sequence, or near main-sequence turnoff, stars in clusters. 

Observational examples supporting this picture are readily forthcoming. 
For instance, CN bimodality persists among giants well down onto 
the subgiant branch in several clusters (\eg, Hesser, Hartwick, \& McClure 
1977\nocite{hes77}, Smith \& Norris 1982\nocite{smi82}, Norris \& Smith 
1984\nocite{nor84}, Cannon \etal\ 1998\nocite{can98}, 
Smith 2002a,b\nocite{smi02a,smi02b}).  
C and N abundance inhomogeneities have been observed among subgiant and even 
main-sequence stars in M5 and NGC 6752 (Suntzeff \& Smith 1991\nocite{sun91}, 
Cohen \etal\ 2002\nocite{coh02}) and M71 (Cohen 1999\nocite{coh99}). 
Moreover, variations involving other light elements among such relatively
unevolved stars have recently come to light, \eg, the correlation 
of CN band strengths and Na in main sequence stars of 47 Tuc (Briley 
\etal\ 1996\nocite{bri96}), and the anticorrelation of O and Na as well as 
Al and Mg among main-sequence stars of NGC 6752 (Gratton \etal\ 
2001\nocite{gra01}). 
O and Na also appear to be anticorrelated among subgiants in M71 
(Ramirez \& Cohen 2002\nocite{ram02}). 
Evidence of proton-capture synthesis of Ne into Na and Mg into Al 
has been found among a small sample of turnoff stars in M92 
(King \etal\ 1998\nocite{kin98}). 
Large ranges in Na in the case of M5 have been found to persist all the way 
from the giants to the main sequence (Ramirez \& Cohen 2003\nocite{ram03}), 
and a substantial range in C abundance has been found among main-sequence 
stars in M13 (Briley \etal\ 2003\nocite{bri03}). 
Evidence favoring a primordial source for the variations among these 
light elements therefore seems well established.

But there is also evidence that the stars we presently observe reshuffle the 
abundances of many light elements as a result of deep mixing during the 
ascent of the red giant branch. 
Material of the stellar envelope is exposed to CN, ON, and in some cases 
NeNa and possibly even MgAl cycling as it is circulated through regions near 
or even in the CNO-burning shell of an evolving low-mass, metal-poor star. 
Among field metal-poor ([Fe/H]~$\lesssim$~--1.0) halo stars, 
Gratton \etal\ (2000)\nocite{gra00} found a progressive decrease in total 
C and \carbiso\ ratios, with corresponding enhancement of N, as stars 
evolve up the first giant branch. 
These observed abundance changes far exceed those predicted from 
conventional evolutionary models. 
Among globular cluster giants having [Fe/H]~$<$~--1.0, these changes in 
C and N are even more marked (Langer \etal\ 1986\nocite{lan86}, Bellman
\etal\ 2001\nocite{bel01}, Trefzger \etal\ 1983\nocite{tre83}, 
Briley \etal\ 1990\nocite{bri90}, Briley \etal\ 2003\nocite{bri03}) and 
contrary to the case of field stars, often involve also the ON cycle, so
that many cluster giants exhibit depletion of O (Pilachowski 
1988\nocite{pil88}, Kraft \etal\ 1997\nocite{kra97}, Sneden \etal\ 
1997\nocite{sne97} and references therein) and corresponding
enhancement of N. 
At the same time,  in many clusters (e.g., NGC 288, NGC 362, M5, M10, M13, 
M15) there is an O versus Na anticorrelation, explicable as evidence
of NeNa cycling in a region experiencing ON cycling (Langer \& Hoffman 
1995\nocite{lan95}, Cavallo \etal\ 1996\nocite{cav96}). 

It is not known if the O versus Na anticorrelation is a result of 
primordial events or a result of deep mixing in the giants we observe. 
Even more controversial is the origin of the correlation between Al and Na 
observed among giants in several clusters (Shetrone 1996\nocite{she96a}), 
since CNO-burning shell temperatures in low-mass, metal-poor red giants 
do not reach temperatures high enough to permit significant operation 
of the \iso{24}{Mg}/\iso{27}{Al} cycle (Cavallo \etal\ 1998\nocite{cav98},
Powell 1999\nocite{pow99}, Langer \etal\ 1997\nocite{lan97}). 
Shetrone also discovered that the M13 giants of lowest oxygen abundance had 
anomalously high ratios of (\iso{25}{Mg}+\iso{26}{Mg})/\iso{24}{Mg}, and a 
similar result has recently been found in a large sample of giants in NGC 6752
(Yong \etal\ 2003\nocite{yon03}). 
Such overabundances of the lesser isotopes of Mg can readily be produced 
in the evolution of low-metallicity 3 to 6 solar mass AGB stars 
(Siess \etal\ 2002\nocite{sie02}), and this strongly suggests that a 
pollution scenario is in operation. 
However, the possibility of enhancement of \iso{27}{Al} through 
proton-captures on \iso{25,26}{Mg}, operating in a region favorable to 
NeNa cycling, seems possible in the case of giants in M4 
(Ivans \etal\ 1999\nocite{iva99}).  
Could such a process also be in operation in M13 and in NGC 6752? 
Finally, as Cavallo \& Nagar (2000)\nocite{cav00} have most recently 
emphasized (see also Pilachowski \etal\ 1996\nocite{pil96b}, Kraft 
\etal\ 1997\nocite{kra97}), the mean Al and Na abundances of M13 giants 
sharply increase in the last 0.4 mag before the red giant tip. 
This is hard to understand unless it is the result of deep mixing which 
brings up the ashes of \iso{22}{Ne} to \iso{23}{Na} and \iso{25}{Mg} or 
\iso{26}{Mg} to \iso{27}{Al} processing in the low-mass giants we 
presently see.  
Finally, in indirect support of at least a modest degree of deep mixing 
is the finding that although main-sequence stars in M13 exhibit a wide 
range in [C/Fe] ratios, many of which have [C/Fe] well below --0.3 
(Briley \etal\ 2003\nocite{bri03}), the mean [C/Fe] ratio is much lower
among giants than dwarfs. 
It is also true that seven out of 30 M13 giants have [O/Fe]~$<$~--0.4 
(Kraft \etal\ 1997\nocite{kra97}), an additional result
that is hard to understand on a primordial ``pollution'' scenario.

A growing consensus emerges that most clusters have stars whose
abundance ratios reflect primordial variations, but that these variations
are further modified by deep mixing (see, \eg, Briley \etal\ 
1994\nocite{bri94}, Cavallo \& Nagar 2000\nocite{cav00}, Briley \etal\ 
2003\nocite{bri03}) as stars evolve up the giant branch. 
Of special interest is the contention that deep mixing may also play a 
role as a driver of the so-called ``second parameter'' effect (Sweigart 
1997\nocite{swe97}) in which clusters of closely the same metallicity 
may nevertheless have quite different horizontal branch morphologies. 
Critical pairs of clusters exhibiting these differences, for example, 
are M3 versus M13 and NGC 362 versus NGC 288. 
Many investigators have argued, based on the study of cluster
color-magnitude diagrams, that age is the most likely driver of the second
parameter (\eg, Buonanno \etal\ 1994\nocite{buo94}, 
Sarajedini \etal\ 1997\nocite{sar97}, Rey \etal\ 2001\nocite{rey01}). 
Others have argued that in at least some of these pairs, 
particularly M3/M13, the age difference inferred 
from the relative location of the cluster turnoff luminosity is too 
small to account for the difference in HB morphology (Stetson \etal\ 
1996\nocite{ste96}, Johnson \& Bolte 1998\nocite{joh98}). 

Mixing, if it is deep enough to penetrate the hydrogen burning shell,
will no doubt increase the He/H ratio in the convective
envelopes of the red giant progenitors, so the resulting HB branch 
descendants are likely to be bluer than would be the case in the 
absence of deep mixing (Sweigart 1997\nocite{swe97}, Kraft \etal\ 
1998\nocite{kra98}). 
Since deep mixing appears to be present to a significant degree among 
M13 giants close to the tip of the first giant branch (Kraft \etal\ 
1997\nocite{kra97}, Cavallo \& Nagar 2000\nocite{cav00}), we would
expect M13 to possess a very blue HB, and indeed it does in comparison
with M3. 
At the same time, evolved M3 giants show a much smaller degree 
of deep mixing than do comparable giants in M13 (Kraft \etal\ 
1992\nocite{kra92}, Cavallo \& Nagar\nocite{cav00}). 
However, owing to the fact that M3 is considerably farther from the Sun 
than is M13, the stars are fainter and less amenable to abundance analysis 
at high spectral resolution. 
Thus the published M3 high resolution sample is much smaller than is 
the case for M13 (10 versus 35). Would more evidence of deep mixing be found
in M3, if a larger sample of giant star spectra could be obtained? 
In addition, Na abundances based on lower resolution spectra are available
for 95 additional M13 giants, and of these Mg abundances have been derived
for 74 (Pilachowski \etal\ 1996\nocite{pil96b}). 

In this paper, we examine more fully the [X/Fe] ratios in M3 based on
high resolution analysis of 28 M3 giants.
We confirm the tentative earlier result that the Na versus O, Mg versus Na, 
and Mg versus Al diagrams of M3 differ significantly from those of M13. 
We compare these diagrams for M3 and M13 giants with those of metal-poor 
field halo giants. 
We address also the abundances of $r$- and $s$-process elements Eu, La and Ba. 
We finally return to a discussion of these results in connection with the 
second parameter problem.

\section{OBSERVATIONS AND REDUCTIONS}

\subsection{General Considerations and Single-slit Spectroscopy}

This M3 abundance survey combines analyses of 23 stars based on new
high resolution spectra with re-analyses of seven stars studied
by Kraft \etal\ (1992, 1995)\nocite{kra92,kra95}.
Two stars (vZ~1397 and AA) are in common between the two samples, having
been re-observed in the present program.
Therefore the total number of M3 giants considered here is 28.
The names of the stars adopted here, their von Zeipel (1908)\nocite{von08}
numbers where available, visual magnitudes $V$, and photometric
indices $B-V$, $V-I$, and $V-K$ are given in Table~1 for the 23 newly
observed stars (plus three more from the observing run that are excluded 
from the abundance analysis).
Stars I-21, IV-77, IV-101, A, and AA were named by Sandage 
(1953, 1970\nocite{san53,san70})\footnote{
$V,B-V$ photometry for the first six stars listed in Table~1 are
taken from Cudworth (1979)\nocite{cud79a}. 
Values of $V$ for all other stars are those of Dorman (\S2.2). 
For these stars, we estimated $B-V$ by comparing the Dorman and Cudworth
color-magnitude arrays at a given value of $V$.}
and the B1.1--B4.5, F2.4 
designations are new to this work.
We employed single slit spectroscopy for six stars: I-21, IV-77, IV-101, A, 
AA, and vZ~1397, and multi-slit spectroscopy for the other stars.
For each of the ``multi-slit'' stars, we list also the equinox J2000 
offset in arcsec (Dorman 1998\nocite{dor98}) from the cluster center 
(von Zeipel 1908\nocite{von08}).

All new spectra were obtained with the HIRES spectrograph of the Keck~I 
telescope (Vogt \etal\ 1994\nocite{vog94}).
For single-slit observations, the spectrograph entrance slit was set to 
a width of 0.86$\arcsec$, which yielded a spectral resolving power of 
R~=~$\lambda/\Delta\lambda$~$\simeq$ 45,000 with the Tektronix 
2048$\times$2048 CCD detector.
A slit length of $7''$ was used; this avoids any overlap in the echelle orders.
In the yellow-red wavelength region, the free spectral ranges of
HIRES orders are larger than can be captured by its CCD detector,
and so spectral coverage is not continuous.
The echelle grating and cross-disperser tilts for our observations were set 
cover the wavelength range $\lambda\lambda 5200$--6800\AA\
to capture the same spectral features that we have employed in the 
earlier papers of this series (\eg, Sneden \etal\ 1997\nocite{sne97}, 
Ivans \etal\ 2001\nocite{iva01}).

The stellar observations were accompanied by quartz and Th-Ar
hollow cathode lamp integrations. 
Spectra of hot, rapidly-rotating stars were also obtained to aid in
the cancellation of telluric (O$_{\rm 2}$ and H$_{\rm 2}$O) absorption
features that contaminate the program-star spectra.
Most of the data reduction tasks employed standard IRAF\footnote{
IRAF is distributed by the National Optical
Astronomy Observatories, which are operated by the Association of
Universities for Research in Astronomy, Inc., under cooperative agreement
with the National Science Foundation.}
$ccdproc$ and $echelle$ packages, for bias subtraction, flat-field
quartz lamp division, cosmic ray excision, scattered light correction, 
individual spectral order extraction, and wavelength calibration.
Final reduction steps (\eg, continuum normalization, bad pixel removal)
were performed with specialized software code SPECTRE (Fitzpatrick \& Sneden 
1987\nocite{fit87}). 

In Table~1 we also list the exposure times of the observations, 
along with estimated signal-to-noise S/N of the fully-reduced spectra
in a couple of relatively stellar-line-free regions near 6400~\AA.
Single integrations were obtained for all stars observed in the single-slit 
mode.

The SPECTRE code was also used to measure equivalent widths (EWs) of stellar
absorption lines.
The input line list was that of Ivans (1999,2001)\nocite{iva99,iva01},
and both Gaussian fits and direct integrations over the
line profiles were employed in the EW computations.
For each line, the EWs for all 23 M3 stars are given in Table~2, 
along with their excitation potentials and transition probabilities.
We defer to a later section discussion of the EWs obtained from our
previous M3 abundance studies.

\subsection{Multi-slit Echelle Spectroscopy}

The combination of echelle spectroscopy and multi-slit masks used in this
paper is fairly unusual and we describe it in detail here.  
There are a few general conditions that must be met in order for 
multi-slit echelle spectroscopy with HIRES to be feasible and worthwhile.  
A high surface density of targets is essential, so that one can ideally 
find 4--5 targets within the $\approx 10\arcsec\times15\arcsec$ area of 
the slitmask (see below).  
This represents a significant multiplexing advantage over the normal
mode of operation in which a single star, or even a pair of stars separated 
by $\lesssim 10\arcsec$, can be observed simultaneously on a single slit.  
A much higher surface density of targets, however, can present problems 
in typical ground-based seeing conditions: blending/crowding, and light 
loss and inadequate blank sky on short slitlets.  
Only relative astrometry is required, accurate over scales of a few tens 
of arcseconds, since mask alignment is carried out interactively.  
A multi-slit 2D echelle spectrogram is significantly more complex than a 
standard single-slit one, so that data reduction requires extra care, but the 
gains in observing efficiency from multiplexing often makes this worthwhile.  
This is especially true if the targets are much brighter than the night sky 
(short slitlets make sky subtraction difficult but accurate sky subtraction 
is not as critical for such targets) and the exposure times long (the added 
overhead of mask alignment is then a negligible perturbation).
These conditions are ideally realized in the central regions of
many globular clusters.

The HIRES mask parameters were constrained by several other design 
considerations, which affect all cross-dispersed echelle spectrographs.
The overall slit length, the sum of the lengths (in the spatial direction) 
of all the slitlets on a given mask, must be less than some critical value 
to prevent adjacent echelle orders from overlapping on the CCD.  
All echelle orders have the same height, as determined by the overall slit 
length, but the inter-order spacing decreases toward shorter wavelengths.  
Thus, the shortest wavelength of interest determines the maximum allowable 
slit length.  
For our M3 HIRES masks, we have used slit lengths in the range 
$8\arcsec$--$9\arcsec$.
The masks with the longer slit lengths have slight order 
overlap at the blue end of our chosen wavelength setting.  
The maximum allowed width (in the dispersion direction) of the region of the
slitmask over which slitlets can be arranged is $15\arcsec$ and this is set 
by the maximum width of the slit-jaw opening.  
Each slitlet is forced to be at least $1\arcsec$ in length for the reasons 
listed above.  
The lengths of all the slitlets are maximized while maintaining a minimum 
spatial gap of $0.3\arcsec$ between ends of successive slitlets to prevent 
overlap, and within the bounds of the overall slit length for the mask.
Each slitlet is assigned a width of $0.86\arcsec$, identical to the width 
of the standard HIRES single slit.

\begin{figure}
\epsscale{0.40}
\plotone{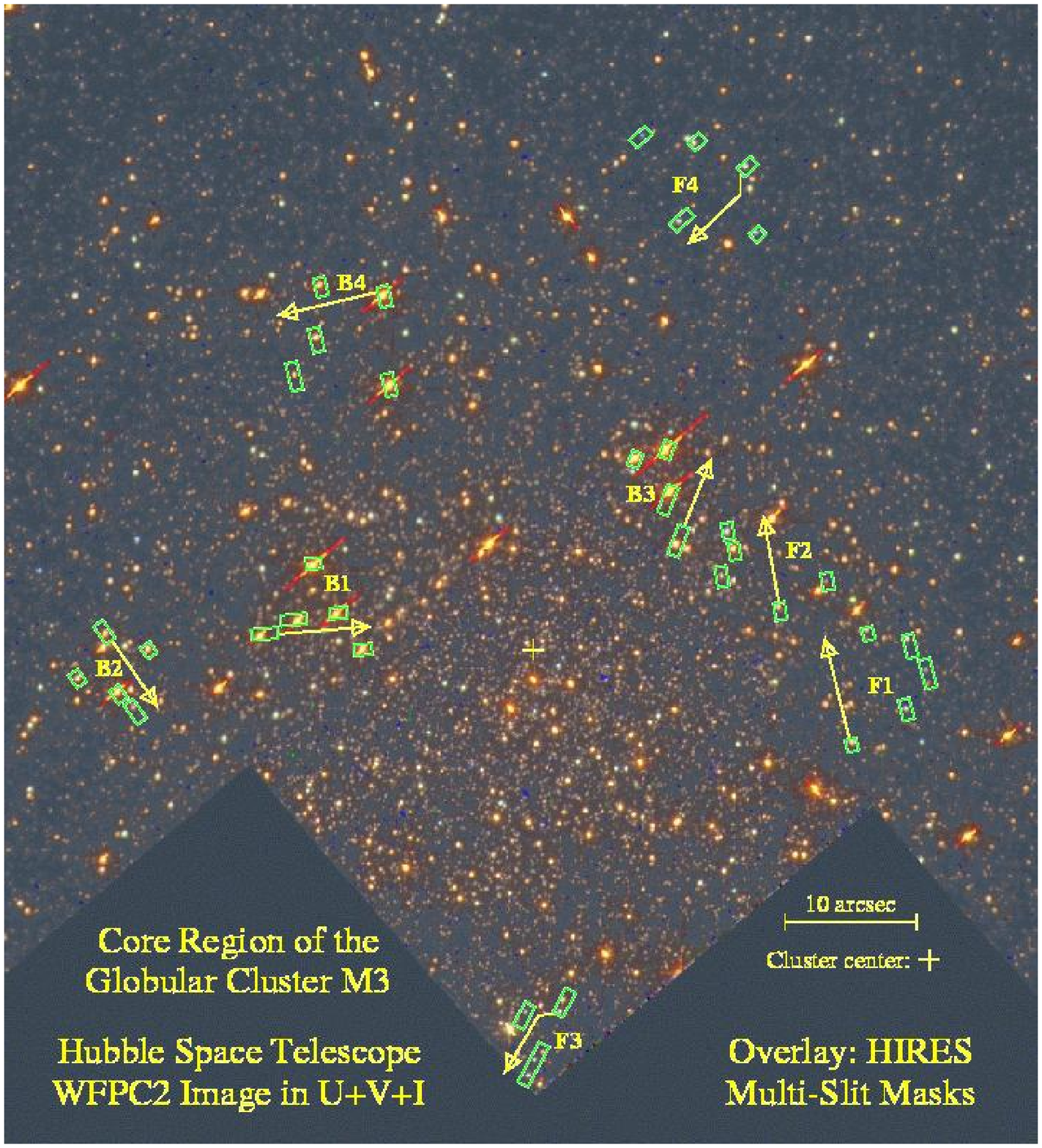}
\caption{
A portion of an HST/WFPC2 image of the core of M3.  
This color composite is based on archival F336W ($U$), F555W ($V$), and 
F814W ($I$) exposures.  
Eight HIRES multi-slit masks are overlaid, B1--B4 targeting bright stars 
and F1--F4 targeting mostly faint stars.  
For each mask, the arrow originates from the slitlet designated \#1 
(\eg,~B1.1, B2.1, etc.) and points in the direction of increasing slitlet 
number for the rest of the slitlets in that mask.  
North is up and east is left.  
Saturation bleeds are visible in the $I$- and $V$-band images for the 
brightest red giants.  
The seeing FWHM was about $0.8\arcsec$ during the HIRES observations, 
whereas the resolution of the WFPC2 image is $0.1\arcsec$. 
Thus, crowding and contamination are more problematic in the spectroscopic 
data than is apparent from this image.
The bright blue spots and streaks correspond to cosmic ray hits. 
The exposure time was longest in F336W so this image has the largest number 
of such hits. 
The PC1 CCD appears to have the highest surface density of hits but that is
simply because it has a higher pixel density on this color image than the WF
CCDs.
\label{f1}}
\end{figure}

\begin{figure}
\epsscale{0.85}
\plotone{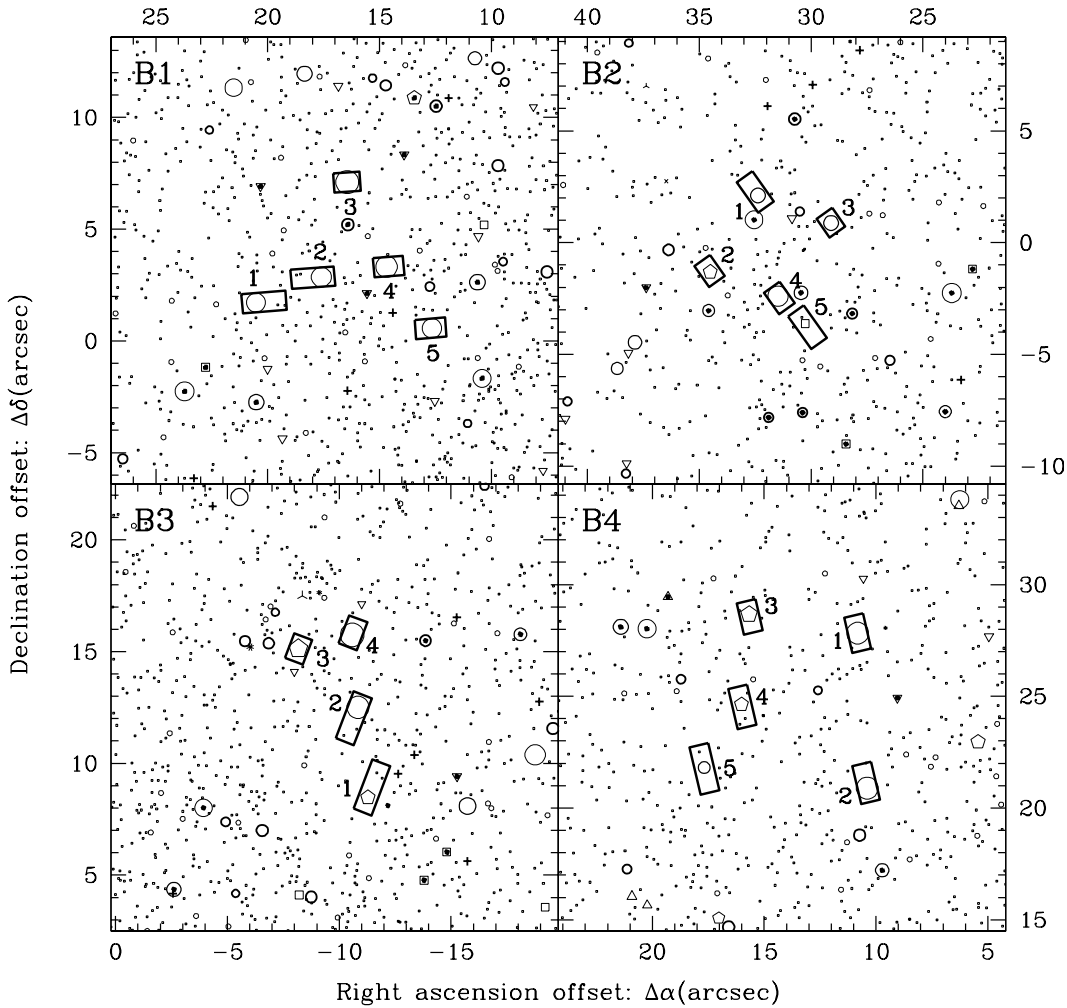}
\caption{
Schematic representation of the HIRES bright star masks B1--B4
superimposed on the M3 star field.  
Each target/slitlet is assigned a unique number as indicated.  
The schematic star field image is derived from a photometry list 
provided by B.~Dorman and collaborators, and should bear a
good resemblance to the color image shown in Figure~1 except in rare
instances where the photometry list is incomplete.  
The symbol size is roughly proportional to apparent stellar brightness. 
The symbols indicate different stellar types: circle~=~RGB; 
pentagon~=~AGB; square~=~red HB; $+$~=~RR~Lyrae; 
up-triangle$\rightarrow$down-triangle$\rightarrow$3-pointed
starred symbol~=~increasingly bluer HB stars;
$\times$/asterisk~=~extreme/moderate blue straggler; dot~=~subgiant/main
sequence.  
The small bold dots superimposed on the stellar symbols mark the
locations of non-reflective marks etched on the shiny front face of each
multi-slit mask plate. 
Mask alignment was carried out by registering these marks with the 
corresponding reference bright stars on the HIRES guider camera image.
\label{f2}}
\end{figure}

Bright red giant targets in M3 were selected from an archival HST/WFPC2
dataset for which a stellar photometry list in the form of $V$, $V-I$
was kindly provided to us by Ben Dorman.  
Appropriate groups of 4--5 targets were isolated by eye and attempts
were made to place each group on a multi-slit mask by optimizing the position
angle and ($\Delta\alpha$, $\Delta\delta$) translation of the mask design.
Four such mask designs labeled B1--B4 are shown in Figures~\ref{f1} and 
\ref{f2}. 
The F1--F4 mask designs shown in Figure~\ref{f1} contain fainter M3 targets, 
mostly horizontal branch stars, which will be discussed in a later paper.
Only one star observed with the F1--F4 masks, F2.4,  was a red giant bright 
enough to be included in the present study.
For each mask, the targets/slitlets were numbered in order increasing in the
direction of the arrow shown in Figure~\ref{f1};
this numbering scheme led to the star names B1.1--B4.5, F2.4 (Table~1) adopted 
for this paper.
Alignment stars were selected around each mask; no spectra are obtained 
for alignment stars and they are permitted to lie outside the 
$\approx 10\arcsec\times15\arcsec$ slitmask area occupied by spectroscopic 
targets.
Of the 42 M3 stars actually observed in the Keck~I program, 23 were RGB
members recorded with sufficient S/N to warrant abundance analysis:
six single-slit stars, 16 stars on the B1--B4 masks, and one star on the 
F1--F4 masks.

The fabrication of HIRES multi-slit echelle masks was carried out by Bill
Mason at Keck Observatory under the supervision of Jeff Lewis of the Lick
Observatory instrument shops.  
Masks were milled onto an aluminum plate with a highly polished surface,
designed to serve as a flat mirror for the TV guider camera system.
Each slitlet was milled as a rectangle whose corners were rounded because 
of the finite diameter ($0.5\arcsec$) of the milling tool.  
At the location of each alignment star, a non-reflective ``blind'' spot was 
etched into the otherwise reflective surface by drilling the mill tool 
halfway into the thickness of the aluminum plate (small bold symbols 
overlaid on stellar symbols in Figure~\ref{f2}.
The aluminum mask plate can accommodate nine multi-slit masks arranged
side by side, any one of which can be placed in the HIRES light
path by rotating the wheel containing the mask plate.

After a mask was selected, and the telescope pointed at the desired
coordinates at the correct sky position angle, mask alignment was achieved by
ensuring that the blind spots were centered on the corresponding alignment
stars on the TV guider image.  
Only translational offsets were needed for this procedure; fine tuning of 
the position angle was never needed.  
The exposure times for masks B1, B3, and B4 given in Table~1 are the sum of
three individual integrations, that of B2 the sum of two integrations,
and F2 that of a single integration.
A uniform sequence of short TV guider camera frames was saved; coadding
these frames allows us to obtain estimates of the seeing and telescope
pointing/mask alignment during the main spectroscopic exposure.  
For the purposes of wavelength calibration, arc lamp exposures were 
obtained through each mask.  
Moreover, radial velocity standard stars were observed through one or 
more slitlets of each mask.

\subsection{Reduction of the Multi-slit Spectra}

The extraction of stellar spectra in multi-slit echelle data generally
used the same procedure as single-slit data.  
The key differences were in the defining and tracing of the extraction 
apertures of individual stars within a given mask and the lack of sky 
subtraction.

As the red giants in our study are relatively bright, identifying the target
stars in a cross-dispersion plot of the CCD image is straightforward.  
The cross-dispersion plot revealed slight overlap between the adjacent 
slitlets in each order.  
To minimize the effect of contamination between neighboring stars/slitlets, 
the widths of the extraction apertures were hand-defined.
The quartz flat-field frames were reduced using the same apertures, as
regions with cross-contamination would show up in having $\approx2\times$
more counts than normal in the cross-dispersion plots.  
These quartz frames, and later the extracted Th/Ar spectra, were inspected 
to ensure that there was no cross-contamination between apertures.  
Unlike the case of the single-slit, no sky background apertures were 
defined as there were no appropriate blank regions.

For the masks with the longest overall slit lengths ($\sim9''$), 
there was slight overlap between the bluest adjacent echelle orders 
for the end slitlets.
In such cases, the overlapping data in those regions were simply discarded.
Otherwise there were 29 full orders per exposed spectrum, each with four or 
five slitlets, resulting in 116 or 145 extraction apertures per exposure.

Scattered light was removed by fitting to the inter-order regions of the
exposed spectrum.  
Generally the maximum of the subtracted surface was on the order of 30 
counts per pixel, which is a few percent of the counts per
pixel in the stellar continuum.

The final one-dimensional spectra were extracted using a 
variance-weighting scheme.  
The extraction of the Th/Ar arc lamp spectra was accomplished using
the same aperture definitions as the target stars.  
As a check of the aperture definitions, the extracted arc lamp spectra were 
inspected for signs of contamination from adjacent apertures; they 
generally passed this test with the exception of ``bleeds'' from very 
strong, overexposed features.

\section{ABUNDANCE ANALYSIS}

\subsection{Keck~I HIRES Spectra of M3 Giants}

Analysis of these spectra proceeded along the lines of earlier papers
in the series by the Lick-Texas group (LTG) (see, \eg, Kraft \etal\ 
1997\nocite{kra97}, Sneden \etal\ 1997\nocite{sne97}), but modified in 
accordance with practices introduced in the LTG paper on M5 
(Ivans \etal\ 2001\nocite{iva01}) and discussed extensively in
the recent review article by Kraft \& Ivans (2003)\nocite{kra03}. 
Briefly, the observed
colors provided a first estimate of \teff, making use of the
\BmV0\ or \VmK0\ (when available) versus \teff\ scales of Alonso \etal\ 
(1999\nocite{alo99}, Table 6), which are in turn based on the IR-Flux 
method (Blackwell \etal\ 1990)\nocite{bla90}. 
Values of \logg\ were estimated by assigning to each giant a mass
of 0.80\Msun, assuming an M3 true distance modulus \mM~= 15.02 
(Kraft \& Ivans 2003\nocite{kra03}), applying Worthey's (1994)\nocite{wor94} 
bolometric corrections, and calculating the surface gravity from the 
fundamental relationship $g$~$\sim$~$\mathcal{M}$\teff$^4$/L.

As before, we employed MARCS models (Gustafsson \etal\ 1975\nocite{gus75}) 
and the current version of the MOOG line analysis code 
(Sneden 1973\nocite{sne73}) to compute abundances from the EWs on a line 
by line basis. 
The resulting plot of log~$\epsilon$(Fe) versus the excitation potential of 
each \ion{Fe}{1} line (hereafter called the ``\ion{Fe}{1}  excitation plot'') 
was then used iteratively to ``correct'' the preliminary estimates of \teff, 
and the models run again until the \ion{Fe}{1} excitation plot 
showed a zero slope, \ie, the value of log~$\epsilon$(Fe) showed no 
dependence on \ion{Fe}{1} excitation potential. 
Values of the microturbulent velocity, \vt, were estimated in the usual way 
by inspection of a plot of log~$\epsilon$(Fe) versus EWs of the \ion{Fe}{1} lines.

We proceeded as just outlined out of concerns expressed in the recent
literature that estimates of metallicity based on \ion{Fe}{1} might 
not be reliable because of the possible overionization of Fe inherent 
in the atmospheres of metal-poor stars (Thevenin \& Idiart 
1999\nocite{the99}, Asplund \& Garcia Perez 2001\nocite{asp01}). 
Fe is mostly singly-ionized in the atmospheres of globular cluster giants,
so ``Fe overionization'' has little effect on the abundance of \ion{Fe}{2}. 
In addition, \ion{Fe}{2} abundances have been shown to remain virtually 
unaffected if 3D calculations replace 1D calculations (Asplund \& 
Garcia Perez 2001); this may not be the case for \ion{Fe}{1} 
(Nissen \etal\ 2002). 
For these reasons, we have based our ``metallicity estimates'' on 
log~$\epsilon$(Fe), derived from the \ion{Fe}{2} lines in the spectra. 
We make no attempt to derive values of \logg\ by forcing equality between 
log~$\epsilon$(Fe) from \ion{Fe}{1} and log~$\epsilon$(Fe) from \ion{Fe}{2}. 
Rather we fix \logg\ from the estimated value of \teff, mass,
and cluster distance modulus, \ie, from evolutionary considerations, with
the consequence that log~$\epsilon$(Fe) from \ion{Fe}{1} and \ion{Fe}{2} 
may differ. 

Despite the concerns about \ion{Fe}{1}, Kraft \& Ivans (2003) found that 
the value of \teff\ derived from the \ion{Fe}{1} excitation plot agreed 
closely with the value of \teff\ obtained from the \teff\ versus color 
scales of Alonso \etal\ (1999\nocite{alo99}).
This is an empirical result, obtained from the analysis of giants in seven
globular clusters having small reddening [$E(B-V)$~$\leq$~0.10]. 
An offset toward higher \teff's by $\sim$50-100~K would be required in the 
case of the Houdashelt \etal\ (2000)\nocite{hou00} scale. 
Adoption of this alternate scale would reduce the abundance of Fe based 
on \ion{Fe}{2} by $\simeq$0.04~dex. 
Since in this paper we are concerned largely with the comparison of 
abundances of Fe in M3 and M13, we adopt the Alonso \etal\ scale, the 
difference in abundance being independent of the choice of \teff. 

In Table~3 we list the adopted model parameters, \ion{Fe}{1} and 
\ion{Fe}{2} abundances and radial velocities for the 23 M3 giants observed 
in the Keck~I HIRES program.
Also listed are photometric parameters and radial velocities for three
stars that were excluded from the abundance analysis: B2.2, a warmer
AGB star with poor S/N in its spectrum; B2.5 a much hotter star,
for which no photometric parameter estimation was attempted; and
B3.1, whose spectrum was compromised by that of a second star displaced
slightly blueward.
The columns labeled \teff(spec) and \logg(spec) are values found by adjusting 
the photometric estimates in accordance with the \ion{Fe}{1} excitation plot. 
The radial velocities were obtained with reference to the night sky O$_2$ 
doublets, as discussed in the paper on NGC~7006 (Kraft \etal\ 
1998\nocite{kra98}, Appendix A). 
These will not be discussed further here, as we will later include  
measurement of velocities derived from the remaining 18 Keck~I spectra
in a separate paper. 
Values of [Fe/H] derived from \ion{Fe}{1} and \ion{Fe}{2} do not include 
the small adjustments (+0.05~dex and +0.02~dex, respectively) in log~$gf$ 
that Kraft \& Ivans (2003)\nocite{kra03} found necessary to reproduce the 
solar Fe abundance [log~$\epsilon$(Fe)~= 7.52], 
using Kurucz models without convective overshoot (Castelli \etal\ 
1997\nocite{cas97}, Peterson \etal\ 2003\nocite{pet03}) plus the Oxford 
log~$gf$ values (Blackwell \etal\ 1980\nocite{bla80}). 
Again our concern is the comparison of M3 with M13; we adopt the same 
set of log~$gf$ values
for all stars in both clusters. 
The mean values of [Fe/H] derived from \ion{Fe}{1} and \ion{Fe}{2}, 
respectively, are --1.58 ($\sigma$~=~0.06) and --1.45 ($\sigma$~=~0.03),
which implies an \ion{Fe}{1} ``underabundance factor'' of --0.13~dex. 
We assume that the ``metallicity'' of M3 is given by the abundance of 
Fe based on \ion{Fe}{2}.

In Table~4 we list the values of [X/Fe] for elements from O to Eu
again from the Keck~I data; mean values for this sample are given at
the bottom of the table.
The star-to-star abundance scatters of elements with atomic weights
$A$~$\geq$~$A$(Si) provide reasonable internal uncertainty estimates
of these abundance values.
For more detailed discussion of overall uncertainties the reader is
referred to Sneden \etal\ (1997)\nocite{sne97} and Ivans \etal\ 
(1999,2001)\nocite{iva99,iva01}.

The abundance of O is derived from spectrum synthesis of the 
[\ion{O}{1}] doublet at 6300, 6364~\AA. 
Allende Prieto, Lambert, \& Asplund (2001) have re-considered the
solar O abundance from the 6300.3~\AA\ line, pointing out that a 
high-excitation (E.P.~= 4.27~eV) \ion{Ni}{1} line is a substantial 
contaminant.  
But in our cool giant stars, the \ion{Ni}{1} contribution 
to the [\ion{O}{1}] is very small, and can be ignored.  
Test calculations suggest that typically EW(\ion{Ni}{1})~$\leq$ 0.5~m\AA,
while EW([\ion{O}{1}]) ranges from $\sim$20 to $\sim$70~m\AA.

Since O is entirely in the neutral state but Fe is mostly ionized, we 
normalized the [O/Fe] ratio to [Fe/H] derived from \ion{Fe}{2} 
(Kraft \& Ivans 2003\nocite{kra03}). 
For consistency with the earlier studies of M13 and halo field giants, we
retained the Anders \& Grevesse (1989)\nocite{and89} ``traditional'' 
solar oxygen abundance of log~$\epsilon$(O)~= 8.93 instead of the revised 
value of 8.69 recommended by Allende Prieto \etal\ (2001)\nocite{all01}. 
All other elements, from Na through Eu, are mostly in the singly ionized 
state in the atmospheres of globular cluster giants, but only Sc, Ti, Ba, 
La and Eu abundances are actually derived from lines in that state. 
We therefore normalized those elements to the abundance of Fe based on 
\ion{Fe}{2}. 

The other elements make their appearance in the spectrum in lines of the 
neutral state, and presumably also suffer from ``overionization'' if 
indeed Fe is so affected.
It therefore seems best to normalize those to [Fe/H] based on 
\ion{Fe}{1}\footnote{ 
Since the term schemes of these elements differ from one another,
and differ also from \ion{Fe}{1}, it is is not clear that all have the same
``overionization'' factors, so these [X/Fe] ratios must be considered
approximate (see discussion in Ivans \etal\ 1999\nocite{iva99}).}
All [Na/Fe] ratios are derived from spectrum synthesis of the \ion{Na}{1} 
doublet at 5682, 5688~\AA\ and EW measurements of the doublet at 6154, 
6160~\AA.

Finally, we also list in Table~4 [Na/Fe] ratios taking into account the
non-local thermodynamic equilibrium (nLTE) versus \logg\ corrections 
recommended by Gratton \etal\ (2000)\nocite{gra00}, from a study of 
metal-poor halo field stars.

\subsection{Lick Hamilton Spectra of M3 Giants}

Somewhat lower resolution (R~$\sim$~30000) Lick 3.0m Hamilton spectra of
10 M3 giants had already been analyzed by members of LTG and reported in
the literature (Kraft \etal\ 1992, 1993, 1995\nocite{kra92,kra93,kra95}). 
We reanalyzed seven of these spectra, assigning slightly revised values of 
\teff\ and \logg\ based on the precepts outlined above in \S 3.1. 
The earlier work did not include estimates of [X/Fe] for Sc (based on 
\ion{Sc}{1}), Mn, Ba and Eu; these are added here. 
We also derive estimates of [Na/Fe] taking into account nLTE effects.
We compute V, Mn, and Ba abundances with full accounting for hyperfine 
substructure, following the recipe outlined in Ivans 
\etal\ (2001)\nocite{iva01} for \ion{V}{1} and \ion{Ba}{2} lines, and 
adopting the hyperfine data of Kurucz (1995)\nocite{kur95}\footnote{
see http://cfaku5.cfa.harvard.edu/}
for the \ion{Mn}{1} 6021.79~\AA line. 
Table~5 contains the adopted models and values of [Fe/H] derived from 
\ion{Fe}{1} and \ion{Fe}{2}. 
Table~6 has derived [X/Fe] ratios for all elements except Mg, Al, Ti 
(based on Ti II) and La, the lines of which fell in the interstices 
between the echelle orders and thus were not recorded. 
Three stars (MB1, vZ~297, III-28) of the Lick sample were dropped, the S/N 
of the spectra having been judged inadequate.

Two stars, AA and vZ~1397, were analyzed from both Lick and Keck~I spectra.
These stars were used to compare the EWs derived from the two investigations.
It was found that the Keck~I HIRES spectra produced lines that, on the
average, were 7\% smaller in EW compared with Lick Hamilton spectra; before
re-analyzing the Lick data, we reduced the EWs of all lines by that amount.
We also used these two stars to study any possible offsets in abundance.
In Tables~5 and 6 we interline for these two stars the Keck~I data taken
from Tables~3 and 4. 
For AA, the agreement in [Fe/H] for both \ion{Fe}{1} and \ion{Fe}{2} is
excellent, as are the common [X/Fe] ratios, the largest difference being
0.09~dex for [Sc/Fe] derived from \ion{Sc}{1} (based on one line). 
For vZ~1397, the agreement is less good, and it is driven principally by the 
0.11~dex difference in [Fe/H] derived from \ion{Fe}{2}. 
Since the S/N and spectral resolution of the Keck~I spectra exceed that 
of the Lick Hamilton spectra, we give precedence to the abundances derived 
from the former.

The mean abundance ratios for the seven Lick stars are given at the bottom of
of Table~6. 
The mean values of [Fe/H] derived from \ion{Fe}{1} and \ion{Fe}{2}
are --1.55 ($\sigma$~= 0.02) and --1.46 ($\sigma$~= 0.05) respectively, 
in excellent agreement with the mean values determined from the Keck~I 
data (Table~3). 
Inclusion of the Lick Hamilton stars brings the total number of M3 giants 
available for discussion to 28.

\subsection{Revised Abundances for M13 Giants}

Again using the precepts described in \S 3.1, we re-assigned values of 
\teff\ and \logg\ to 18 of the 21 M13 giants that were observed
either with the post-1995 version of the Lick Hamilton spectrograph
(R~$\sim$~50000) or the HIRES spectrograph of the Keck~I telescope 
(R~$\sim$~45000) (Kraft \etal\ 1997\nocite{kra97}); the three stars 
omitted had spectra of relatively low S/N.  
As before, we adopted the Alonso \etal\ (1999)\nocite{alo99} \teff\ scale 
based on \BmV0\ (Cudworth \& Monet 1979\nocite{cud79b}) and/or \VmK0\ 
(Cohen \etal\ 1978\nocite{coh78}) when available, and took a
distance modulus for M13 of \mM~= 14.42 (discussed by Kraft \& Ivans 
2003\nocite{kra03}).
Adjustments to [Fe/H] determined from both \ion{Fe}{1} and \ion{Fe}{2} 
lines and to [X/Fe] values were determined from the offsets in Table~3 of 
Ivans \etal\ (2001); the offsets themselves remain nearly constant with 
small changes in metallicity (in this case $\sim$0.3 dex) (see also 
Shetrone 1996\nocite{she96a}). 
We also determined abundances of Ba, La and Eu, elements that had not been
included in the earlier analysis. 
Results are listed in the top portion of Table~7, but are restricted to 
elements of particular interest in the comparison of M13 and M3 (in this 
case, O, Na, Mg, Al, Ba, La and Eu).
In the lower portion of Table~7, we exhibit corresponding results for
the remaining 16 giants that had been studied earlier, based entirely
on Hamilton spectra of lesser resolution (R~$\sim$~30000). 
For these, only O and Na abundances (in addition to [Fe/H]) are given.

Only one star, L835, is common to both the Keck~I and Lick samples.
In this case, the agreement in the value of [Fe/H] based on \ion{Fe}{1} is
disappointingly poor, but the situation is much better for \ion{Fe}{2}. 
The values of [X/Fe] for O and Na, however, are in excellent agreement.
As with M3, we prefer the values based on the superior S/N and spectral 
resolution of the Keck I observations. 

Mean values of the abundances are found at the bottom of Table~7.
For the Keck I sample alone, the values of $\rm\langle[Fe/H]\rangle$ based on 
\ion{Fe}{1} and \ion{Fe}{2} lines are --1.62 ($\sigma$~= 0.06) 
and --1.55 ($\sigma$~= 0.09), respectively.
Corresponding values using the Lick sample only are --1.57 ($\sigma$~= 0.07)
and --1.50 ($\sigma$~= 0.09). 
The small differences most likely spring from the lower resolution and 
generally poorer S/N of the Lick, compared with Keck~I spectra. 
In any case, the Fe I ``overionization'' deficiency appears to be 
$\sim$0.08 dex, based on this analysis procedure.

\section{DISCUSSION}

The major concern of this paper centers on the remarkable contrast of
certain [X/Fe] ratios among giants in M3 in comparison with comparably-evolved
giants in M13, on the one hand, and the halo field giants on the other.
We divide the discussion into the familiar groups by nucleosynthetic 
origin, and discuss M3 stars along with their M13 and halo field surrogates
within each group.
We do not, in any one of these cases, have a ``complete'' sample
of stars selected by \MV0.
Brighter than \MV0~=~--1.5, we have high resolution analyses for 19
giants in M3 and 22 giants in M13.
Total sample sizes in M3 and M13 are 28 and 32, respectively; in each
cluster, our sample includes a few stars having \MV0~$>$~--0.5.
We therefore regard our two samples, comparable in size, as at least
representative, and therefore anticipate that larger samples will not
lead to unexpected surprises.

\subsection{Fe-peak and Heavier Elements}

\begin{figure}
\epsscale{0.95}
\plotone{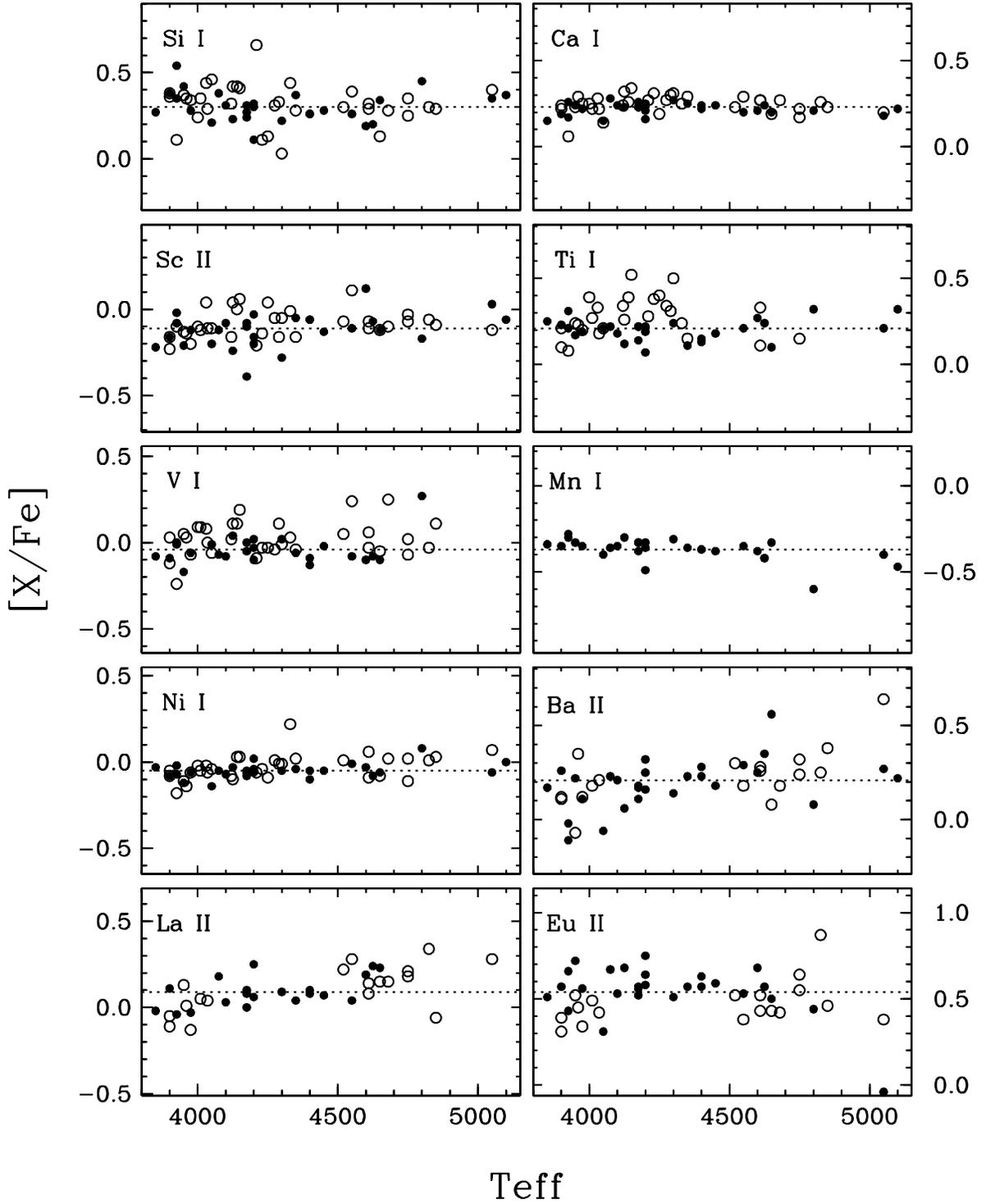}
\caption{
Abundance ratios for silicon and heavier elements in M3 (filled
circles) and M13 (open circles) as functions of \teff.
The ordinate of each panel has the same extent (1.2~dex), and the midpoint
(indicated with a dotted line) is the mean abundance ratio of the
particular element in M3.
\label{f3}}
\end{figure}

Existing research on cluster abundances indicates that there is little
star-to-star intracluster variation in [X/Fe] ratios for elements 
having atomic weights $A$~$\geq$~$A$(Si).
There are, of course, exceptions, most notably among $r$-process species
in the very metal-poor cluster M15 (Sneden \etal\ 1997\nocite{sne97},
Sneden \etal\ 1999\nocite{sne00}).
There are also cases in which $\rm\langle[X/Fe]\rangle$ differs 
significantly from one cluster to another at a similar value of [Fe/H], 
as for example, the substantial differences in Si, Ba, La and Eu that exist 
when one compares M4 with M5 (Ivans \etal\ 2001\nocite{iva01}).
These variations emerge when stellar sample sizes are moderately
large (N~$\gtrsim$~20), and since our M3 and M13 samples exceed this
number, we begin by exhibiting in Figure~\ref{f3} the [X/Fe] ratios for
Si and heavier elements versus \teff\ for the M3 stars of
Table~4 and 6, and M13 stars of Table~7.
Horizontal dashed lines refer to the mean values of [X/Fe] for M3 stars
only.

Inspection of Figure~\ref{f3} and examination of the accidental errors
tabulated at the feet of Tables~4, 6, and 7 suggests that the spread in 
[X/Fe] is for the most part small and attributable to a paucity of lines 
(in the cases of, \eg, \ion{Sc}{2}, \ion{Ba}{2}, \ion{La}{2}, \ion{Eu}{2}) 
and/or line weakness (EW~$<$~10~m\AA, \eg, for \ion{V}{1}, \ion{La}{2}, 
\ion{Eu}{2}).  
Exceptions appear to be the very high abundances of La and Ba abundances in
two different M13 giants, and the very low value of Eu in a single M3 giant
(B4.4).
None of these stars exhibit unusual abundances of other elements.
The data of Figure~\ref{f3} also show that the mean values of
[X/Fe] in M3 and M13 differ no more than $\sim$0.1~dex.
In addition, the pattern also follows that of field halo stars with
metallicities [Fe/H]~$\sim$~--1.5 (Fulbright 2002\nocite{ful00}),
with the possible exception of Ba, which on average appears to be about
0.15~dex more Ba-rich in both clusters compared with the field.
On the other hand, [Eu/Fe] in both clusters is quite close to the mean
value for field stars.

\subsection{Interrelated C, N, and O Abundances}

Carbon and nitrogen abundances were obtained for a substantial sample
of CN-weak and CN-strong giants in both M3 and M13 in a pioneering study
by Suntzeff (1981).\nocite{sun81}
A number of these stars were reanalyzed by Smith \etal\ (1996)\nocite{smi96}
who, coupling these results with what were then freshly determined
values of [O/Fe] (Kraft \etal\ 1992,1993\nocite{kra92,kra93}), examined
whether the relative abundance of these species was consistent with CNO
tricycle processing.
Their cyanogen-band index S(3839) was shown to be anticorrelated with [C/Fe]
and [O/Fe], and correlated with [N/Fe]; they concluded that the strength
of the cyanogen band was driven largely by the abundance of N.
Further modeling led them to conclude that [C/Fe] and [O/Fe] were
correlated, whereas [N/Fe] remained anticorrelated with both [C/Fe]
and [O/Fe] (their Figure~5).
They also found that the total abundance C+N+O was, within
the errors, essentially the same for all stars in both M3 and M13.
These results support the view that the behavior of C, N, and O
is compatible with CNO tricycle processing.

\begin{figure}
\epsscale{0.95}
\plotone{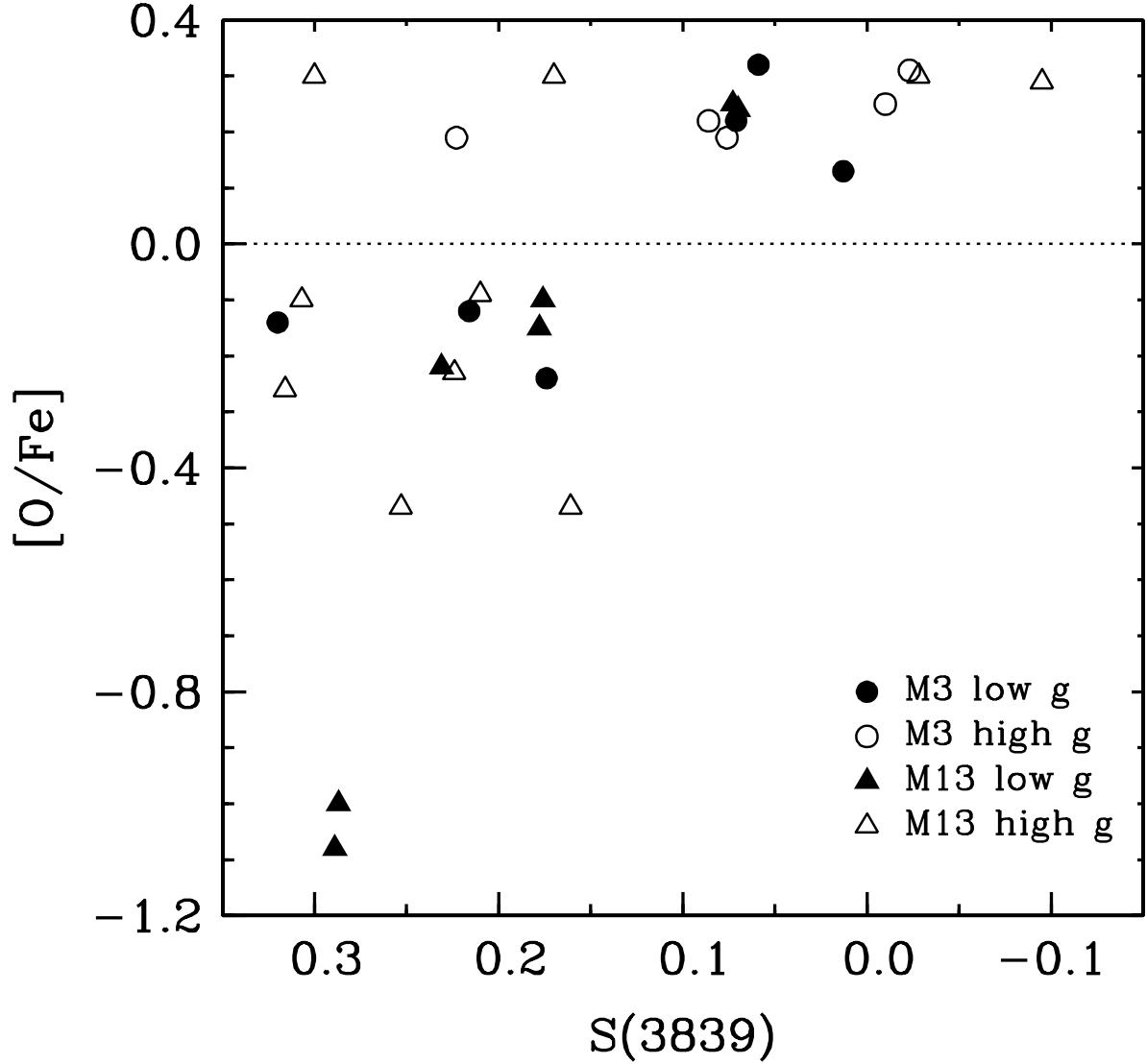}
\caption{
Correlation of the CN band strength index S(3839) and oxygen abundances
of M3 and M13 stars.
In this and all succeeding figures, M3 stars will be plotted as circles and
M13 stars as triangles;
the symbols will be filled ones for lower gravity stars (\logg~$\leq$~0.8, 
corresponding approximately to \MV0~$\leq$~--1.7), and open ones for
higher gravity stars.
Whenever possible, in this and succeeding figures dotted horizontal and
or vertical lines will be drawn to denote the solar abundance ratios
of various elements.
\label{f4}}
\end{figure}

The new M3 oxygen abundances reported here add only a small amount of
additional information, largely because the newly analyzed stars, lying
as they do mostly near the center of M3, are not the ones studied previously
for C and N abundances.
Included among stars with newly derived [O/Fe]$_{\rm II}$ ratios\footnote{
Hereafter in the text we employ the notation [O/Fe]$_{\rm II}$ to
emphasize that this ratio has been formed between abundances
derived from [\ion{O}{1}] and \ion{Fe}{2} lines.
The notation [Na, Mg, or Al/Fe]$_{\rm I}$ is used because these
ratios have been formed between neutral-species abundances of these
elements and \ion{Fe}{1} abundances.
Carbon abundances, taken without alteration from the literature,
are written simply as [C/Fe].
The ``I'' and ``II'' subscripts are not written in the figures.}
are I-21, IV-77, IV-101, and A.
Although they have known S(3839) indices, only I-21, IV-77, and IV-101
have freshly derived carbon abundances (Pilachowski \etal\
(2003).\nocite{pil03}
In addition, oxygen abundances for M3 giants previously analyzed by
Kraft \etal\ (1992,1993\nocite{kra92,kra93}) have been somewhat revised
in this paper.
In Figure~\ref{f4} we plot both the revised and newly derived abundances of
[O/Fe]$_{\rm II}$ against S(3839).
The anticorrelation is morphologically the same as that found previously
by Smith \etal\ (1996)\nocite{smi96}. 
The reader should note that M3 and M13 stars lie along the same locus, but 
in contrast with M13, there are no M3 stars with very low oxygen abundance.

\begin{figure}
\epsscale{0.95}
\plotone{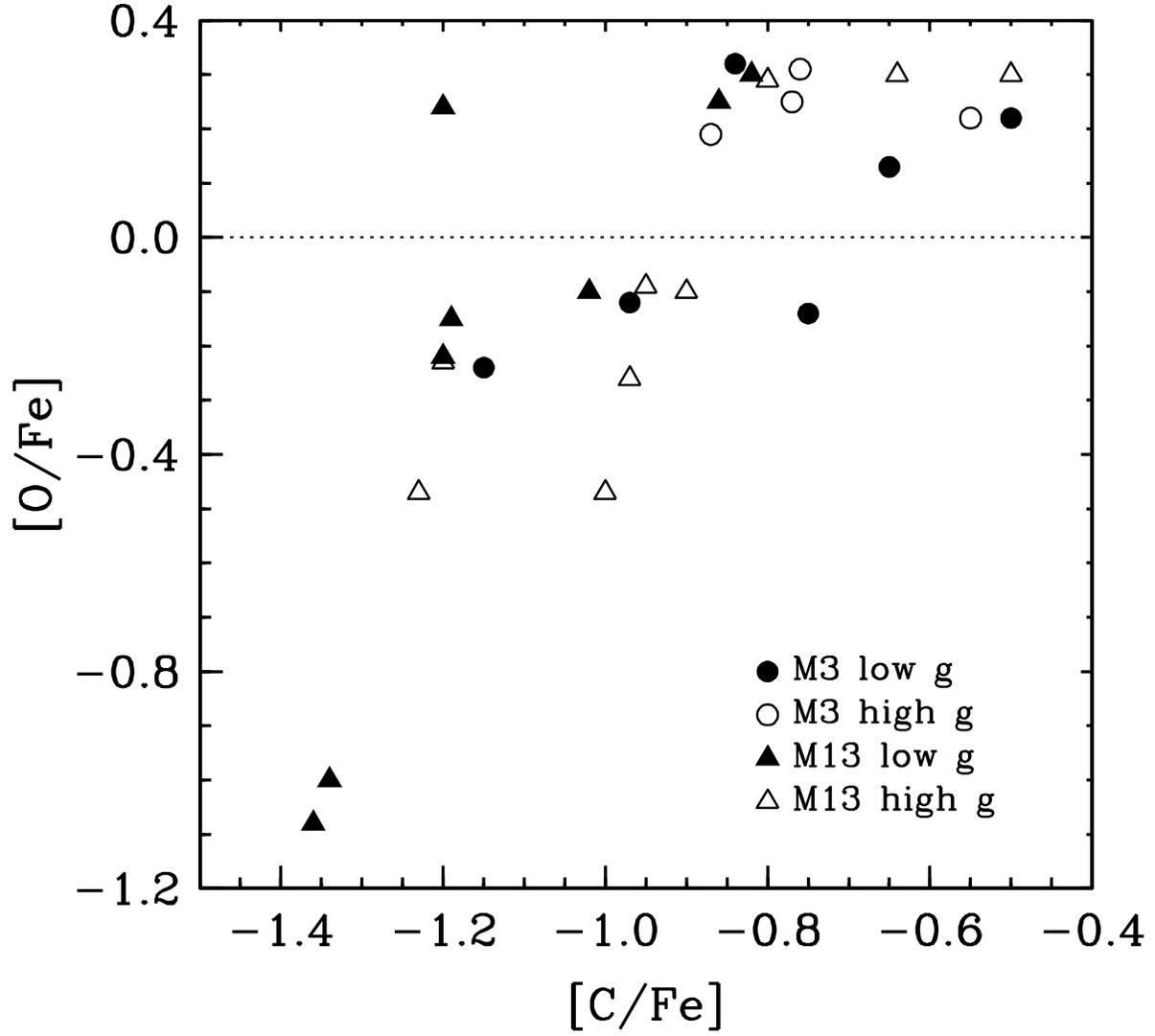}
\caption{
Correlation of the carbon and oxygen abundances of M3 and M13 stars.
Symbols and lines are as in Figure~\ref{f4}.
\label{f5}}
\end{figure}

In Figure~\ref{f5}, we plot also the oxygen versus carbon correlation,
following Smith \etal\ (1996)\nocite{smi96}, but with the revised oxygen
abundances tabulated here, plus the new carbon abundances for the three
M3 stars noted above.
The results are in the expected direction: whenever oxygen is depleted
via CNO processing, carbon also is depleted, and the abundances are
correlated.
There is a slight hint that at a given [O/Fe]$_{\rm II}$, 
[C/Fe] is slightly smaller in M13 than in M3.

\begin{figure}
\epsscale{0.95}
\plotone{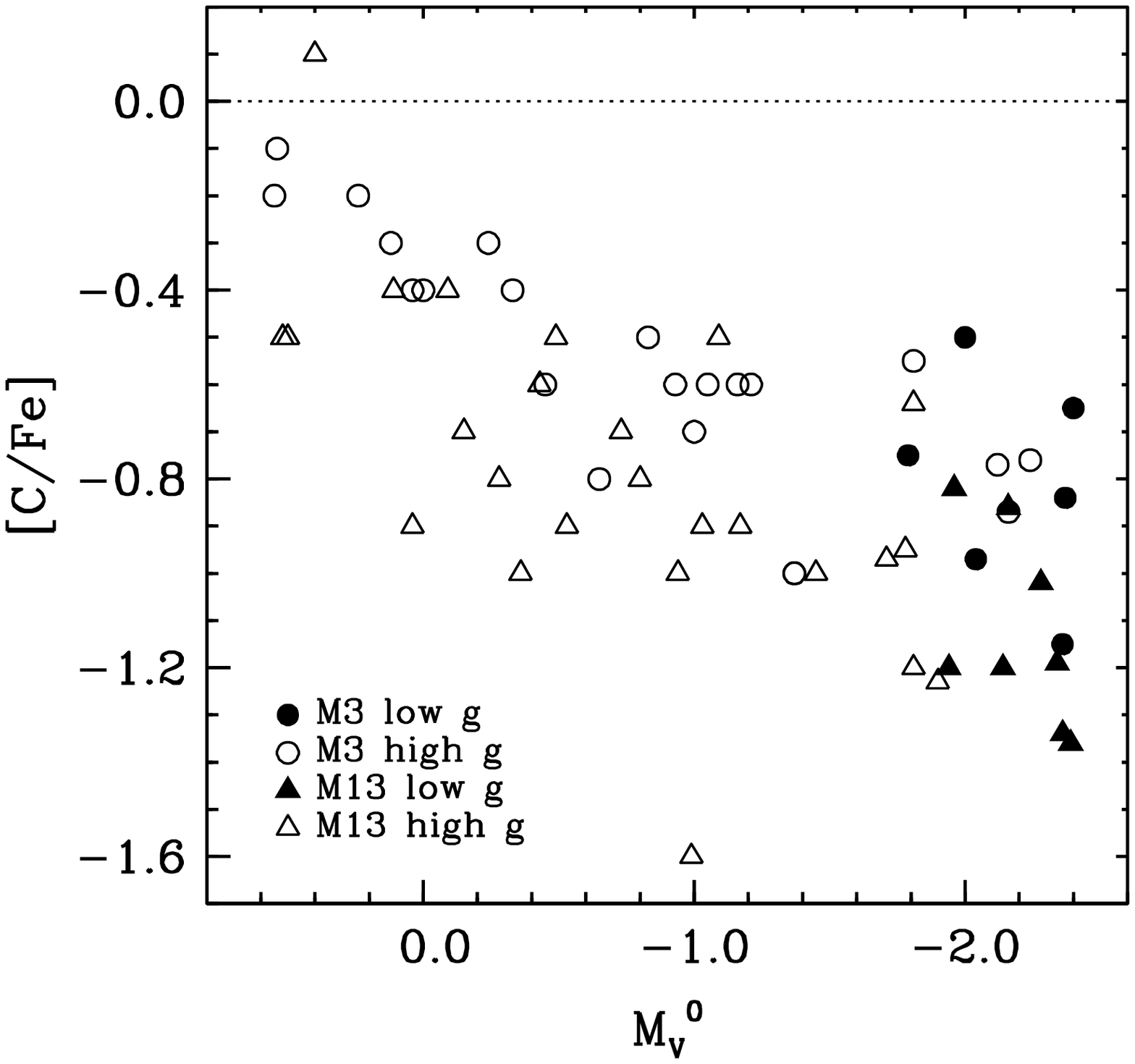}
\caption{
Correlation of the carbon abundances of M3 and M13 stars with their absolute 
magnitudes.
Symbols and lines are as in Figure~\ref{f4}.
\label{f6}}
\end{figure}

The conclusion that the relationship between C, N, and O abundances
is consistent with CNO tricycle processing does not, reveal the
site of the corresponding nucleosynthetic reshuffling.
However, Suntzeff's (1981)\nocite{sun81} early study showed that
$\rm\langle[C/Fe]\rangle$ declined with evolutionary state along the first giant
branches of both M3 and M13.
In the case of M3, this result was strengthened owing to a recent study by
Smith \etal\ (2002a).\nocite{smi02a}
These authors found that $\rm\langle[C/Fe]\rangle$ declines from $\sim$--0.2 at
$M_V$~$\sim$~+0.6 to $\sim$--1.1 at $M_V$~$\sim$~--2.3 (the red giant tip).
They also detected a slight offset in carbon abundances between CN-weak and  
CN-strong stars: $\rm\langle[C/Fe]\rangle$ was --0.14 lower in the CN-weak 
group.  Plotting Suntzeff's M13 data on top of the M3 data (Figure~\ref{f6}), 
we find a similar decline in $\rm\langle[C/Fe]\rangle$ (albeit with 
larger scatter) over the same luminosity range. 
Additionally, the M13 giants achieve even lower carbon abundances 
than do their M3 counterparts (by $\sim$0.2~dex).
Even more striking is the result of Briley \etal\ (2003).\nocite{bri03}
They found a range of [C/Fe] values from --0.8 to 0.0 among main 
sequence turnoff stars in M13, but at the same time found that 
$\rm\langle[C/Fe]\rangle$ declined from $\sim$--0.4 at the main sequence 
to the value of $\sim$--1.3 obtained earlier by Suntzeff for stars at 
the red giant tip.

Owing to stellar flux limitations, nothing is known of oxygen and nitrogen
abundances among main sequence members of either M3 or M13.
However, the spread in [C/Fe] among these stars in M13 suggests that
CNO processing must have occurred in material that was incorporated in
these stars before they achieved their present evolutionary state.
Nevertheless, there is no escaping the fact that, on the average, 
$\rm\langle[C/Fe]\rangle$ declines as M13 and M3 stars evolve above 
$M_V$~$\sim$~+0.6, the luminosity level at which the hydrogen shell 
burns through the molecular weight discontinuity left behind from 
the deepest penetration of the outer convection zone 
(the so-called red giant ``bump'').

Among halo field giants, $\rm\langle[C/Fe]\rangle$ also begins to decline at 
the same red giant ``bump'' luminosity (Gratton \etal\ 2000\nocite{gra00}),
but contrary to the situation in M3 and M13, the degree of carbon depletion
at its most severe does not exceed --0.7, and averages about --0.5.
This limit is strikingly similar to the highest [C/Fe] ratios seen
among giants in M3 and M13 (see Figure~\ref{f5}).
It is therefore perhaps not surprising to find that, among halo field giants,
Gratton \etal\ (2000) found no change in [O/Fe] with advancing
evolutionary state, fixed at a value of $\rm\langle[O/Fe]\rangle$~=~+0.34.

\begin{figure}
\epsscale{0.95}
\plotone{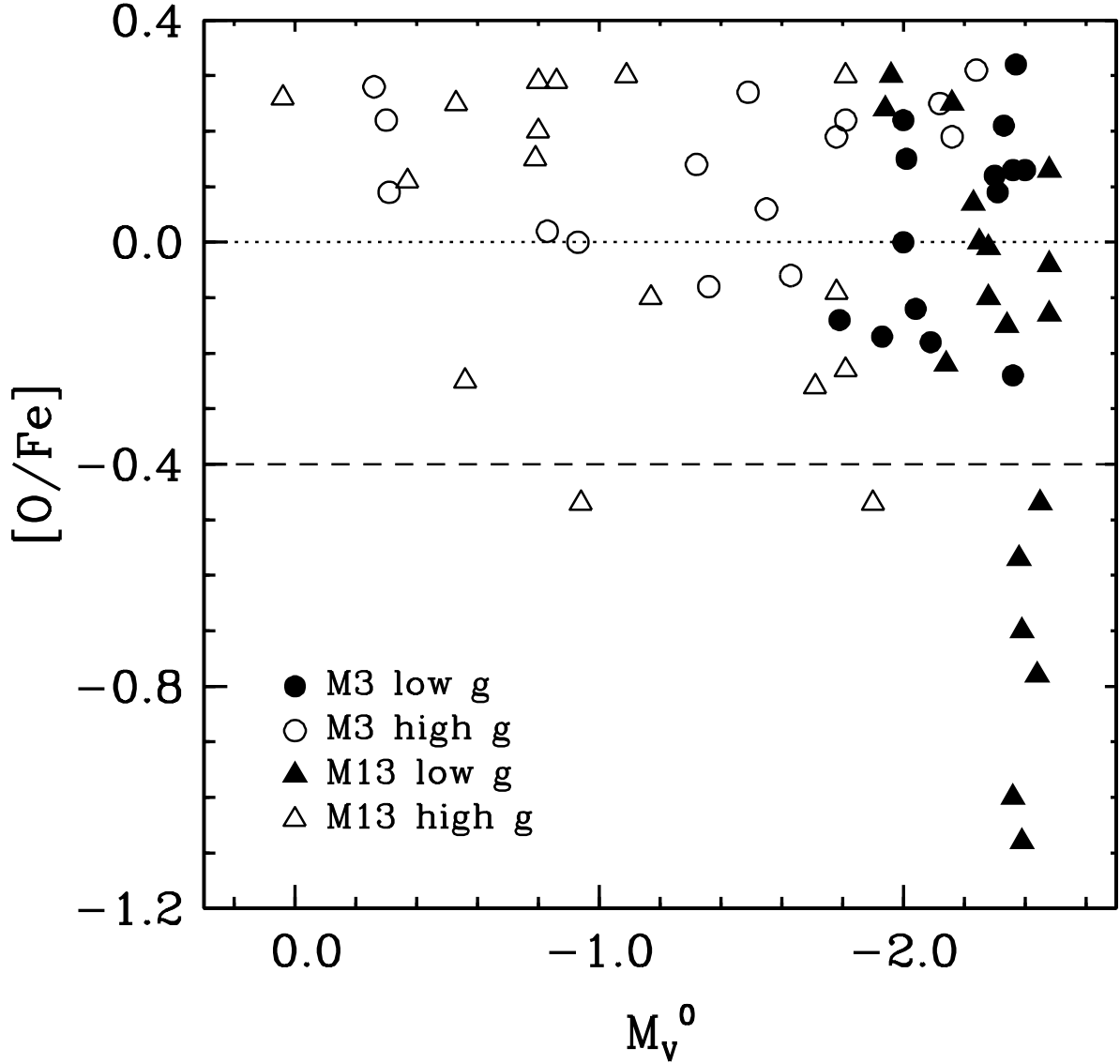}
\caption{
Correlation of the oxygen abundances of M3 and M13 stars with their absolute
magnitudes.
Symbols and the dotted line are as in Figure~\ref{f4}.
The dashed horizontal line represents the dividing line between
those stars designated as ``oxygen-poor'' and those we call 
``super oxygen-poor''.
\label{f7}}
\end{figure}

In Figure~\ref{f7} we illustrate [O/Fe]$_{\rm II}$ as a function of 
\MV0\ in M3 (Tables~4 and 6) and M13 (Table~7).
Our samples contain no stars with \MV0~$\gtrsim$~+0.1, so we cannot test
the behavior of [O/Fe]$_{\rm II}$ at luminosities below that of the 
red giant bump.   
We can describe the distribution of the stars brighter than \MV0~=~+0.1,
by dividing them into two groups near log~$g$~=~0.8 (\MV0~$\sim$~--1.7),
and by introducing three levels of oxygen ``deficiency''.
Typically among halo field giants near [Fe/H]~=~--1.5, [O/Fe]$_{\rm II}$ 
lies in the range 0.0 to +0.5 (average near +0.3), and we speak of these as
``oxygen-normal''.
Those in the clusters with 0.0~$>$~[O/Fe]$_{\rm II}$~$>$~--0.4, we call 
``oxygen-poor''.
Finally we refer to those in the range --0.4~$>$~[O/Fe]$_{\rm II}$~$>$~--1.2 
as ``super oxygen-poor''.

Of the 11 M3 stars having \MV0~$>$~--1.7, nine (82\%) are O-normal
and two (18\%) are O-poor; there are no super-O-poor stars in this range.
The absence of super-O-poor stars is also a characteristic of the 16
brighter M3 stars. 
The split among normal and O-poor stars in this group is 72\% to 28\%,
respectively, roughly similar to the lower luminosity group.
(We arbitrarily assigned one star, B.4.2, having 
[O/Fe]$_{\rm II}$~=~0.00 to the ``normal'' group.)
The situation in M13 is not so different for the 11 stars in the fainter 
group: 73\% are O-normal, 18\% are O-poor, and one star (I-12) is 
super-O-poor.
But among the stars of the brighter group, the situation is dramatically
different. 
Of the 23 brighter stars, only seven (30\%) are O-normal, nine 
(39\%) are O-poor, and seven (30\%) are super-O-poor.
(Again we arbitrarily assign one star, L261, having 
[O/Fe]$_{\rm II}$~=~0.00 to the ``normal'' group).
Six of the nine super-O-poor stars are brighter than \MV0~=~--2.3, and
are therefore essentially at the tip of the giant branch.

We summarize the results of this section as follows:

\begin{enumerate}

\item Metal-poor halo stars, whether in the field or in such clusters as
M3 and M13, all show a decline in $\rm\langle[C/Fe]\rangle$ starting 
when stars evolve through the red giant bump.    
This is superimposed on top of a pre-existing spread in [C/Fe] exhibited
among main sequence and turnoff stars at least in M13, and one would not
be surprised if this were true also in M3 (although this is not yet
observationally demonstrated).
A spread in [C/Fe] is found also among field halo subdwarfs (Gratton
\etal\ 2000\nocite{gra00}), but it appears to be smaller than that found
among main sequence and turnoff stars in M13.

\item The decline is driven by stellar evolution and presumably results
from the deep mixing of envelope material through the region just ahead
of the H-burning shell where conversion of C to N takes place.
The deficiency of $\rm\langle[C/Fe]\rangle$ among halo giants is typically 
--0.5 dex, but is larger in clusters, reaching --0.7 to --0.9~dex 
in M3 and M13.

\item We divide (somewhat arbitrarily) our sample of giants into three
groups depending on the degree of oxygen deficiency: O-normal stars
have [O/Fe]$_{\rm II}$~$>$~0.0, O-poor stars lie in the range 
0.0~$>$~[O/Fe]$_{\rm II}$~$>$~--0.4,
and super-O-poor stars have [O/Fe]$_{\rm II}$~$<$--0.4.
Among halo field stars, there are no O-poor or super-O-poor objects, and
giants appear to have on the average the same oxygen abundances as main
sequence subdwarfs (see also Gratton \etal\ 2000\nocite{gra00}).
In M3 there are no super-O-poor stars, and of the 28 giants in the sample,
about one-third are O-poor.
If this is attributable to O to N conversion in the material making up the
envelope of some giants in M3 (either pre-existing processing or processing
within the stars themselves), we conclude that both the degree and the
frequency of the effect is rather mild.
The situation in M13 is dramatically different: fewer than half the giants
are O-normal, and of the remaining, 30\% are super-O-poor.
Most of these lie near the red giant tip.
Indeed of the giants brighter than \MV0~=~--2.3, 60\% are super O-poor.
It is hard to escape the conclusion that the majority of the brightest
M13 giants mix envelope material deep enough into the H-burning shell
that significant processing of O to N occurs within the stars themselves.

\end{enumerate}

\subsection{Interrelationships Among the Light Elements: O, Na, Mg, Al}

\subsubsection{The Na versus O Anticorrelation}

\begin{figure}
\epsscale{0.95}
\plotone{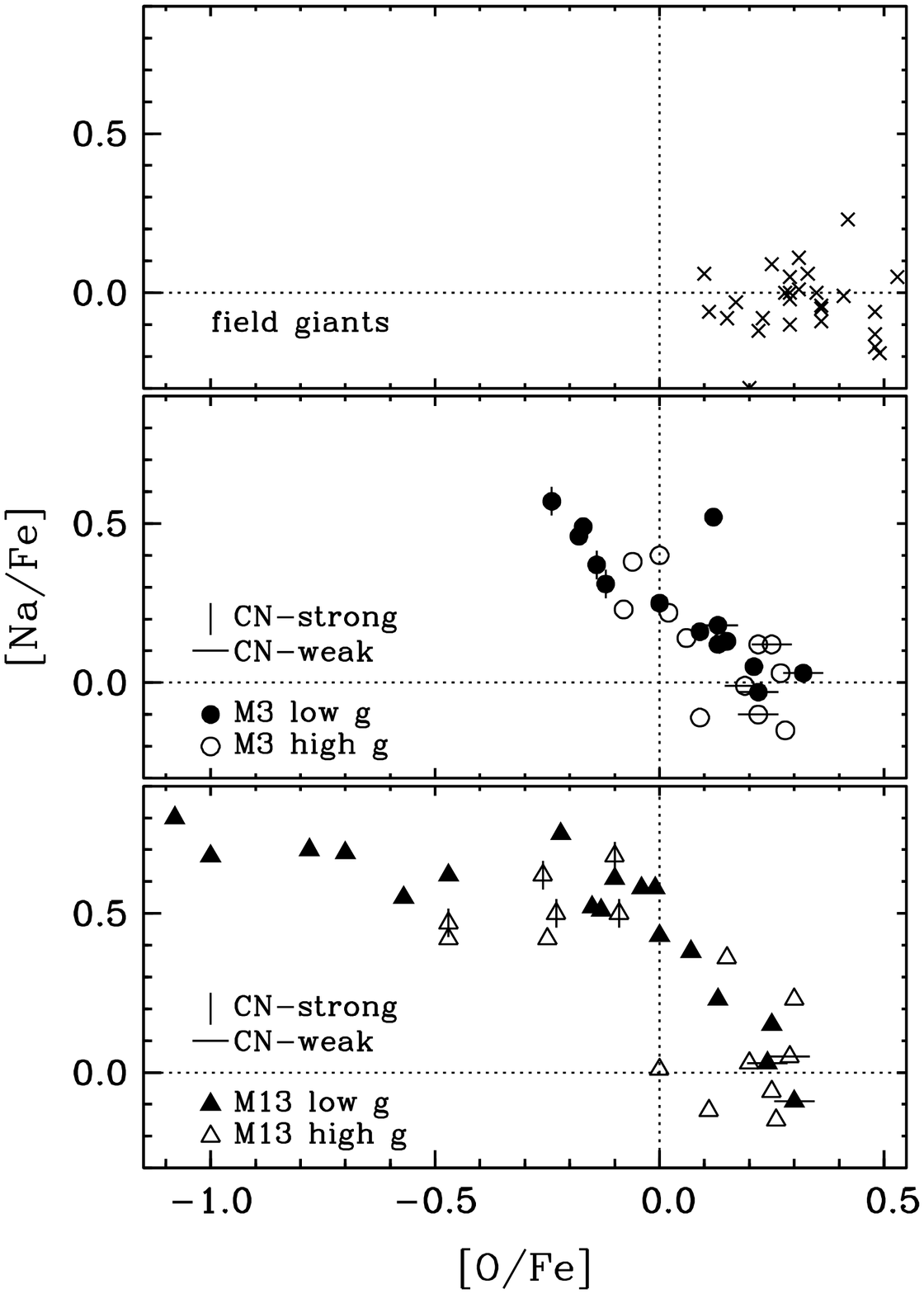}
\caption{
Correlation of the sodium and oxygen abundances in field giants
(top panel, ``$\times$'' symbols), M3 giants (middle panel), and M13 giants
(bottom panel).
The dotted lines are as in Figure~\ref{f4}, and so are the symbols
for M3 and M13 stars.
Vertical and horizontal lines through the M3 and M13 points denote
CN-strong and CN-weak stars, respectively.
\label{f8}}
\end{figure}

Giants in most low metallicity globular clusters exhibit an anticorrelation
of [Na/Fe] and [O/Fe], and M3 is no exception.
In Figure~\ref{f8}, we compare our samples of giants in M3 (Tables~4 and 6)
with those of M13 (Table~7), as well as giants of the halo field
(Shetrone 1996\nocite{she96a}, Gratton \etal\ 2000), the last-named group
consisting of stars with [Fe/H]~$<$~--1.0.
Small vertical and horizontal lines drawn through the points denote
stars known to be CN-strong or CN-weak, respectively 
(Smith 2002a,b\nocite{smi02a,smi02b}); most field halo giants are
CN-weak (Hesser \etal\ 1977\nocite{hes77}; Langer \etal\ 1992\nocite{lan92}).
In contrast with earlier plots of this kind (\eg, Kraft 1994\nocite{kra94},
his Figure~8), we have included the nLTE Na corrections recommended by
Gratton \etal\ (2002), which have the effect of enhancing values of [Na/Fe]
with increasing luminosity.
We also normalize [O/Fe] to [Fe/H] based on \ion{Fe}{2} and [Na/Fe] to
[Fe/H] based on \ion{Fe}{1}, in accordance with precepts argued by Kraft \&
Ivans (2003)\nocite{kra03}.  Inspection of the three panels leads us to
the following conclusions.

\begin{enumerate}

\item Halo field giants show no sign of a Na versus O anticorrelation.
There is a considerable range of values of [Na/Fe]$_{\rm I}$ 
(--0.5 to +0.3), and a smaller range in [O/Fe]$_{\rm II}$ (0.0 to +0.6), 
a result noted earlier by Kraft \etal\ (1993), Hanson \etal\ (1998), 
and Gratton \etal\ (2000).  
Mean values of [O/Fe]$_{\rm II}$~= +0.3 and [Na/Fe]$_{\rm I}$~=~--0.1 are
generally regarded as numbers characteristic of Type~II supernovae ejecta.

\item M13 giants exhibit a range of [Na/Fe]$_{\rm I}$ (a factor of 
$\sim$8) and a range of [O/Fe]$_{\rm II}$ (a factor of $\sim$25) 
that is larger than that seen in any other cluster (although the 
range in $\omega$~Cen may be comparable, see Norris \& DaCosta 
1995\nocite{nor95}), and the anticorrelation is decidedly marked.
Na and O abundances in the most Na-poor and O-rich giants of M13 are
comparable to what is seen in halo field giants.
As expected, these stars are CN-weak; CN-strong stars are those with
deficiencies of O.
In the present sample, there is a tendency for giants that are less
advanced in evolutionary state (\logg~$>$ 0.8) to favor higher
[O/Fe]$_{\rm II}$ and lower [Na/Fe]$_{\rm I}$ than are found in stars
nearer the red giant tip (\logg~$<$ 0.8).

\item M3 giants exhibit an anticorrelation that is ``intermediate'' 
between the halo field and M13 stars.
M3 stars follow quite closely the anticorrelation defined by the M13
stars, but ``cutoff'' near [O/Fe]$_{\rm II}$~= --0.2, 
[Na/Fe]$_{\rm I}$~= +0.5.
There are no super-O-poor stars in M3, as already noted.
It is also the case that the distribution of stars in the diagram is
essentially independent of evolutionary state, in contrast to the situation
in M13.
The shape and boundaries of the M3 Na versus O distribution are, in fact,
quite similar to those found recently in NGC~6752 (Yong \etal\
2003\nocite{yon03}), a point to which we will return in Section \S 4.1.1.

\end{enumerate}

\subsubsection{[Mg/Fe]$_{\rm I}$ versus [O/Fe]$_{\rm II}$}

\begin{figure}
\epsscale{0.95}
\plotone{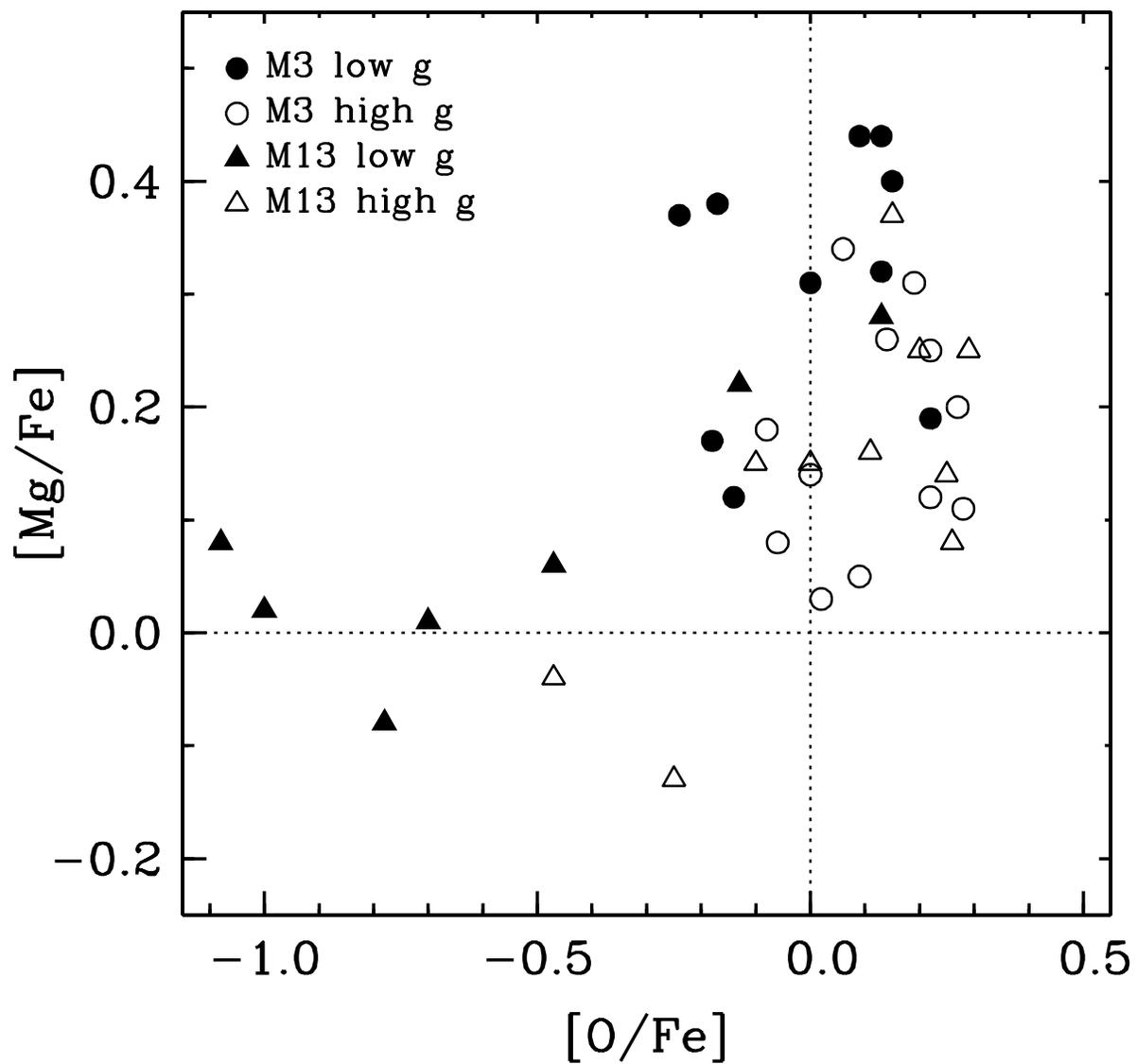}
\caption{
Correlation of the oxygen and magnesium abundances of M3 and M13 stars.
Symbols and lines are as in Figure~\ref{f4}.
\label{f9}}
\end{figure}

Previous work (Figure~9 of Kraft \etal\ 1997\nocite{kra97}) indicated 
that M13 giants with lower than average [O/Fe] also had lower 
than average [Mg/Fe]. 
In the present Figure~\ref{f9}, we plot [Mg/Fe]$_{\rm I}$ versus 
[O/Fe]$_{\rm II}$ for the sample of M13 giants of that study observed 
with either the Keck~I HIRES spectrograph or the post-1995 version of 
the Lick Hamilton spectrograph. 
As before, we find $\rm\langle[Mg/Fe]_I\rangle$~=~+0.22 ($\sigma$~=~$\pm$~0.09) 
for giants with [O/Fe]$_{\rm II}$~$>$~--0.2 whereas 
$\rm\langle[Mg/Fe]_I\rangle$~=~--0.01 ($\sigma$~=~0.07) for giants with 
[O/Fe]$_{\rm II}$~$>$~--0.2.
In Figure~\ref{f9}, we superimpose the O and Mg abundances for 
22 of the 23 giants in our M3 sample (Table~4).  
One star has [O/Fe]$_{\rm II}$~$>$~--0.2, and it does not have a low Mg 
abundance; on the other hand, the one M3 giant (B4.5) having a very low, 
indeed negative, value of [Mg/Fe]$_{\rm I}$ does not have a derivable 
oxygen abundance (the S/N of the spectrum is too low).
However, it is clearly the case that our M3 sample contains few, if any,
stars with the low Mg abundances characteristic of many of the brightest
giants in M13. 
We address later implications of the change in the abundance ratio of Mg 
isotopes with [O/Fe]$_{\rm II}$ among giants in M13 (Shetrone 1996b).

\subsubsection{[Al/Fe]$_{\rm I}$ versus [Mg/Fe]$_{\rm I}$}

\begin{figure}
\epsscale{0.95}
\plotone{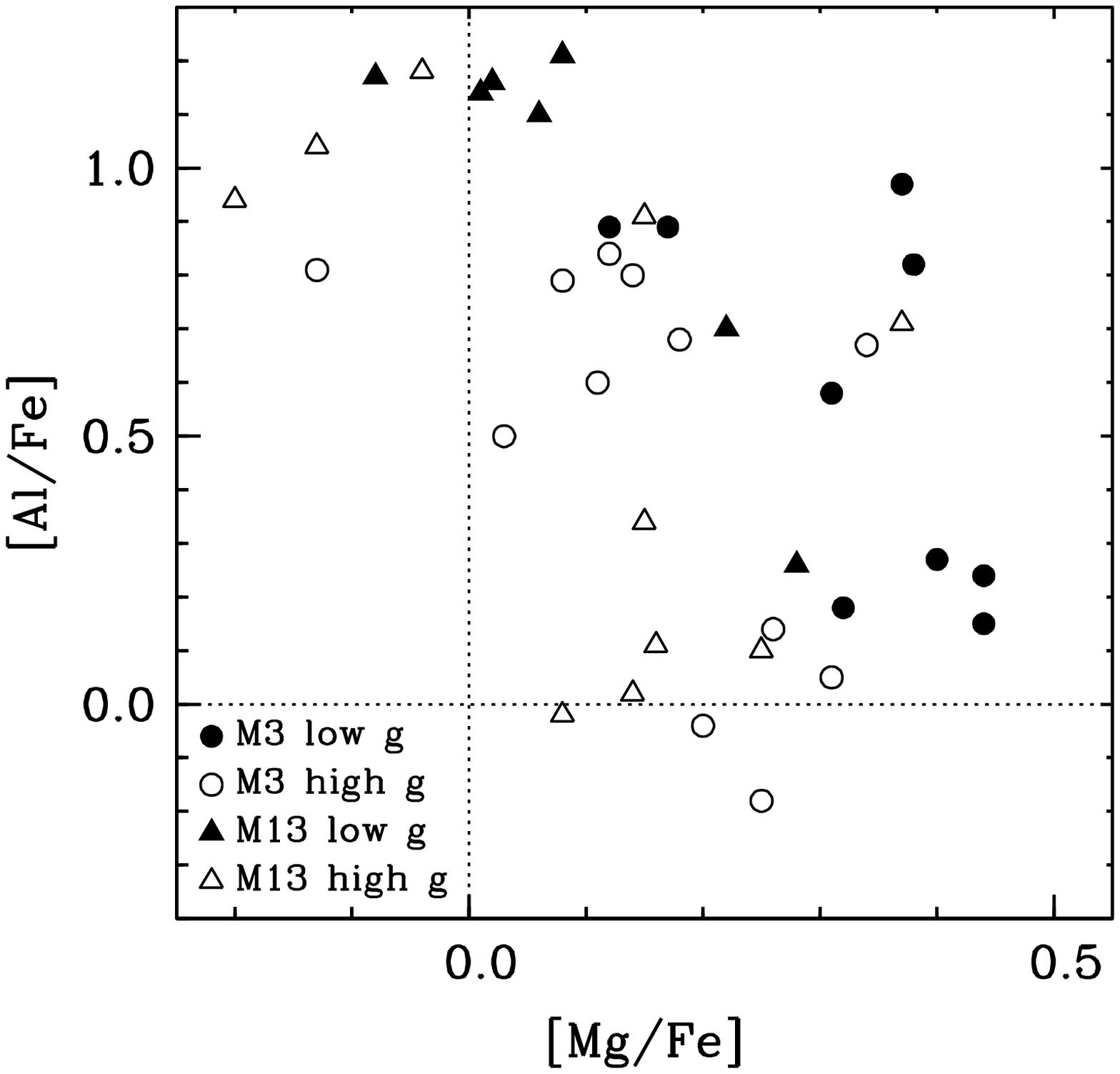}
\caption{
Correlation of the magnesium and aluminum abundances of M3 and M13 stars.
Symbols and lines are as in Figure~\ref{f4}.
\label{f10}}
\end{figure}

The expectation that Al may arise owing to proton captures on the
various isotopes of Mg (\eg, Kraft \etal\ 1997\nocite{kra97}, 
Shetrone 1996b\nocite{she96b}, Ivans \etal\ 1999\nocite{iva99}) 
prompts an investigation of [Al/Fe]$_{\rm I}$ versus [Mg/Fe]$_{\rm I}$
which is shown in Figure~\ref{f10}.
All data are taken from Tables~4 and~7. 
We divide the stars at \logg~=~+0.8 into two groups by evolutionary 
state, as we did previously in the discussion of the Na versus O 
anticorrelation (\S 4.2).
Although the sample sizes are small, the distribution suggests that in 
each cluster there are two groups with distinctly differing Al abundances. 
Whereas the higher and lower luminosity giants are distributed among the 
two groups about equally in the case of M3, the distribution appears to 
be quite different in the case of M13, with most of the brighter giants 
lying in the group with high Al abundances. 
The highest Al abundances in M13 also appear to be about a factor of two 
higher than the highest Al abundances in M3.

Generally speaking, the observational accuracy of our Al abundances, which 
are based on two lines of moderate EW and excitation potential, is likely 
higher than that of our Mg abundances, which are based on one or two rather 
strong lines of rather high excitation potential. 
This may in part account for the rather displeasing result that the M3 
giants of higher luminosity seems to have [Mg/Fe]$_{\rm I}$ ratios that are on 
average about 0.1~dex higher than the stars of lower luminosity (omitting 
the anomalous star (B4.5) with a negative [Mg/Fe]$_{\rm I}$ ratio). 
The result suggests the presence of a systematic error in our adopted 
\teff\ or \logg\ scales, as a function of luminosity.  
If an arbitrary adjustment to correct the effect in M3 were also applied
to the sample of stars in M13, it would strengthen the conclusion that
the more luminous M13 giants have a lower Mg abundance than the less
luminous stars. However, the same conclusion would be reached even if
such a correction were not applied to the M13 sample.

\subsubsection{[Al/Fe]$_{\rm I}$ versus [Na/Fe]$_{\rm I}$}

\begin{figure}
\epsscale{0.95}
\plotone{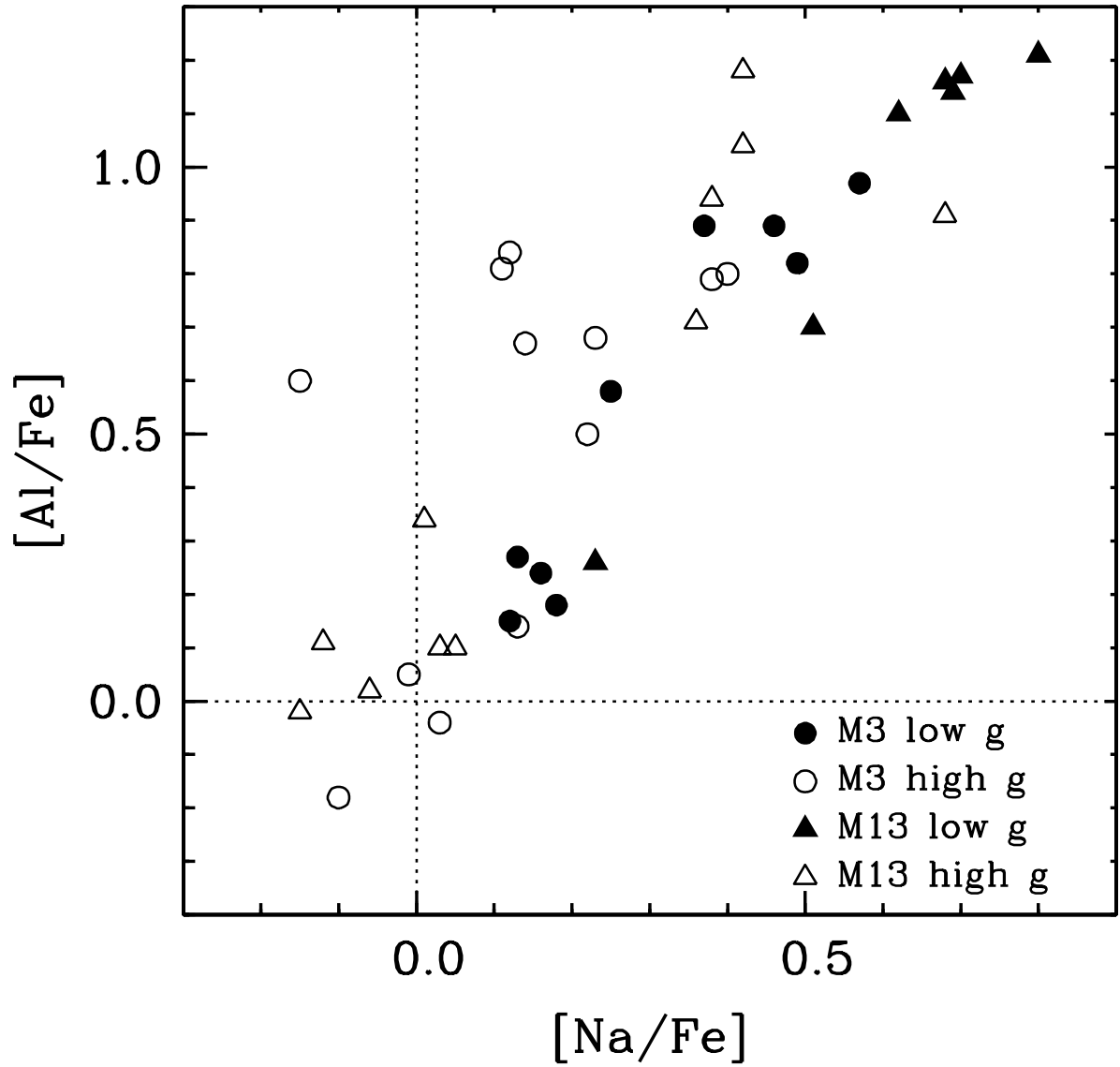}
\caption{
Correlation of the sodium and aluminum abundances of M3 and M13 stars.
Symbols and lines are as in Figure~\ref{f4}.
\label{f11}}
\end{figure}

As expected, these two ``odd-elements'' are correlated in both M3 and M13, 
but in each case, there appear to be two well-separated groups 
(Figure~\ref{f11}). 
M3 and M13 giants satisfy essentially the same correlation, but the M13 giants 
extend toward higher values of both Na and Al by a factor of about two. 
The population of M13 stars having the highest values of Na and Al 
abundances is dominated by stars in the most advanced evolutionary state 
(\logg~$<$~0.8). 
All halo field giants considered here lie among the group of stars 
with lowest Na and Al abundances.

\subsubsection{``Odd'' versus ``Even'': [Na/Fe]$_{\rm I}$ versus 
[Mg/Fe]$_{\rm I}$}

\begin{figure}
\epsscale{0.95}
\plotone{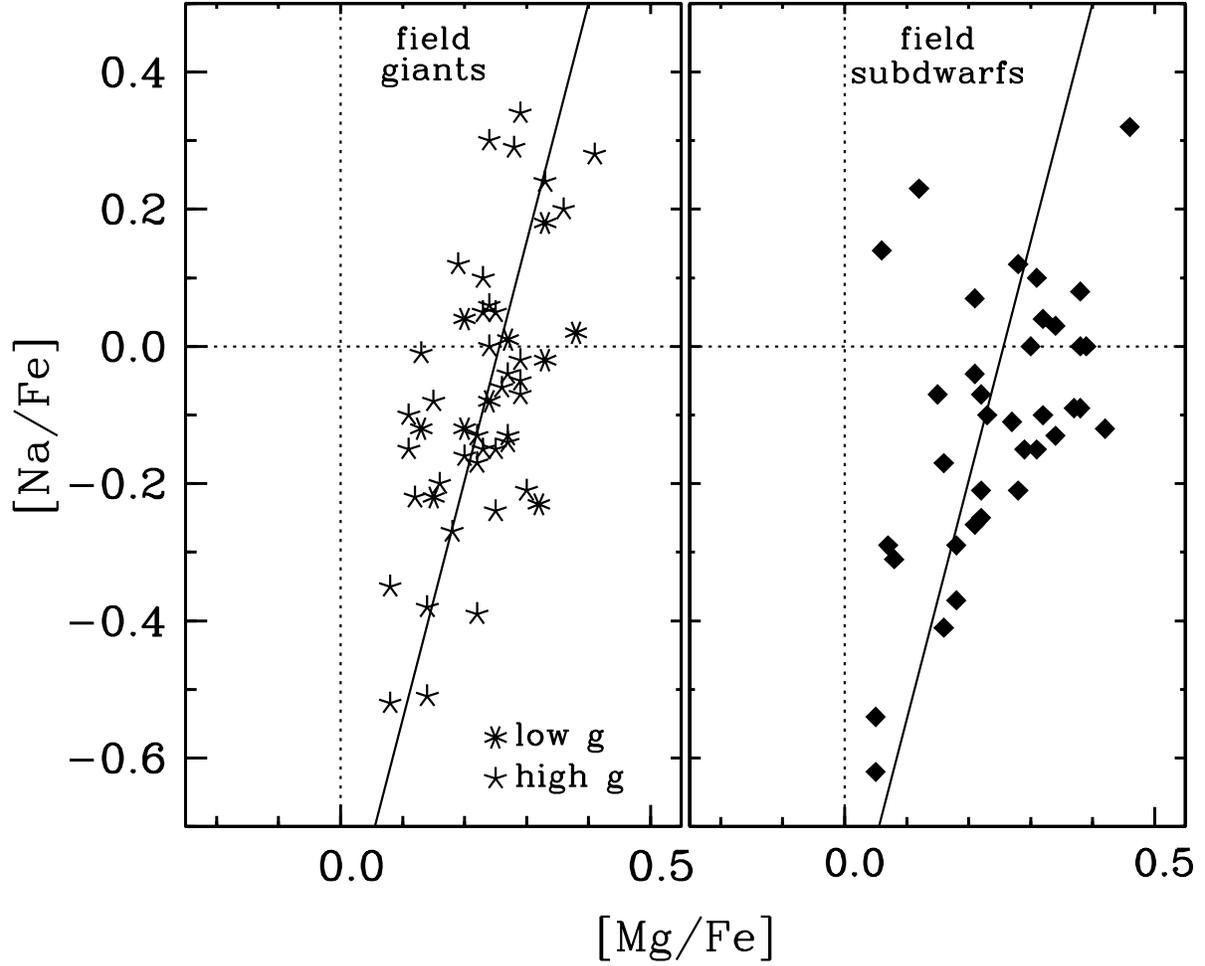}
\caption{
Correlation of the sodium and magnesium abundances of field giants (left-hand
panel), and field subdwarfs (right-hand panel).
Different symbols are used to represent low gravity (\logg~$\leq$~0.8)
and high gravity, while one symbol represents all subdwarfs.
The dotted lines are as in Figure~\ref{f4}.  
The slanting solid lines represent the mean trend of the field giant
data.
\label{f12}}
\end{figure}

Hanson \etal\ (1998)\nocite{han98}, using the the observational 
results of Pilachowski \etal\ (1996)\nocite{pil96a}, noted that there 
exists a Na versus Mg correlation among halo field giants having 
[Fe/H]~$<$~--1.0.
In Figure~\ref{f12} (left-hand panel), we plot [Na/Fe]$_{\rm I}$ versus 
[Mg/Fe]$_{\rm I}$ in these giants,
the data having been ``corrected'' for the nLTE effect in Na. 
In the right hand panel of Figure~\ref{f12}, we plot the results
of Fulbright (2002)\nocite{ful02} for halo field subdwarfs, the straight 
line taken from a fit (by eye) of the correlation for the halo giants. 
The Mg abundances of Fulbright have been modified, first to eliminate all
Mg lines except $\lambda$5528 and/or $\lambda$5711 (the lines used
exclusively for studies of the giants), and second to allow for
differences in the adopted log~$gf$ values assumed by Fulbright for
the 5528/5711 pair. 
The two panels should therefore be on the same system of Mg and Na abundances.

Inspection of the two panels of Figure~\ref{f12} suggests that, (1) there 
is no change in the correlation whether the samples are drawn from the
main sequence or from the giants, and (2) there is no change in the
distribution whether one deals with giants having gravities greater or
less than \logg~=~0.8. 
In other words, there is no dependence of the correlation on evolutionary 
state among metal-poor halo field stars.

Sodium and magnesium are near neighbors in the periodic table and
are believed to arise during carbon and neon burning in Type~II SNe, 
giving rise to the so-called ``odd-even'' effect. 
The correlation suggests that there is a range in the [Na/Mg] ratio at a given
metallicity, since all stars regardless of metallicity below [Fe/H]~=~--1.0 
seem to satisfy the same correlation. 
Thus along the correlation, the [Na/Mg] ratio varies from --0.6 to -0.1, 
with a mean near --0.4; this is in quite good agreement with Arnett's 
(1971) hydrostatic and explosive carbon-burning models, in which 
[Na/Mg] ranges from --0.8 to --0.2 for stars having [Fe/H]~$<$~--1.0. 
The more recent models reported by Timmes \etal\ (1995)\nocite{tim95} suggest 
a lower mean [Na/Mg] ratio, near --0.7, for such low-metallicity stars.
We note that the stars with [Na/Mg] ratios closest to the theoretical
value of Timmes \etal\ are the one shown by Fulbright (2002)\nocite{ful02} 
to favor Galactic orbits having relatively high values of orbital energy
and (absolute) angular momentum.

\begin{figure}
\epsscale{0.95}
\plotone{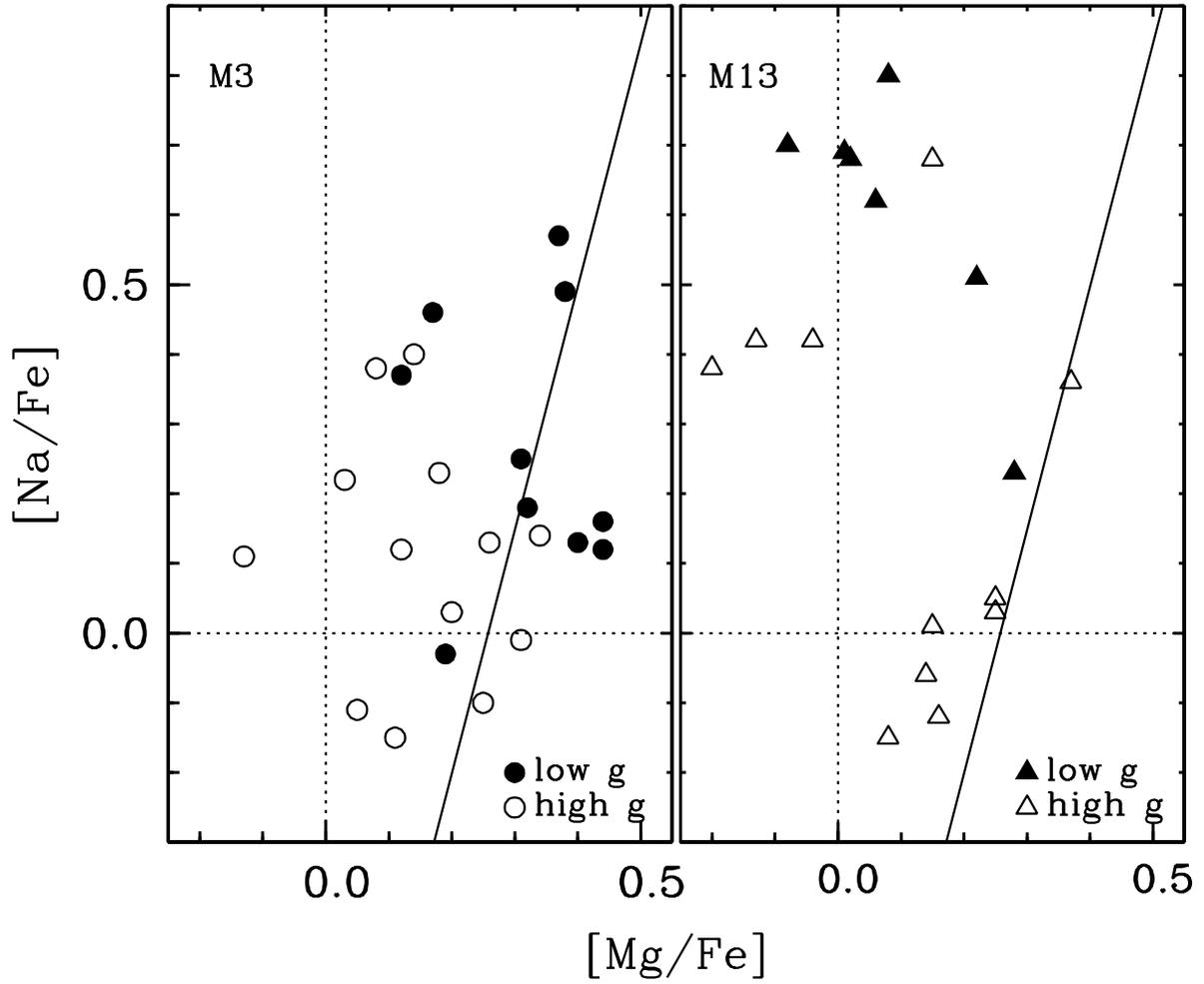}
\caption{
Correlation of the sodium and magnesium abundances of M3 and M13 stars.
Symbols and dotted lines are as in Figure~\ref{f4}.
The slanting solid lines are repeated from Figure~\ref{f12}.
\label{f13}}
\end{figure}

Corresponding plots of [Na/Fe]$_{\rm I}$ versus [Mg/Fe]$_{\rm I}$ for 
M3 and M13 are shown respectively in the left and right hand panels 
of Figure~\ref{f13}, the data being taken from Tables~4, 6, and 7. 
The straight line is the same as that in the two panels of Figure~\ref{f12}. 
Again we see the somewhat unpleasing result that in M3 the stars of low 
luminosity have systematically lower Mg abundances than the stars of higher 
luminosity (\logg~$<$~0.8), but otherwise the two kinds of giants are 
essentially intermixed. 
The giants of M13 present a rather different picture.  
Here the more luminous giants prefer a location well off the halo-defined 
line, in the direction of low Mg, high Na, whereas the lower luminosity 
giants, except for three exceptions (out of a sample of 10 stars) prefer 
a location near the normal halo line.

\begin{figure}
\epsscale{0.95}
\plotone{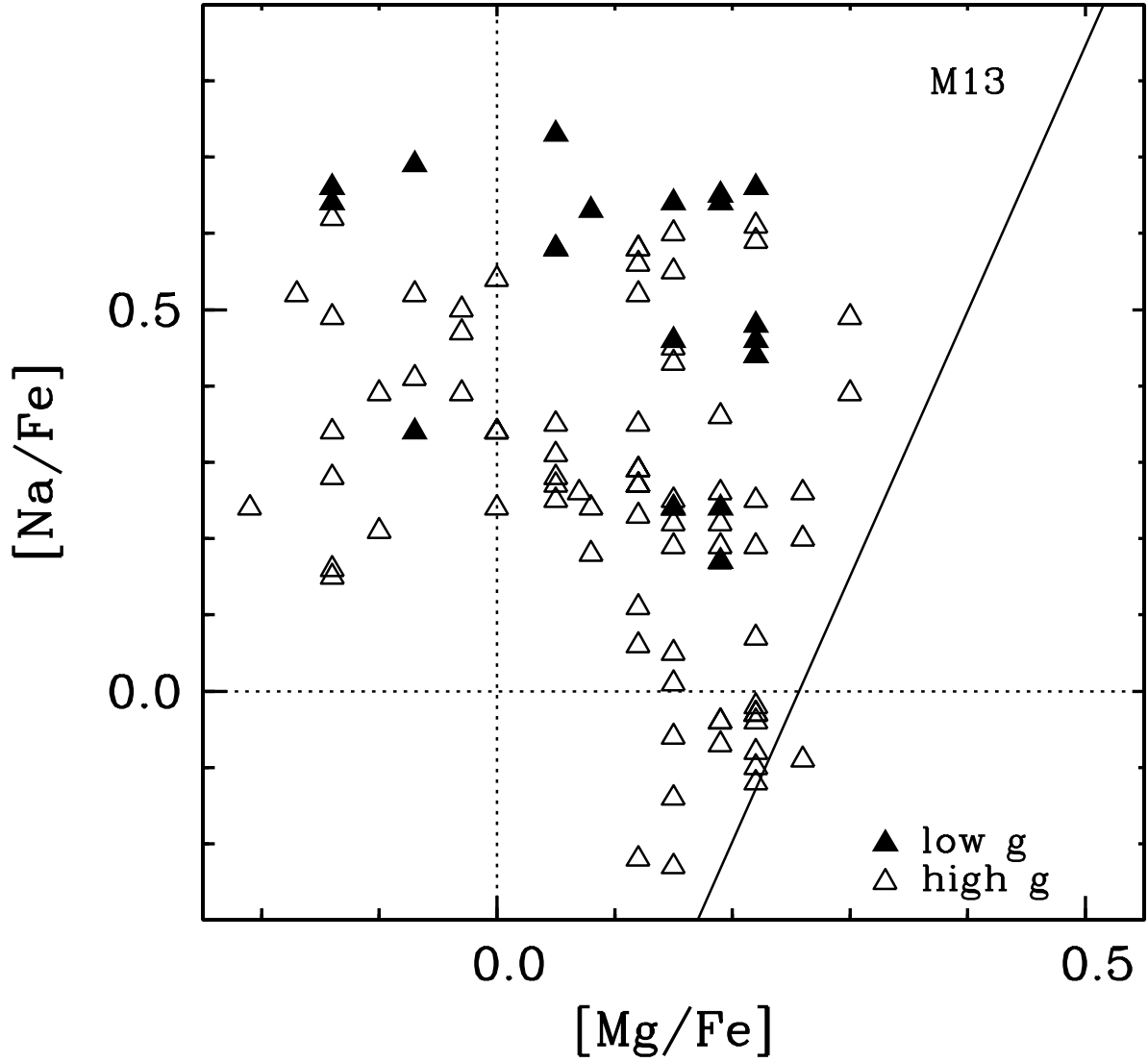}
\caption{
Correlation of the sodium and magnesium abundances based on the Pilachowski 
\etal\ (1996) sample of M13 giants. 
The diagram is morphologically the same as the right-hand panel of 
Figure~\ref{f13}. 
The slanting solid line is that of Figure~\ref{f12}.
\label{f14}}
\end{figure}

The effect is illustrated more dramatically in Figure~\ref{f14}, in
which [Na/Fe]$_{\rm I}$ is plotted against [Mg/Fe]$_{\rm I}$ for M13, 
this time using the very much larger sample of M13 giants studied 
by Pilachowski \etal\ (1996b)\nocite{pil96b}. 
The Mg abundances were adjusted slightly to conform to the system of Mg 
abundances adopted here (by plotting [Mg/Fe]$_{\rm I}$ values of
Table~7 against the [Mg/Fe] values of Pilachowski \etal\ for common
M13 giants), and the Na abundances adjusted for nLTE as explained earlier. 
Again we see that M13 giants of high luminosity are confined
to the upper left-hand part of the diagram, although they do not have
that domain to the exclusion of stars having \logg~$>$~0.8.

\subsection{Implications of the Interrelationships among O, Na, Mg, and Al}

\subsubsection{Deep Mixing versus Primordial (or Pollution) Scenarios}

In all of the interrelationships illustrated in \S 4.3, M13
giants inhabit a more ``extreme'' domain than do M3 giants, when the stars
are compared at essentially the same evolutionary state. 
As discussed in \S 4.3.1, most M13 giants near the red giant tip have, 
in comparison with corresponding M3 giants, very low O and very high Na 
abundances, suggesting that first ascent M13 giants immediately prior to 
the He core flash undergo a period of deep mixing not experienced by 
similar giants in M3. 
But a recent study of giants in NGC 6752 (Yong \etal\ 2003\nocite{yon03}), 
a cluster having [Fe/H] essentially the same as that of M3 and M13, casts 
doubt on such an interpretation. 
Thus, in what follows, we describe similarities and differences 
among giants in NGC 6752, M13 and M3.

From a sample of 20 (mostly) first ascent NGC~6752 giants with
--1.0~$\leq$~$M_{\rm V}$~$\leq$~--2.5, Yong \etal\ (2003)\nocite{yon03} 
discovered that the abundances of the minor isotopes \iso{25}{Mg} and 
\iso{26}{Mg}, relative to the dominant \iso{24}{Mg}, were ofter higher 
than in the solar system (\iso{24}{Mg}:\iso{25}{Mg}:\iso{26}{Mg}~= 79:10:11, 
Lodders 2003\nocite{lod03} and references therein) and normal stars. 
The minor Mg isotopes were also very much more abundant than the values 
predicted for the ejecta of Type~II SNe (Timmes \etal\ (1995)\nocite{tim95}.
For NGC~6752, Yong \etal\ found a range of abundance ratios 
running from 63:08:30 to 84:08:08, with one NGC~6752 giant (star 702) 
having an unusually low ratio of 53:09:39.
Hydrogen-burning shell temperatures in first ascent NGC~6752
giants are too low to allow efficient operation of the reaction 
\iso{24}{Mg}(p,$\gamma$)\iso{25}{Al} (Powell \etal\ 1999\nocite{pow99}), 
and thus the chain of proton captures leading to the destruction of 
\iso{24}{Mg} and the production of \iso{25}{Mg}, \iso{26}{Mg} and 
\iso{27}{Al}. 
Therefore Yong \etal\ concluded that deep mixing was ruled out as a 
source of the anomalous Mg abundance ratios as well as the enhancement 
of \iso{27}{Al}.  
They proposed instead a scenario in which the abundance ratios in NGC~6752
giants actually reflect those found in the ejecta of 3-6~\Msun\ cluster 
giants that had already evolved: either the low mass giants we presently 
see formed from such secularly ejected material, or their envelopes were 
seriously polluted or even ablated and replaced by this material. 
A scenario like this is supported both observationally and theoretically. 
On the one hand, most of the abundance anomalies in Na, Mg and Al found 
among NGC~6752 giants, that had earlier been attributed to deep mixing, 
were found to exist already in their main sequence progenitors 
(Gratton \etal\ 2001\nocite{gra01}), which points to a pre-existing origin.
In addition, models which predict the abundance ratios in the ejected
envelope material of intermediate-mass low-metallicity AGB stars (Siess
\etal\ 2002\nocite{sie02}, Karakas \& Lattanzio 2003\nocite{kar03}) show 
that \iso{25}{Mg} and \iso{26}{Mg} are overproduced relative to \iso{24}{Mg}. 
For example, Karakas \& Lattanzio found that in the envelope of a 4~\Msun\ 
AGB star having Z~=~0.004, there can develop a situation in which the 
abundances of \iso{24}{Mg}, \iso{25}{Mg} and \iso{26}{Mg} are approximately 
equal; in a 6~\Msun, Z~=~0.004 AGB star, \iso{24}{Mg} virtually disappears 
relative to the other isotopes. 
At the same time, hot bottom CNO cycling in the same stars can lead to an 
overabundance of N at the expense of O and excess production of \iso{23}{Na}. 
Qualitatively these are just the sorts of anomalies found in the spectra 
of present-day NGC 6752, M3 and M13 giants.

\begin{figure}
\epsscale{0.95}
\plotone{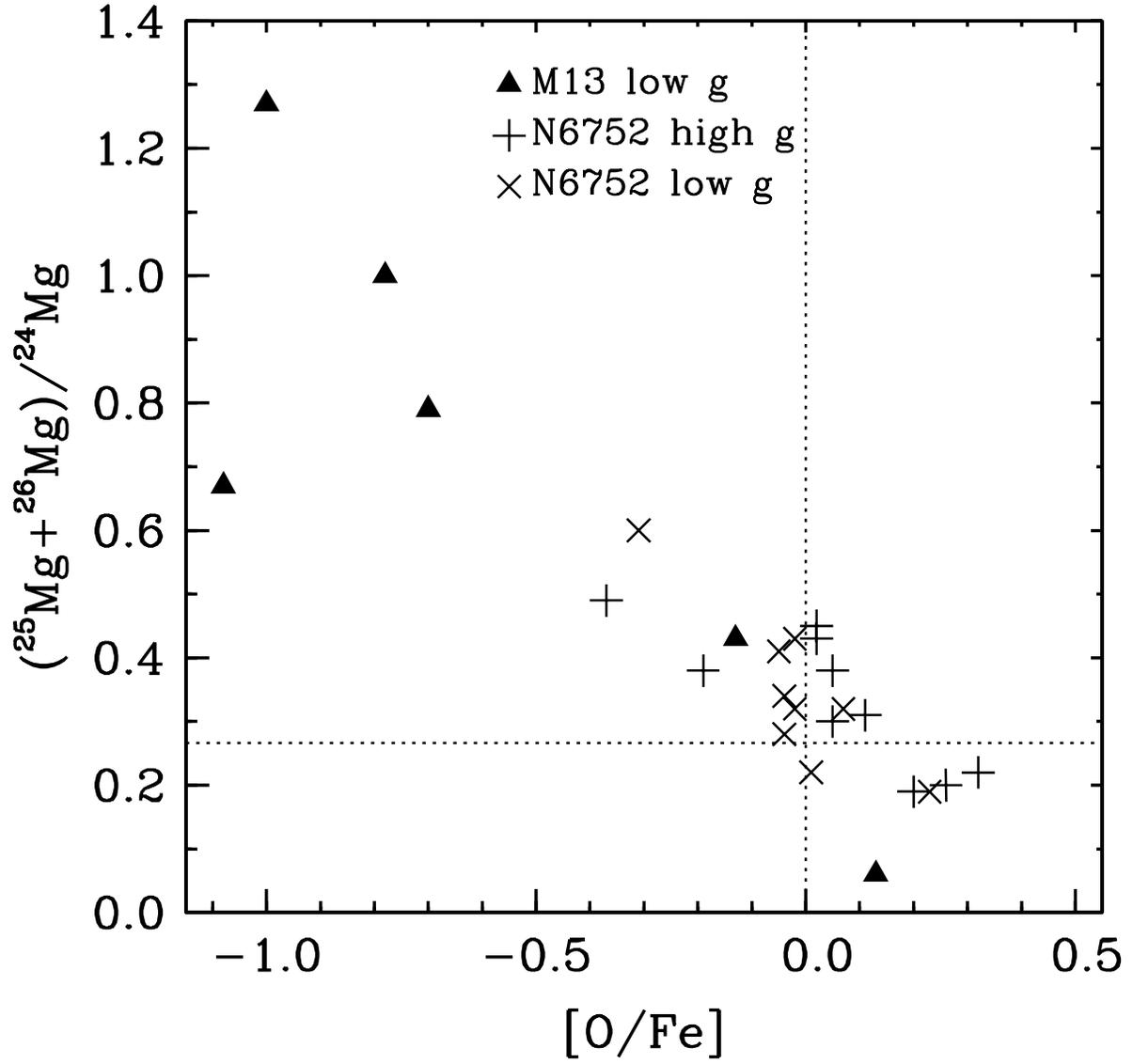}
\caption{
Correlation of [O/Fe]$_{\rm II}$ with (\iso{25}{Mg}+\iso{26}{Mg})/\iso{24}{Mg} 
among giants of M13 and NGC~6752.
The horizontal dotted line corresponds to the solar ratio of the sum of 
the rare Mg isotopes to \iso{24}{Mg} (\eg., Lodders 2003 and references 
therein).
\label{f15}}
\end{figure}

So far, we have discussed recent results found in giants of NGC 6752.
What bearing does this have on M3 and/or M13? 
Our M3 spectra do not have sufficient S/N near the MgH bands at 
$\lambda$5135 to permit discussion of the Mg isotopes. 
But a previous study of the MgH bands in six M13 giants 
(Shetrone 1996b\nocite{she96b}) permits us to compare ratios of 
(\iso{25}{Mg} + \iso{26}{Mg})/\iso{24}{Mg} in M13 with NGC 6752. 
Shetrone's spectra, of somewhat lower resolution than those of Yong \etal
(2003)\nocite{yon03}, did not permit the separation of lines of 
\iso{25}{Mg}H from \iso{26}{Mg}H, so we resort to discussing the quantity
(\iso{25}{Mg}+\iso{26}{Mg})/\iso{24}{Mg}). 
In Figure~\ref{f15} we plot [O/Fe]$_{\rm II}$ versus the quantity 
(\iso{25}{Mg}+\iso{26}{Mg})/\iso{24}{Mg} for 19 of the 20 giants studied by 
Yong \etal, plus the six M13 giants of Shetrone. 
The missing NGC 6752 star is No. 702, the one with the lowest abundance 
of \iso{24}{Mg}.
Yong \etal\ were unable to measure [O/Fe]$_{\rm II}$ for this star, 
owing to lower than average S/N for this spectrum.

Inspection of Figure~\ref{f15} shows that there is an anticorrelation
between oxygen abundance and the overabundance of the rare isotopes
of Mg relative to \iso{24}{Mg}. 
Further, the plot indicates that whatever is the nucleosynthetic process 
that depletes oxygen and enhances the rarer isotopes of Mg, NGC~6752 and 
M13 share in that process. 
Moreover, if a primordial ablation or pollution scenario is responsible
for what is seen in NGC~6752, the same scenario must have operated
to any even greater extent in M13. 
Finally, the existence of this anticorrelation means that the extreme 
oxygen depletion seen in M13 giants near the red giant tip is probably 
not a result of deep mixing.

\begin{figure}
\epsscale{0.95}
\plotone{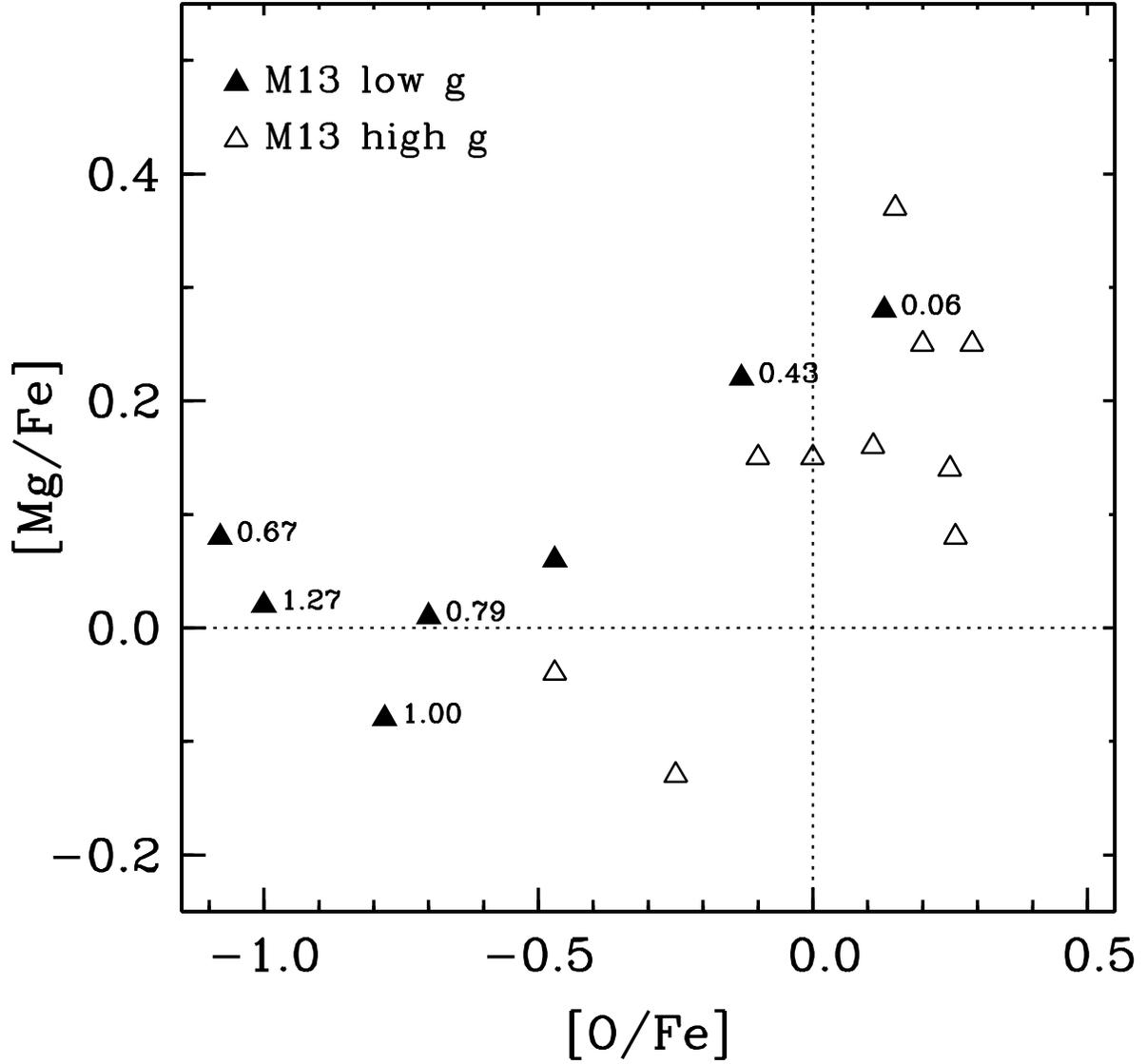}
\caption{
A plot of [Mg/Fe]$_{\rm I}$ vs [O/Fe]$_{\rm II}$ for M13 giants. 
The numbers are the values of (\iso{25}{Mg}+\iso{26}{Mg})/\iso{24}{Mg}, 
according to Shetrone (1996b).
The giants with low oxygen abundances have Mg isotopic abundances higher 
than those exhibited by giants with higher oxygen abundances.
\label{f16}}
\end{figure}

Parallel support for this conclusion is found Figure~\ref{f16}, which
is a replot of Figure~\ref{f9}, this time excluding the giants of M3, 
but noting the Mg abundance ratios of the six stars 
studied by Shetrone (1996a,b)\nocite{she96a,she96b}. 
(Because Shetrone could not resolve \iso{25}{Mg} from \iso{26}{Mg}, he simply 
assigned them equal weight in the blended spectra.)
The stars of lowest oxygen abundance also have the lowest abundance
of \iso{24}{Mg} as well as the highest abundances of the rarer isotopes,
as already noted by Shetrone. 
If the current nuclear reaction rates are correct, deep mixing again cannot 
be responsible for the depletion of \iso{24}{Mg} found among most of the 
M13 giants of lowest oxygen abundance and highest luminosity.

What about M3? 
Even though we cannot report on the Mg isotopic ratios in this cluster, 
we can compare some key abundance-ratio plots in M3 with those of NGC 6752. 
The comparison is somewhat clouded by the possibility of systematic 
differences between investigators. 
For example, Yong \etal\ (2003)\nocite{yon03} employ a completely different 
set of \ion{Fe}{1} and \ion{Fe}{2} lines compared with the set adopted 
in our analyses of M3 and M13.
NGC~6752 stars have not been investigated by our group; as Yong \etal\ note, 
previous investigations yielded [Fe/H] ranging from --1.42 to --1.62.
It seems best then, in comparing abundance ratio diagrams of NGC 6752
with those of M3 and M13, to concentrate on the morphology and recognize
the possibility of zero-point differences of 0.1 or 0.2~dex in values
of [X/Fe].

\begin{figure}
\epsscale{0.95}
\plotone{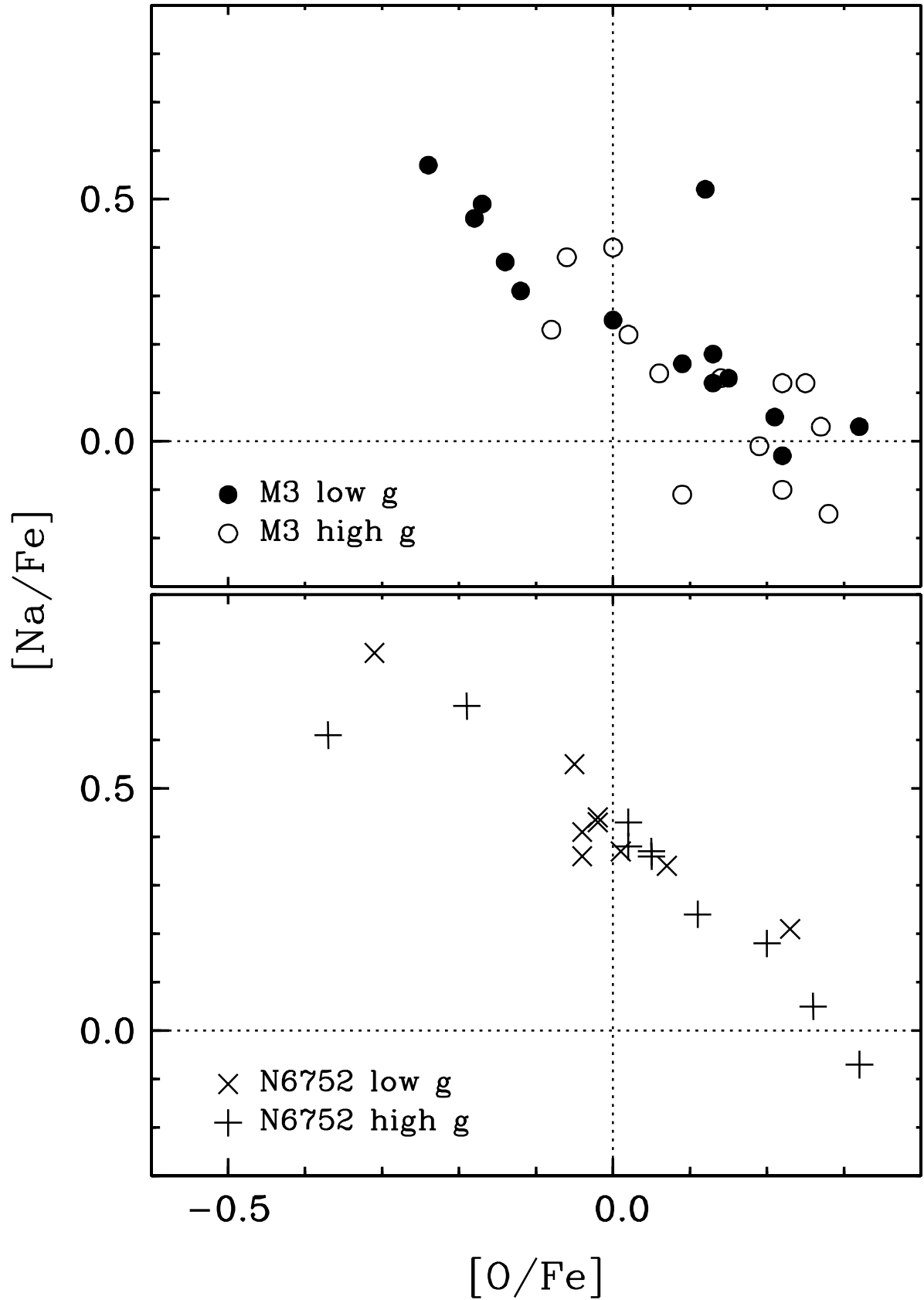}
\caption{
[Na/Fe]$_{\rm I,nLTE}$ versus [O/Fe]$_{\rm II}$ for giants in M3 and NGC 6752.
The diagrams are morphologically quite similar.
\label{f17}}
\end{figure}

\begin{figure}
\epsscale{0.95}
\plotone{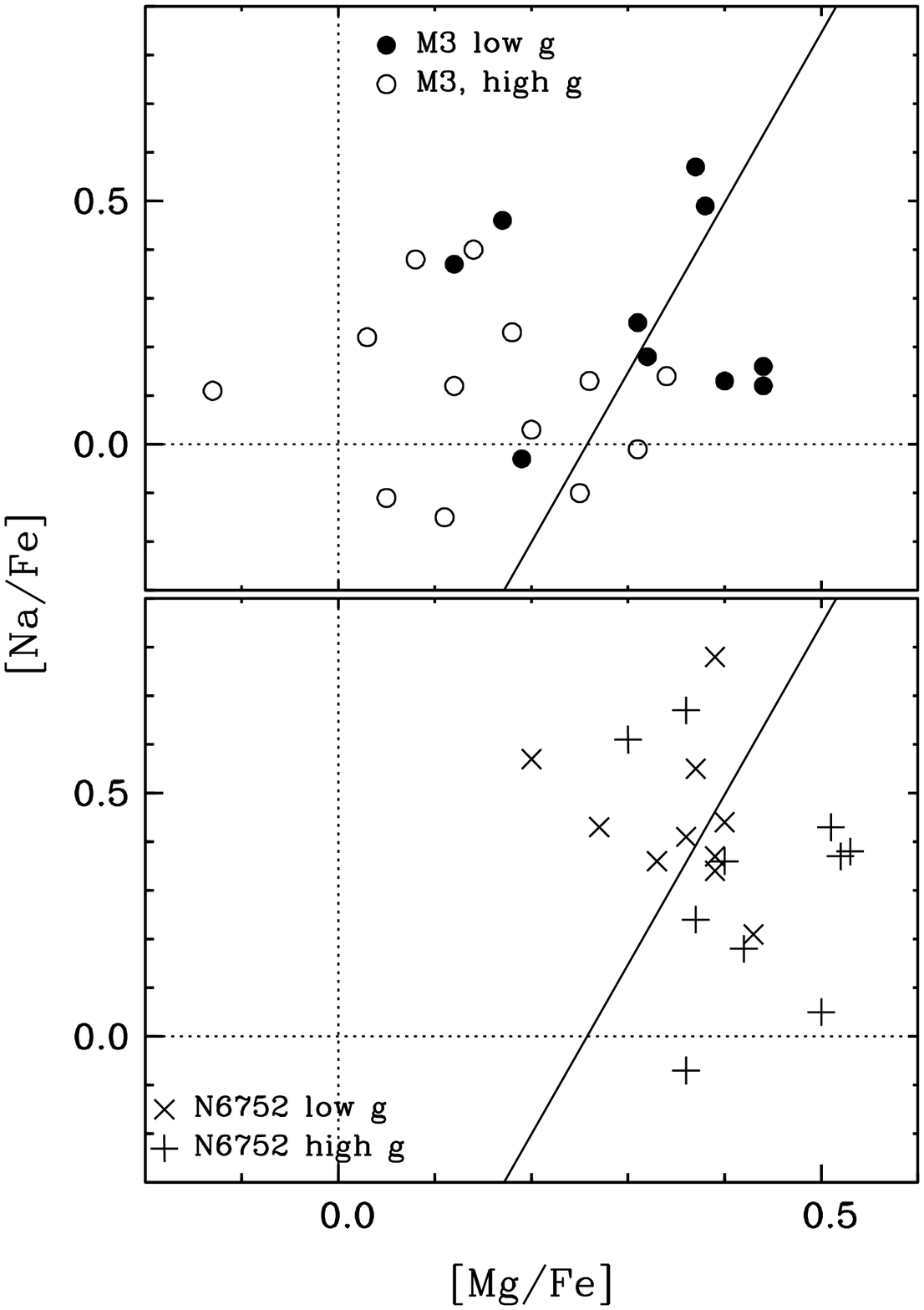}
\caption{
[Na/Fe]$_{\rm I,nLTE}$ versus [Mg/Fe]$_{\rm I}$ for giants in M3 and NGC 6752.
The M3 giants show a considerably wider range of Mg abundances for
a given Na abundance. 
Although differences in abundance zero-points may exist, neither 
distribution is morphologically similar to the distribution in M13 
(Figure~\ref{f14}).
\label{f18}}
\end{figure}

In the two panels of Figure~\ref{f17} we compare the 
[Na/Fe]$_{\rm I,nLTE}$ versus [O/Fe]$_{\rm II}$ diagrams of M3 and NGC~6752. 
The two diagrams are morphologically quite similar, and both are remarkably 
different from the corresponding M13 diagram.  
In Figure~\ref{f18}, we compare the [Na/Fe]$_{\rm I,nLTE}$ versus 
[Mg/Fe]$_{\rm I}$ diagrams of M3 and NGC~6752, again noting that both 
are morphologically dissimilar to the corresponding diagram for M13 
(Figure~\ref{f14}).
M3 and NGC 6752 do not match perfectly, as one can easily see:
M3 giants seem to have a larger scatter of Mg abundances. 
The main point is that neither cluster has the large population of giants
lying far from the field giant relationship in the low Mg versus high 
Na regime, such as one finds in M13.

\subsubsection{HB Morphology}

The main difference in the morphology of the color-magnitude diagrams for 
these three clusters is the distribution of stars along the HB, a phenomenon 
usually referred to as ``the second parameter problem''. 
The Lee \etal\ (1994)\nocite{lee94} second parameter index is 
nearly the same for M13 vs NGC 6752 (0.97 versus 1.00), which indicates
that the HB is very blue, whereas the value of the index for M3
is 0.08, which corresponds to a uniformly populated HB. 
Although age is rather widely believed to be the factor responsible for
the second parameter (\eg, Sarajedini \etal\ 1997\nocite{sar97}, 
Rey \etal\ 2001\nocite{rey01}), other possibilities have been invoked. 
For example, if deep mixing were to significantly penetrate into the 
hydrogen shell burning region, evolution to the red giant tip would be 
slowed and the He/H ratio in the envelopes of evolving red giants would 
be increased (Sweigart 1997\nocite{swe97}). 
The HB descendants would then be bluer than their unmixed counterparts. 
Another possibility would involve an ``ab initio'', or ``primordial'' 
difference in He abundance between clusters (Johnson \& Bolte 
1998\nocite{joh98}), perhaps brought about if the low mass giants we 
presently see formed from the He-enriched ejecta of 3-6~\Msun\ cluster 
giants (Ventura \etal\ 2002\nocite{ven02}). 
Qualitatively at least, an increase in the He/H ratio would favor the 
development of a blue HB, all other factors being equal.

But if deep mixing were responsible for the blue HB's of M13 and
NGC 6752, and if seriously depleted O were the most significant
``symptom'' of the deep mixing phenomenon, then one might expect NGC~6752
to exhibit a Na vs O relationship as ``advanced'' as that of M13. 
This is clearly not the case. 
As we have seen, in both the Na versus O and Na versus Mg plots, NGC~6752 
is much more like M3, a cluster with an ``intermediate'' HB morphology. 
However, these remarks apply only to the case in which an increase in the 
He/H ratio is a result of deep mixing. 
It does not rule out a picture in which NGC~6752 and M13 share the property 
of a ``primordial'' overabundance of He, relative to M3. 
It would clearly be of interest to explore the Mg isotopic ratios in M3 as 
a test (Ventura \etal\ 2002\nocite{ven02}) of the He/H ratio 
among M3 giants close to the red giant tip.

\section{SUMMARY}

Previous high resolution studies, based on a small sample of stars,
suggested that M3 giants had a smaller degree of chemical ``peculiarity''
than was the case among M13 giants, this despite the fact that the two
clusters have about the same metallicity ([Fe/H]~$\approx$~--1.5). 
The present study, based largely on high resolution Keck~I spectra of 23 
M3 giants, confirms the earlier conclusion vis-a-vis M13 giants. 
Compared with M13, M3 giants exhibit only modest deficiencies of O and 
enhancements of Na, less extreme enhancements of Al, fewer stars with low 
Mg and correspondingly high Na, and no evidence that O depletions and 
Na enhancements are dependent on an advancing evolutionary state as appears 
to be the case in M13. 
Field halo giants show correlated behavior of Mg and Na, consistent with an 
origin in supernova Type II sites either in hydrostatic or explosive carbon 
burning. 
Both M3 and M13 giants, however, also show evidence of proton capture 
synthesis of \iso{23}{Na} from \iso{22}{Ne} and \iso{27}{Al} from 
proton-capture chains starting with \iso{24}{Mg}. 
M3 stars appear to represent a case of proton-capture synthesis intermediate 
between the field and M13.

We find that giants in NGC 6752 and M13 satisfy the same anticorrelation 
of [O/Fe]$_{\rm II}$ with the ratio (\iso{25}{Mg}+\iso{26}{Mg})/\iso{24}{Mg}, 
which is a measure of the relative contribution of rare to the abundant 
isotope of Mg. 
As Yong \etal\ (2003)\nocite{yon03} argue, such a ratio could only have come
about in nuclearly processed material ejected from cluster stars in the 
3-6~\Msun\ range (Siess \etal\ 2002\nocite{sie02}, Karakas \& Lattanzio
2003\nocite{kar03}). 
This points to a scenario in which the peculiar abundance ratios of stars 
in M13 and NGC 6752 arose in such ejected material and was then either 
incorporated into (or ablated) the atmospheres of the low-mass stars we now 
see in these clusters, or else these low-mass stars were formed from 
such ejecta. 
The process seems to have advanced to a much greater degree in M13 than 
in NGC~6752. 
This in turn suggests that the very low O abundances seen in M13 giants
near the red giant tip are not, after all, the result of deep mixing driven 
by stellar evolution, but have their origin in hot-bottom CNO burning in 
the same 3-6 solar mass giants responsible for the overabundances of the 
rare isotopes of Mg. 
If deep mixing is not responsible for the severe O-depletion among M13 giants, 
then it cannot be called upon to increase the He/H ratio in the
atmospheres of such stars. 
This suggests that the excessive blueness of the M13 HB branch is not a 
result of an excess He abundance due to deep mixing. 
It does not, of course, rule out excess He from a primordial or pollution 
source (Johnson \& Bolte 1998\nocite{joh98}, 
Ventura \etal\ 2002\nocite{ven02}).

The Mg isotopic ratios among M3 giants are not known. 
However, the abundance anticorrelation of O and Na, and the Na vs Mg diagrams
of NGC 6752 and M3 tend to be rather similar and both are unlike 
the same diagrams in M13.

The most interesting conclusion of this work then presents a curious 
conundrum. 
The anticorrelation of O and the Mg isotopic abundance ratio suggests 
that M13 giants in a very small interval of initial mass preferentially 
underwent severe pollution or ablation, or alternatively, were 
preferentially formed from material secularly ejected earlier from more 
massive cluster giants. 
These stars only just at this moment arrived essentially at the red giant 
tip, exactly where the effect of deep mixing, if it indeed exists, would 
likely be most easily manifest. 
It is a most remarkable coincidence, if indeed that's what it is.

\acknowledgments

We thank Ben Dorman for communicating positional and photometric data on 
some M3 stars to us, Steve Vogt and Mike Keane for technical assistance 
with the design of HIRES multi-slit masks, Bill Mason at Keck and Jeff 
Lewis at Lick for fabrication of the slitmasks, John Lattanzio for 
conveying useful results in advance of publication, and the referee
for helpful comments on the manuscript.
This research has been supported in part by grants from the U.S. National
Science Foundation, in particular AST-0098453 to RPK, AST-9618351 to RCP, 
and AST-0307495 to CS.


\newpage
\tablenum{1}
\tablecolumns{10}
\tablewidth{0pt}

\begin{deluxetable}{lrrrrrrrrr}

\tablecaption{Photometry, Coordinates and Observations of M3 Giants}
\tablehead{
\colhead{Star}                             &
\colhead{vZ\tablenotemark{a}}              &
\colhead{$V$}                              &
\colhead{$B-V$}                            &
\colhead{$V-I$}                            &
\colhead{$V-K$}                            &
\colhead{$\Delta\alpha$\tablenotemark{b}}  &
\colhead{$\Delta\delta$\tablenotemark{b}}  &
\colhead{Exp\tablenotemark{c}}             &
\colhead{S/N\tablenotemark{d}}             \\
\colhead{}                                 &
\colhead{}                                 &
\colhead{}                                 &
\colhead{}                                 &
\colhead{}                                 &
\colhead{}                                 &
\colhead{$\arcsec$}                        &
\colhead{$\arcsec$}                        &
\colhead{min}                              &
\colhead{}                                  
}
\startdata
I-21    &    1352 & 13.11 & 1.35 & \nodata &    3.10 & \nodata & \nodata & 15 &  65 \\
IV-77   &     334 & 13.11 & 1.32 & \nodata & \nodata & \nodata & \nodata & 15 &  75 \\
IV-101  &     265 & 13.18 & 1.35 & \nodata & \nodata & \nodata & \nodata & 15 & 110 \\
A       &     640 & 13.27 & 1.26 & \nodata & \nodata & \nodata & \nodata & 15 &  75 \\
AA      &     238 & 12.69 & 1.57 & \nodata &    3.43 & \nodata & \nodata & 15 &  85 \\
\nodata &    1397 & 12.65 & 1.56 & \nodata &    3.49 & \nodata & \nodata & 15 &  75 \\
B1.1    &     911 & 13.69 & 1.32 &    1.22 & \nodata &  +20.52 &   +1.72 & 90 & 130 \\
B1.2    &     898 & 13.50 & 1.22 &    1.22 & \nodata &  +17.60 &   +2.86 & 90 & 140 \\
B1.3    &     893 & 12.74 & 1.50 &    1.58 & \nodata &  +16.44 &   +7.11 & 90 & 150 \\
B1.4    &     887 & 13.12 & 1.34 &    1.31 & \nodata &  +14.68 &   +3.32 & 90 & 150 \\
B1.5    &     874 & 13.65 & 1.22 &    1.23 & \nodata &  +12.65 &   +0.57 & 90 & 150 \\
B2.1    & \nodata & 14.79 & 0.92 &    1.05 & \nodata &  +32.38 &   +2.11 & 60 &  85 \\
B2.2    & \nodata & 14.72 & 0.77 &    0.94 & \nodata &  +34.51 &  --1.31 & 60 &  65 \\
B2.3    & \nodata & 14.74 & 0.88 &    1.02 & \nodata &  +29.11 &   +0.88 & 60 &  40 \\
B2.4    &     964 & 13.42 & 1.25 &    1.26 & \nodata &  +31.48 &  --2.39 & 60 & 105 \\
B2.5    & \nodata & 15.69 & 0.12 &    0.58 & \nodata &  +30.27 &  --3.62 & 60 &  65 \\
B3.1    & \nodata & 14.69 & 0.96 &    0.88 & \nodata & --11.29 &   +8.48 & 90 & 110 \\
B3.2    &     729 & 12.96 & 1.41 &    1.40 & \nodata & --10.84 &  +12.51 & 90 & 125 \\
B3.3    &     746 & 13.56 & 1.10 &    1.12 & \nodata &  --8.18 &  +15.13 & 90 & 100 \\
B3.4    &     733 & 12.61 & 1.63 &    1.68 & \nodata & --10.58 &  +15.75 & 90 & 130 \\
B4.1    &     860 & 13.04 & 1.43 &    1.41 & \nodata &  +10.82 &  +27.83 & 85 & 145 \\
B4.2    &     855 & 13.05 & 1.41 &    1.40 & \nodata &  +10.39 &  +20.89 & 85 & 130 \\
B4.3    & \nodata & 14.12 & 0.93 &    1.06 & \nodata &  +15.67 &  +28.67 & 85 &  90 \\
B4.4    & \nodata & 14.75 & 0.76 &    0.94 & \nodata &  +16.01 &  +24.64 & 85 &  75 \\
B4.5    & \nodata & 15.44 & 0.76 &    0.94 & \nodata &  +17.68 &  +21.81 & 85 &  70 \\
F2.4    & \nodata & 14.22 & 1.02 &    1.17 & \nodata & --15.74 &   +8.09 & 60 &  50 \\
\enddata

\tablenotetext{a}{von Zeipel 1908}

\tablenotetext{b}{RA and DEC offsets (J2000) relative to the cluster 
center (von Zeipel 1908)}

\tablenotetext{c}{Observation exposure time}

\tablenotetext{d}{Signal-to-noise near 6400~\AA}

\end{deluxetable}

\tablenum{2}
\tablecolumns{27}
\tablewidth{0pt}

\begin{deluxetable}{@{\extracolsep{-0.12in}}lcrrrrrrrrrrrrrrrrrrrrrrrrr}

\rotate

\tabletypesize{\scriptsize}
\tablecaption{Equivalent Widths of M3 Giants}
\tablehead{
\colhead{$\lambda$\tablenotemark{a}}       &
\colhead{ID}                               &
\colhead{E.P.\tablenotemark{b}}            &
\colhead{log $gf$}                         &
\colhead{(1)\tablenotemark{c}}             &
\colhead{(2)}                              &
\colhead{(3)}                              &
\colhead{(4)}                              &
\colhead{(5)}                              &
\colhead{(6)}                              &
\colhead{(7)}                              &
\colhead{(8)}                              &
\colhead{(9)}                              &
\colhead{(10)}                             &
\colhead{(11)}                             &
\colhead{(12)}                             &
\colhead{(13)}                             &
\colhead{(14)}                             &
\colhead{(15)}                             &
\colhead{(16)}                             &
\colhead{(17)}                             &
\colhead{(18)}                             &
\colhead{(19)}                             &
\colhead{(20)}                             &
\colhead{(21)}                             &
\colhead{(22)}                             &
\colhead{(23)}                              
}
\startdata
6300.31 & [O I] & 0.00 &  --9.75 &     54 &     44 &     26 &     46 &     34 &     70 &     18 &     26 &     69 &     24 &     30 &     18 &      9 &     22 &     25 &     37 &     72 &     57 &     42 &     15 &     13 &\nodata &      9 \\
6363.79 & [O I] & 0.02 & --10.25 &     22 &     22 &     15 &     20 &     11 &     37 &      7 &     11 &     33 &     11 &      8 &      5 &      8 &      6 &      9 &     11 &     36 &     27 &     16 &\nodata &\nodata &\nodata &      2 \\
6154.23 &  Na I & 2.10 &  --1.56 &     11 &      7 &     20 &     10 &     28 &     18 &     12 &     12 &     17 &     22 &      6 &      4 &     13 &     13 &     24 &      7 &     15 &     13 &     15 &      9 &\nodata &\nodata &\nodata \\
6160.75 &  Na I & 2.10 &  --1.26 &     17 &     11 &     30 &     20 &     43 &     29 &     18 &     17 &     29 &     37 &     15 &      6 &\nodata &     22 &     39 &      5 &     27 &     23 &     27 &     18 &      4 &     10 &     16 \\
5528.42 &  Mg I & 4.35 &  --0.36 &    165 &    160 &    174 &    168 &    189 &    197 &    148 &    162 &    205 &    175 &    153 &    121 &    104 &    156 &    178 &    151 &    203 &    187 &    173 &    125 &    103 &     94 &    126 \\
5711.09 &  Mg I & 4.34 &  --1.63 &\nodata &\nodata &\nodata &    100 &    105 &    101 &     73 &     84 &    111 &     98 &     70 &\nodata &\nodata &\nodata &\nodata &\nodata &\nodata &    100 &     93 &     56 &     40 &     27 &     49 \\
6696.03 &  Al I & 3.14 &  --1.57 &\nodata &\nodata &\nodata &\nodata &\nodata &     26 &\nodata &\nodata &\nodata &\nodata &      8 &\nodata &\nodata &     33 &     59 &\nodata &\nodata &\nodata &\nodata &\nodata &\nodata &     12 &     17 \\
6698.67 &  Al I & 3.14 &  --1.89 &      5 &      5 &     35 &      7 &     45 &     12 &     17 &     13 &     15 &     28 &      5 &     10 &\nodata &     18 &     34 &      3 &     12 &     11 &     20 &     11 &      8 &     14 &     10 \\
5793.08 &  Si I & 4.93 &  --2.06 &\nodata &\nodata &\nodata &\nodata &     20 &     22 &     20 &     17 &     24 &     19 &     19 &     14 &     14 &     18 &\nodata &     19 &     21 &     25 &     23 &     13 &      8 &\nodata &     13 \\
6142.49 &  Si I & 5.62 &  --1.48 &     10 &      7 &\nodata &      9 &     10 &     10 &      9 &      7 &     13 &     10 &     10 &      5 &      9 &      6 &      7 &      8 &      8 &     13 &     10 &      7 &      5 &\nodata &      8 \\
6145.02 &  Si I & 5.61 &  --1.37 &     12 &      8 &     13 &     13 &     11 &     14 &     11 &      9 &     12 &     11 &     12 &\nodata &\nodata &     11 &     11 &     12 &     10 &     15 &     11 &     11 &\nodata &      7 &      7 \\
6243.82 &  Si I & 5.61 &  --1.27 &     22 &     16 &     21 &     17 &     12 &\nodata &     12 &     14 &\nodata &     14 &     11 &\nodata &      9 &     15 &     17 &     13 &     13 &     15 &     14 &      8 &      9 &     12 &     11 \\
6244.48 &  Si I & 5.61 &  --1.27 &     12 &     10 &     16 &     19 &     14 &     17 &     14 &     12 &     17 &     16 &     12 &     11 &\nodata &     17 &     17 &     11 &      6 &     21 &     18 &\nodata &\nodata &     13 &     13 \\
5590.12 &  Ca I & 2.51 &  --0.71 &\nodata &\nodata &\nodata &     91 &    106 &    117 &\nodata &     74 &    114 &     93 &     74 &     60 &     48 &     77 &\nodata &\nodata &    115 &    106 &     95 &     58 &     48 &     44 &     61 \\
5867.56 &  Ca I & 2.92 &  --1.57 &\nodata &\nodata &\nodata &     20 &     21 &     28 &     11 &     10 &     27 &     16 &     11 &      9 &\nodata &     11 &\nodata &      9 &     24 &     23 &     18 &\nodata &\nodata &\nodata &\nodata \\
6161.30 &  Ca I & 2.52 &  --1.27 &     68 &     62 &     67 &     67 &     81 &     93 &     50 &     39 &     95 &     67 &     47 &     34 &     22 &     44 &     76 &     41 &     85 &     77 &     71 &     28 &     17 &     20 &     39 \\
6166.44 &  Ca I & 2.52 &  --1.14 &     74 &     69 &     75 &     77 &     86 &     99 &     53 &     49 &    101 &     73 &     53 &     34 &     30 &     54 &     73 &     48 &     89 &     84 &     76 &     34 &     20 &     21 &     42 \\
6169.04 &  Ca I & 2.52 &  --0.80 &     94 &     89 &     92 &     91 &    113 &    125 &     76 &     69 &    119 &     93 &     72 &     54 &     42 &     73 &     96 &     67 &    115 &    105 &     96 &     48 &     30 &     39 &     56 \\
6169.56 &  Ca I & 2.52 &  --0.48 &    106 &    103 &    110 &    107 &    128 &    136 &     90 &     88 &    136 &    107 &     87 &     74 &     57 &     86 &    112 &     82 &    130 &    120 &    111 &     65 &     46 &     49 &     71 \\
6455.60 &  Ca I & 2.52 &  --1.29 &     58 &     57 &     61 &     59 &     77 &     89 &     45 &     40 &     85 &     60 &     43 &     30 &     21 &     41 &     63 &     33 &     79 &     69 &     65 &     19 &     12 &     18 &     37 \\
6471.66 &  Ca I & 2.51 &  --0.69 &\nodata &\nodata &\nodata &    103 &    119 &    132 &     81 &     81 &    131 &    103 &     78 &     68 &     55 &     85 &\nodata &     75 &    127 &    112 &    107 &     56 &     34 &     38 &     66 \\
6499.65 &  Ca I & 2.52 &  --0.82 &    100 &     89 &     92 &     99 &    116 &    126 &     77 &     69 &    122 &     95 &     73 &     55 &     49 &     74 &     96 &     65 &    121 &    103 &     98 &     53 &     29 &     37 &     62 \\
6508.84 &  Ca I & 2.51 &  --2.41 &\nodata &\nodata &\nodata &\nodata &     16 &     17 &\nodata &      5 &     14 &     10 &      6 &\nodata &\nodata &      8 &\nodata &\nodata &     17 &     13 &     11 &\nodata &\nodata &\nodata &\nodata \\
6305.67 &  Sc I & 0.02 &  --1.28 &\nodata &\nodata &\nodata &     37 &     90 &    136 &     15 &\nodata &    122 &     38 &\nodata &\nodata &\nodata &\nodata &\nodata &      8 &    136 &     65 &\nodata &\nodata &\nodata &\nodata &\nodata \\
5526.82 & Sc II & 1.77 &  --0.03 &\nodata &\nodata &\nodata &     96 &    106 &    100 &     86 &    100 &    110 &    100 &     90 &     75 &     55 &    100 &\nodata &     85 &    105 &     98 &    101 &     86 &     64 &     45 &     73 \\
6279.74 & Sc II & 1.50 &  --1.16 &     51 &     39 &     41 &     45 &     58 &     54 &     38 &     40 &     56 &     45 &     38 &     30 &     19 &     39 &     49 &     35 &     56 &     53 &     54 &     26 &     17 &     23 &\nodata \\
6309.90 & Sc II & 1.50 &  --1.52 &     13 &     17 &     18 &     27 &     32 &     35 &     24 &     22 &     38 &     24 &     21 &     10 &     10 &     21 &     26 &     21 &     33 &     28 &     27 &     14 &      9 &\nodata &\nodata \\
5866.45 &  Ti I & 1.07 &  --0.84 &\nodata &\nodata &\nodata &    101 &    142 &    163 &     70 &     66 &    165 &     99 &     68 &     44 &     26 &     73 &\nodata &     63 &    179 &    122 &    113 &     31 &     18 &     21 &     51 \\
5922.11 &  Ti I & 1.05 &  --1.47 &\nodata &\nodata &\nodata &     70 &    104 &    125 &     42 &     33 &    122 &     68 &     34 &     20 &     10 &     37 &\nodata &     24 &    130 &     87 &     80 &     10 &\nodata &     11 &     30 \\
5965.83 &  Ti I & 1.88 &  --0.41 &\nodata &\nodata &\nodata &     57 &     91 &    100 &     36 &     30 &    102 &     58 &     31 &     17 &     15 &     29 &\nodata &     27 &     94 &     71 &     66 &     11 &\nodata &\nodata &     29 \\
5978.54 &  Ti I & 1.87 &  --0.50 &\nodata &\nodata &\nodata &     52 &     74 &     91 &     28 &     22 &     86 &     47 &     25 &     15 &\nodata &     25 &\nodata &     22 &     85 &     65 &     56 &     12 &\nodata &      8 &     24 \\
6064.63 &  Ti I & 1.05 &  --1.94 &\nodata &\nodata &\nodata &     42 &     71 &     90 &     17 &     14 &     85 &     39 &     19 &     11 &      7 &     17 &\nodata &     13 &     93 &     56 &     46 &\nodata &\nodata &\nodata &     10 \\
6303.77 &  Ti I & 1.44 &  --1.57 &     33 &     22 &     38 &     27 &     54 &     68 &     14 &     10 &     68 &     28 &     12 &      5 &      7 &      9 &     37 &      9 &     68 &     39 &     37 &\nodata &\nodata &\nodata &     10 \\
6312.24 &  Ti I & 1.46 &  --1.55 &\nodata &     18 &     43 &     30 &     36 &     73 &     12 &      7 &     64 &     18 &     10 &      9 &\nodata &     12 &     28 &     13 &     69 &     40 &     32 &\nodata &\nodata &\nodata &     12 \\
6606.97 & Ti II & 2.06 &  --2.79 &\nodata &\nodata &\nodata &     25 &     25 &     27 &     16 &     19 &     24 &     23 &     15 &     13 &\nodata &     17 &\nodata &     13 &     23 &     21 &     22 &     16 &\nodata &\nodata &     12 \\
6233.20 &   V I & 0.28 &  --2.00 &     34 &     29 &     34 &     33 &     63 &     86 &     12 &      7 &     85 &     29 &     13 &      5 &      4 &     12 &     38 &\nodata &     90 &     45 &     39 &\nodata &\nodata &      6 &\nodata \\
6274.66 &   V I & 0.27 &  --1.69 &     48 &     40 &     46 &     46 &     91 &    113 &     19 &     15 &    105 &     48 &     19 &\nodata &\nodata &     19 &     58 &     14 &    111 &     68 &     59 &\nodata &      6 &\nodata &\nodata \\
6285.16 &   V I & 0.28 &  --1.56 &     58 &     42 &     56 &     55 &     95 &    114 &     25 &     15 &    113 &     51 &     21 &      7 &\nodata &     19 &     59 &     14 &    119 &     72 &     64 &\nodata &\nodata &\nodata &     11 \\
6292.82 &   V I & 0.29 &  --1.52 &     59 &     45 &     60 &     59 &    100 &    121 &     26 &     19 &    125 &     54 &     25 &      7 &\nodata &     22 &     67 &     15 &    132 &     74 &     68 &      6 &\nodata &      6 &     16 \\
6504.19 &   V I & 1.18 &  --1.23 &     15 &      8 &     17 &     15 &     27 &     41 &\nodata &\nodata &     34 &     15 &      5 &\nodata &\nodata &\nodata &     17 &\nodata &     37 &     21 &     18 &\nodata &\nodata &\nodata &\nodata \\
6021.79 &  Mn I & 3.08 &  + 0.03 &\nodata &\nodata &\nodata &     81 &    100 &    111 &     59 &     54 &    109 &     77 &     57 &     32 &     18 &     57 &\nodata &     51 &    106 &     90 &     87 &     29 &     16 &     16 &     45 \\
5501.48 &  Fe I & 0.96 &  --3.05 &    187 &    183 &    190 &    184 &    233 &    249 &    155 &    168 &    258 &    185 &    151 &    120 &    100 &    164 &    195 &    147 &    259 &    211 &    199 &    127 &     92 &     81 &    121 \\
5506.79 &  Fe I & 0.99 &  --2.79 &\nodata &\nodata &\nodata &\nodata &\nodata &\nodata &\nodata &\nodata &\nodata &\nodata &\nodata &    122 &    106 &    190 &\nodata &    151 &\nodata &\nodata &\nodata &    134 &     94 &     82 &    128 \\
5522.45 &  Fe I & 4.21 &  --1.40 &\nodata &\nodata &\nodata &     30 &     32 &     38 &     21 &     19 &     39 &     26 &     17 &     15 &\nodata &     18 &\nodata &     21 &     34 &     35 &     30 &\nodata &\nodata &\nodata &\nodata \\
5560.21 &  Fe I & 4.43 &  --1.04 &\nodata &\nodata &\nodata &     31 &     37 &     40 &     24 &     25 &     42 &     31 &     25 &     20 &     12 &     24 &\nodata &     24 &     35 &     34 &     37 &     14 &\nodata &     11 &     20 \\
5586.77 &  Fe I & 3.37 &  --0.14 &    160 &    148 &    155 &    152 &    175 &    178 &    135 &    145 &    186 &    154 &    139 &    115 &     95 &    153 &    158 &    138 &    189 &    162 &\nodata &    120 &     90 &     88 &    117 \\
6079.01 &  Fe I & 4.65 &  --0.97 &\nodata &\nodata &\nodata &     25 &     24 &     30 &     19 &     16 &     30 &     26 &     17 &\nodata &     11 &     16 &\nodata &     15 &     26 &     28 &     25 &\nodata &      8 &\nodata &     17 \\
6151.62 &  Fe I & 2.18 &  --3.37 &     90 &     81 &     83 &     84 &    104 &    115 &     63 &     63 &    115 &     84 &     62 &     42 &     24 &     67 &     88 &     55 &    113 &     96 &     91 &     33 &     11 &     17 &     44 \\
6157.73 &  Fe I & 4.07 &  --1.26 &     63 &     57 &     61 &     62 &     71 &     73 &     44 &     45 &     80 &     58 &     46 &     30 &     23 &     47 &     63 &     42 &\nodata &     69 &     65 &\nodata &     15 &     19 &     31 \\
6159.38 &  Fe I & 4.61 &  --1.97 &\nodata &\nodata &\nodata &      5 &\nodata &\nodata &\nodata &\nodata &      7 &      5 &\nodata &\nodata &\nodata &\nodata &      5 &\nodata &\nodata &\nodata &\nodata &\nodata &\nodata &\nodata &\nodata \\
6165.36 &  Fe I & 4.14 &  --1.47 &     35 &     31 &     34 &     34 &     39 &     46 &     24 &     19 &     45 &     33 &     20 &     13 &      9 &     20 &     34 &     16 &     39 &     36 &     37 &     11 &      5 &\nodata &     19 \\
6226.74 &  Fe I & 3.88 &  --2.22 &     22 &     17 &     17 &\nodata &     26 &     28 &     13 &     11 &\nodata &     19 &     14 &\nodata &\nodata &     14 &     23 &     11 &\nodata &\nodata &     19 &      6 &\nodata &\nodata &     10 \\
6229.23 &  Fe I & 2.84 &  --3.00 &     59 &     54 &     53 &     53 &     65 &     74 &     37 &     35 &     76 &     53 &     34 &     18 &\nodata &     35 &     55 &     28 &     67 &     61 &     57 &     14 &\nodata &\nodata &     22 \\
6240.66 &  Fe I & 2.22 &  --3.23 &     85 &     77 &     83 &     80 &    103 &    113 &     61 &     65 &    112 &     83 &     57 &     34 &     22 &     61 &     88 &     49 &    109 &     94 &     90 &     27 &     12 &     14 &     43 \\
6246.32 &  Fe I & 3.60 &  --0.88 &    113 &    102 &    108 &    107 &    126 &    129 &     93 &     98 &    126 &    109 &     90 &     72 &     61 &     96 &    114 &     87 &    121 &    117 &    114 &     75 &     51 &     50 &     75 \\
6280.62 &  Fe I & 0.86 &  --4.37 &    139 &    130 &    138 &    141 &    182 &    193 &    106 &    110 &    190 &    138 &    106 &     78 &     52 &    115 &    147 &     94 &    195 &    153 &    154 &     66 &     33 &     31 &     77 \\
6290.97 &  Fe I & 4.73 &  --0.76 &     33 &     31 &     35 &     38 &     39 &     47 &     22 &     23 &     44 &     34 &     24 &     17 &     12 &     28 &     36 &     22 &     38 &     41 &     36 &\nodata &\nodata &\nodata &     23 \\
6297.80 &  Fe I & 2.22 &  --2.74 &    117 &    116 &    121 &    113 &    137 &    142 &     94 &     99 &    151 &    114 &     90 &     68 &     47 &     97 &    120 &     88 &    146 &    122 &    120 &     63 &     41 &     38 &     68 \\
6301.50 &  Fe I & 3.65 &  --0.72 &    115 &    105 &    115 &    113 &    128 &    127 &     93 &    102 &    130 &    107 &     93 &     70 &     59 &     96 &    112 &     98 &    123 &    117 &    111 &     82 &     58 &     50 &     85 \\
6302.49 &  Fe I & 3.69 &  --1.15 &     88 &     84 &     84 &     83 &    101 &    100 &     73 &     72 &    103 &     83 &     69 &     52 &     39 &     67 &     87 &     67 &     94 &     89 &     86 &     45 &     30 &     32 &     58 \\
6311.51 &  Fe I & 2.83 &  --3.22 &     45 &     33 &     42 &     38 &     57 &     59 &\nodata &     24 &     58 &     40 &     27 &     13 &     11 &     24 &     43 &     19 &     58 &     49 &     43 &     10 &\nodata &\nodata &     17 \\
6355.04 &  Fe I & 2.84 &  --2.29 &     99 &     92 &    104 &     95 &    119 &    127 &     76 &     76 &    129 &     99 &     74 &     58 &     48 &     78 &    103 &     70 &    121 &    106 &    104 &     42 &     21 &     28 &     60 \\
6380.75 &  Fe I & 4.19 &  --1.40 &     44 &     40 &     44 &     36 &     44 &     50 &     29 &     25 &     54 &     42 &     25 &     19 &\nodata &     30 &     43 &     25 &     43 &     46 &     43 &     12 &      9 &\nodata &     22 \\
6419.96 &  Fe I & 4.73 &  --0.24 &     60 &     55 &     61 &     57 &     64 &     68 &     46 &     45 &     69 &     56 &     43 &     37 &     32 &     48 &     62 &     41 &     60 &     60 &     59 &     38 &     19 &     25 &     36 \\
6421.35 &  Fe I & 2.28 &  --1.94 &    155 &    150 &    153 &    150 &    182 &    195 &    132 &    138 &    198 &    155 &    125 &     99 &     96 &    142 &    161 &    123 &    202 &    165 &    162 &    103 &     79 &     61 &    100 \\
6494.98 &  Fe I & 2.40 &  --1.24 &\nodata &\nodata &\nodata &    190 &    224 &    228 &    162 &    169 &    240 &    190 &    156 &    121 &    107 &    172 &\nodata &    157 &    237 &    208 &    197 &    133 &     93 &     80 &    130 \\
6498.94 &  Fe I & 0.96 &  --4.69 &    126 &    114 &    119 &    117 &    151 &    163 &     86 &     89 &    167 &    119 &     82 &     53 &     37 &     91 &    122 &     70 &    167 &    132 &    127 &     48 &     23 &     22 &     64 \\
6574.23 &  Fe I & 0.99 &  --5.02 &    103 &     93 &     95 &     95 &    130 &    142 &     64 &     62 &    143 &     96 &     59 &     32 &     18 &     67 &    100 &     51 &    140 &    113 &    101 &     28 &     13 &\nodata &     45 \\
6593.88 &  Fe I & 2.44 &  --2.42 &\nodata &\nodata &\nodata &    126 &    146 &    147 &     98 &    102 &    154 &    122 &     92 &\nodata &     63 &    103 &\nodata &     94 &    152 &    133 &    127 &     69 &     43 &     48 &     77 \\
6609.12 &  Fe I & 2.56 &  --2.69 &\nodata &\nodata &\nodata &     98 &    113 &    117 &     76 &     78 &    128 &     95 &     73 &     51 &     48 &     77 &\nodata &     64 &    121 &    108 &    102 &     45 &     29 &     22 &     60 \\
6733.15 &  Fe I & 4.64 &  --1.43 &\nodata &\nodata &\nodata &     11 &     17 &     19 &      8 &      7 &     17 &     12 &      6 &\nodata &\nodata &      8 &\nodata &\nodata &     14 &     12 &     11 &\nodata &\nodata &\nodata &      5 \\
6750.15 &  Fe I & 2.42 &  --2.62 &\nodata &\nodata &\nodata &    113 &    135 &    143 &     92 &     93 &    140 &    111 &     88 &     67 &     65 &     98 &\nodata &     79 &    141 &    120 &    123 &     67 &     29 &     35 &     69 \\
6149.25 & Fe II & 3.89 &  --2.72 &     23 &     26 &     22 &     20 &     22 &     23 &     22 &     26 &     22 &     22 &     20 &     21 &     16 &     27 &     25 &     25 &     20 &     22 &     25 &     24 &     23 &     17 &     18 \\
6369.46 & Fe II & 2.89 &  --4.25 &     20 &     20 &     19 &     18 &     15 &     19 &     14 &     18 &     18 &     15 &     16 &\nodata &\nodata &     19 &     17 &     18 &     17 &     16 &     15 &     17 &     13 &\nodata &     12 \\
6416.93 & Fe II & 3.89 &  --2.79 &     16 &     25 &     19 &     19 &     28 &     25 &     22 &     22 &     25 &     24 &     19 &     17 &     16 &     24 &     23 &     22 &     22 &     23 &     23 &     19 &     23 &     16 &     15 \\
6456.39 & Fe II & 3.90 &  --2.08 &     57 &     51 &     55 &     44 &     46 &     39 &     50 &     51 &     45 &     48 &     49 &     46 &     38 &     63 &     48 &     55 &     45 &     50 &     50 &     53 &     44 &     38 &     46 \\
6516.08 & Fe II & 2.89 &  --3.45 &     54 &     45 &     49 &     46 &     48 &     50 &     48 &     52 &     51 &     47 &     45 &     39 &     39 &     56 &     47 &     47 &     42 &     49 &     49 &     50 &     39 &     33 &     40 \\
6175.37 &  Ni I & 4.09 &  --0.53 &     22 &     25 &     28 &     26 &     30 &     34 &     19 &     17 &     30 &     25 &     18 &     10 &      8 &     17 &     27 &     17 &     26 &     28 &     25 &     10 &      7 &\nodata &     17 \\
6176.82 &  Ni I & 4.09 &  --0.53 &     33 &     34 &     36 &     40 &     38 &     44 &\nodata &     26 &     42 &     34 &     27 &\nodata &     14 &     25 &     39 &     33 &     39 &     39 &     37 &     13 &      8 &     12 &     22 \\
6177.25 &  Ni I & 1.83 &  --3.50 &     28 &     24 &     29 &     31 &     41 &     44 &     18 &     17 &     47 &     25 &     17 &      6 &      6 &     13 &     32 &     22 &     44 &     36 &     32 &\nodata &\nodata &\nodata &      9 \\
6378.26 &  Ni I & 4.15 &  --0.89 &\nodata &     16 &     18 &     16 &     17 &     20 &      9 &      9 &     17 &     13 &      8 &\nodata &     13 &     12 &     14 &      8 &     19 &     15 &     16 &      6 &\nodata &\nodata &\nodata \\
6586.33 &  Ni I & 1.95 &  --2.81 &\nodata &\nodata &\nodata &     71 &     84 &     90 &     47 &     44 &     92 &     64 &     44 &     29 &     30 &     50 &\nodata &     40 &     87 &     79 &     71 &     21 &      9 &     13 &     34 \\
6643.64 &  Ni I & 1.68 &  --2.01 &\nodata &\nodata &\nodata &    139 &    162 &    167 &    109 &    113 &    174 &    135 &    105 &     85 &     65 &    119 &\nodata &     99 &    174 &    150 &    144 &     76 &     46 &     44 &     89 \\
6767.78 &  Ni I & 1.83 &  --2.17 &\nodata &\nodata &\nodata &    119 &    138 &    140 &     94 &     99 &    151 &    115 &     93 &     74 &     54 &    102 &\nodata &     86 &    150 &    125 &    122 &     67 &     41 &     42 &     74 \\
5853.69 & Ba II & 0.60 &  --1.01 &    114 &    110 &    123 &    120 &    135 &    135 &     99 &    114 &    156 &    118 &    101 &     89 &     64 &    116 &    123 &    106 &    149 &    127 &    121 &    111 &     73 &     51 &     77 \\
6141.73 & Ba II & 0.70 &  --0.08 &    165 &    166 &    169 &    172 &    201 &    197 &    146 &    178 &    227 &    167 &    152 &    135 &    106 &    171 &    177 &    161 &    218 &    189 &    180 &    167 &    127 &     82 &    120 \\
6496.91 & Ba II & 0.60 &  --0.38 &    170 &    170 &    173 &    170 &    197 &    200 &    148 &    173 &    225 &    170 &    144 &    115 &    110 &    167 &    174 &    145 &    219 &    183 &    179 &    165 &    108 &\nodata &    120 \\
6390.49 & La II & 0.30 &  --1.41 &\nodata &\nodata &\nodata &     33 &     35 &     38 &     19 &     18 &     49 &     27 &     16 &      9 &\nodata &     22 &\nodata &     15 &     41 &     37 &     31 &     14 &\nodata &\nodata &     13 \\
6774.33 & La II & 0.13 &  --1.75 &\nodata &\nodata &\nodata &     28 &     33 &     38 &\nodata &     15 &     51 &     26 &     16 &     12 &\nodata &     15 &\nodata &\nodata &     48 &     40 &     27 &\nodata &\nodata &\nodata &\nodata \\
6645.13 & Eu II & 1.37 &  + 0.20 &     32 &     28 &     39 &     38 &     42 &     38 &     28 &     26 &     48 &     32 &     25 &     15 &      9 &     30 &     32 &     22 &     45 &     42 &     34 &     15 &      4 &     81 &     20 \\
\enddata

\tablenotetext{a}{Wavelength in \AA}

\tablenotetext{b}{Excitation Potential in eV}                  

\tablenotetext{c}{Star Names: (1) = I-21; 
                              (2) = IV-77;
                              (3) = IV-101;
                              (4) = A;
                              (5) = AA;
                              (6) = vZ~1397;
                              (7) = B1.1;
                              (8) = B1.2;
                              (9) = B1.3;
                             (10) = B1.4;
                             (11) = B1.5;
                             (12) = B2.1;
                             (13) = B2.3;
                             (14) = B2.4;
                             (15) = B3.2;
                             (16) = B3.3;
                             (17) = B3.4;
                             (18) = B4.1;
                             (19) = B4.2;
                             (20) = B4.3;
                             (21) = B4.4;
                             (22) = B4.5;
                             (23) = F2.4}

\end{deluxetable}

\newpage
\tablenum{3}
\tablecolumns{9}
\tablewidth{0pt}

\begin{deluxetable}{lrrrrrrrr}

\tablecaption{Model Parameters, Fe Metallicities, and Radial Velocities of 
                     M3 Giants}
\tablehead{
\colhead{Star}                             &
\colhead{T$_{eff}$}                        &
\colhead{log $g$}                          &
\colhead{T$_{eff}$}                        &
\colhead{log $g$}                          &
\colhead{$v_t$}                            &
\colhead{[Fe/H]}                           &
\colhead{[Fe/H]}                           &
\colhead{$v_R$}                            \\
\colhead{}                                 &
\colhead{K,phot}                           &
\colhead{phot}                             &
\colhead{K,spec}                           &
\colhead{spec}                             &
\colhead{km s$^{-1}$}                      &
\colhead{I}                                &
\colhead{II}                               &
\colhead{km s$^{-1}$}                       
}
\startdata
I-21    &    4150 &    0.72 & 4175    & 0.70    &    1.70 &  --1.52 &  --1.42 & --146.9 \\
IV-77   &    4225 &    0.88 & 4300    & 0.85    &    1.80 &  --1.52 &  --1.43 & --147.7 \\
IV-101  &    4150 &    0.72 & 4200    & 0.75    &    1.70 &  --1.50 &  --1.43 & --145.6 \\
A       &    4200 &    0.80 & 4200    & 0.90    &    1.65 &  --1.52 &  --1.44 & --151.1 \\
AA      &    3975 &    0.45 & 3975    & 0.20    &    1.95 &  --1.57 &  --1.45 & --150.3 \\
vZ~1397 &    3950 &    0.40 & 3925    & 0.10    &    2.00 &  --1.53 &  --1.42 & --147.2 \\
B1.1    &    4340 &    1.14 & 4400    & 1.10    &    1.50 &  --1.56 &  --1.47 & --171.9 \\
B1.2    &    4340 &    1.06 & 4400    & 1.10    &    1.80 &  --1.62 &  --1.50 & --152.1 \\
B1.3    &    3950 &    0.40 & 3900    & 0.00    &    2.05 &  --1.54 &  --1.42 & --150.9 \\
B1.4    &    4190 &    0.79 & 4175    & 0.70    &    1.70 &  --1.58 &  --1.46 & --146.6 \\
B1.5    &    4290 &    1.08 & 4350    & 1.10    &    1.50 &  --1.67 &  --1.47 & --147.0 \\
B2.1    &    4750 &    1.83 & 4625    & 1.65    &    1.20 &  --1.61 &  --1.47 & --146.6 \\
B2.2    &    4950 &    1.80 & \nodata & \nodata & \nodata & \nodata & \nodata & --157.0 \\
B2.3    &    4825 &    1.85 & 4800    & 2.10    &    1.40 &  --1.70 &  --1.50 & --149.8 \\
B2.4    &    4275 &    1.05 & 4450    & 1.00    &    1.90 &  --1.59 &  --1.47 & --146.7 \\
B2.5    & \nodata & \nodata & \nodata & \nodata & \nodata & \nodata & \nodata & --147.1 \\
B3.1    &    4620 &    1.60 & \nodata & \nodata & \nodata & \nodata & \nodata & --155.9 \\
B3.2    &    4100 &    0.65 & 4175    & 0.70    &    1.75 &  --1.51 &  --1.43 & --146.5 \\
B3.3    &    4530 &    1.09 & 4550    & 1.30    &    1.65 &  --1.55 &  --1.45 & --149.9 \\
B3.4    &    3900 &    0.34 & 3850    & 0.00    &    2.00 &  --1.62 &  --1.40 & --140.2 \\
B4.1    &    4075 &    0.60 & 4075    & 0.40    &    1.70 &  --1.54 &  --1.46 & --149.7 \\
B4.2    &    4090 &    0.61 & 4100    & 0.40    &    1.70 &  --1.56 &  --1.48 & --144.5 \\
B4.3    &    4740 &    1.44 & 4650    & 1.40    &    1.60 &  --1.75 &  --1.50 & --151.7 \\
B4.4    &    4950 &    1.80 & 5050    & 2.00    &    1.50 &  --1.71 &  --1.48 & --154.9 \\
B4.5    &    4950 &    2.08 & 5100    & 2.40    &    1.00 &  --1.58 &  --1.47 & --147.2 \\
F2.4    &    4550 &    1.72 & 4600    & 1.70    &    1.20 &  --1.54 &  --1.45 & --140.4 \\
\enddata

\end{deluxetable}

\tablenum{4}
\tablecolumns{18}
\tablewidth{0pt}

\begin{deluxetable}{lrrrrrrrrrrrrrrrrr}

\rotate

\tabletypesize{\scriptsize}
\tablecaption{Abundance Ratios in 23 M3 Giants Observed with Keck~I HIRES}
\tablehead{
\colhead{Star}                             &
\colhead{O\tablenotemark{a}}               &
\colhead{Na}                               &
\colhead{Na,rev\tablenotemark{b}}          &
\colhead{Mg}                               &
\colhead{Al}                               &
\colhead{Si}                               &
\colhead{Ca}                               &
\colhead{Sc}                               &
\colhead{Sc}                               &
\colhead{Ti}                               &
\colhead{Ti}                               &
\colhead{V}                                &
\colhead{Mn}                               &
\colhead{Ni}                               &
\colhead{Ba}                               &
\colhead{La}                               &
\colhead{Eu}                               \\
\colhead{}                                 &
\colhead{[I]}                              &
\colhead{I}                                &
\colhead{I}                                &
\colhead{I}                                &
\colhead{I}                                &
\colhead{I}                                &
\colhead{I}                                &
\colhead{I}                                &
\colhead{II}                               &
\colhead{I}                                &
\colhead{II}                               &
\colhead{I}                                &
\colhead{I}                                &
\colhead{I}                                &
\colhead{II}                               &
\colhead{II}                               &
\colhead{II}
}
\startdata
I-21    &  +0.22 & --0.23 & --0.03 &  +0.19 &\nodata &  +0.27 &  +0.23 & --0.14 & --0.39 &  +0.22 &  +0.23 & --0.05 & --0.33 & --0.08 &  +0.11 &   0.00 &  +0.52 \\
IV-77   &  +0.22 & --0.28 & --0.10 &  +0.25 & --0.18 &  +0.22 &  +0.27 & --0.06 & --0.28 &  +0.24 &  +0.25 &  +0.02 & --0.31 & --0.05 &  +0.14 &  +0.09 &  +0.51 \\
IV-101  & --0.14 &  +0.17 &  +0.37 &  +0.12 &  +0.89 &  +0.32 &  +0.21 & --0.10 & --0.16 &  +0.22 &  +0.25 &  +0.02 & --0.33 & --0.06 &  +0.25 &  +0.06 &  +0.64 \\
A       &  +0.19 & --0.19 & --0.01 &  +0.31 &  +0.05 &  +0.30 &  +0.25 & --0.09 & --0.03 &  +0.19 &  +0.33 & --0.03 & --0.36 &  +0.02 &  +0.32 &  +0.25 &  +0.75 \\
AA      & --0.24 &  +0.32 &  +0.57 &  +0.37 &  +0.97 &  +0.28 &  +0.22 &  +0.09 & --0.12 &  +0.19 &  +0.14 & --0.06 & --0.35 & --0.06 &  +0.11 & --0.03 &  +0.56 \\
1397    &  +0.13 & --0.08 &  +0.18 &  +0.32 &  +0.18 &  +0.35 &  +0.26 &\nodata & --0.02 &  +0.31 &  +0.15 &   0.00 & --0.30 & --0.07 &  -0.02 & --0.04 &  +0.43 \\
B1.1    & --0.08 &  +0.06 &  +0.23 &  +0.18 &  +0.68 &  +0.26 &  +0.24 & --0.10 & --0.06 &  +0.15 &  +0.06 & --0.09 & --0.37 & --0.10 &  +0.23 &  +0.10 &  +0.63 \\
B1.2    &  +0.06 & --0.03 &  +0.14 &  +0.34 &  +0.67 &  +0.26 &  +0.22 &\nodata & --0.06 &  +0.13 &  +0.20 & --0.13 & --0.37 & --0.05 &  +0.28 &  +0.08 &  +0.57 \\
B1.3    &  +0.09 & --0.10 &  +0.16 &  +0.44 &  +0.24 &  +0.38 &  +0.19 &  +0.20 & --0.16 &  +0.23 &  +0.04 & --0.09 & --0.35 & --0.08 &  +0.26 &  +0.11 &  +0.57 \\
B1.4    & --0.17 &  +0.29 &  +0.49 &  +0.38 &  +0.82 &  +0.31 &  +0.26 & --0.07 & --0.10 &  +0.14 &  +0.22 & --0.05 & --0.38 & --0.06 &  +0.17 &  +0.10 &  +0.57 \\
B1.5    &  +0.14 & --0.04 &  +0.13 &  +0.26 &  +0.14 &  +0.37 &  +0.25 &\nodata & --0.05 &  +0.11 &  +0.11 & --0.06 & --0.36 & --0.04 &  +0.23 &  +0.04 &  +0.57 \\
B2.1    &  +0.18 & --0.23 & --0.15 &  +0.20 &  +0.60 &  +0.23 &  +0.26 &\nodata & --0.10 &  +0.18 &  +0.26 & --0.18 & --0.42 & --0.09 &  +0.43 &  +0.14 &  +0.51 \\
B2.3    &  +0.09 & --0.17 & --0.11 &  +0.05 &\nodata &  +0.45 &  +0.21 &\nodata & --0.17 &  +0.32 &\nodata &  +0.27 & --0.60 &  +0.08 &  +0.08 &\nodata &  +0.44 \\
B2.4    & --0.06 &  +0.22 &  +0.38 &  +0.08 &  +0.79 &  +0.28 &  +0.24 &\nodata & --0.13 &  +0.18 &  +0.09 & --0.02 & --0.35 & --0.05 &  +0.18 &  +0.07 &  +0.59 \\
B3.2    & --0.18 &  +0.26 &  +0.46 &  +0.17 &  +0.89 &  +0.24 &  +0.23 &  +0.03 & --0.08 &  +0.22 &  +0.14 &   0.00 & --0.33 & --0.05 &  +0.18 &  +0.08 &  +0.56 \\
B3.3    &  +0.27 & --0.11 &  +0.03 &  +0.20 & --0.04 &  +0.26 &  +0.20 & --0.12 & --0.11 &  +0.21 &  +0.03 & --0.08 & --0.35 & --0.01 &  +0.29 &  +0.04 &  +0.53 \\
B3.4    &  +0.13 & --0.15 &  +0.12 &  +0.44 &  +0.15 &  +0.27 &  +0.15 &\nodata & --0.22 &  +0.25 &  +0.04 & --0.08 & --0.34 & --0.03 &  +0.17 & --0.02 &  +0.51 \\
B4.1    &  +0.15 & --0.09 &  +0.13 &  +0.40 &  +0.27 &  +0.38 &  +0.28 &  +0.03 & --0.12 &  +0.22 &  +0.12 & --0.07 & --0.36 & --0.05 &  +0.23 &  +0.18 &  +0.67 \\
B4.2    &   0.00 &  +0.03 &  +0.25 &  +0.31 &  +0.58 &  +0.31 &  +0.24 &\nodata & --0.08 &  +0.18 &  +0.13 & --0.08 & --0.35 & --0.07 &  +0.21 &  +0.03 &  +0.53 \\
B4.3    & --0.10 &  +0.34 &  +0.42 &  +0.22 &  +0.83 &  +0.38 &  +0.22 &\nodata & --0.09 &  +0.06 &  +0.22 & --0.18 & --0.33 & --0.06 &  +0.63 &  +0.13 &  +0.43 \\
B4.4    &  +0.22 &  +0.14 &  +0.12 &  +0.12 &  +0.84 &  +0.35 &  +0.18 &\nodata &  +0.03 &  +0.21 &\nodata &\nodata & --0.40 & --0.06 &  +0.27 &\nodata & --0.04 \\
B4.5    &\nodata &  +0.11 &  +0.11 & --0.13 &  +0.81 &  +0.37 &  +0.22 &\nodata & --0.06 &  +0.32 &\nodata &\nodata & --0.47 &   0.00 &  +0.22 &\nodata &\nodata \\
F2.4    &  +0.02 &  +0.09 &  +0.22 &  +0.03 &  +0.50 &  +0.19 &  +0.21 &\nodata &  +0.12 &  +0.27 &  +0.18 & --0.10 & --0.38 & --0.03 &  +0.25 &  +0.19 &  +0.68 \\
\multicolumn{18}{c}{Cluster Mean Abundances} \\
$<>$    &  +0.15 &  +0.01 &  +0.18 &  +0.22 &  +0.51 &  +0.30 &  +0.23 & --0.03 & --0.11 &  +0.21 &  +0.16 & --0.04 & --0.37 & --0.05 &  +0.21 &  +0.09 &  +0.54 \\
$\pm$   &   0.03 &   0.04 &   0.04 &   0.03 &   0.08 &   0.01 &   0.01 &   0.03 &   0.02 &   0.01 &   0.02 &   0.02 &   0.01 &   0.01 &   0.02 &   0.02 &   0.03 \\
$\sigma$&   0.15 &   0.19 &   0.20 &   0.15 &   0.36 &   0.06 &   0.03 &   0.11 &   0.10 &   0.06 &   0.09 &   0.08 &   0.06 &   0.04 &   0.11 &   0.09 &   0.15 \\
No.     &     22 &     23 &     23 &     23 &     21 &     23 &     23 &     11 &     23 &     23 &     20 &     21 &     23 &     23 &     23 &     20 &     22 \\
\enddata

\tablenotetext{a}{All abundance ratios are of the form [X/Fe], where
                  the Fe abundance is that of \ion{Fe}{1} when X is
                  an abundance determined from neutral species lines
                  (except O), or that of \ion{Fe}{2} for O and elements
                  X determined determined from ionized species lines.}

\tablenotetext{b}{Na abundance after correction for the suggested
                  NLTE corrections of Gratton et al. (2000).}

\end{deluxetable}

\tablenum{5}
\tablecolumns{6}
\tablewidth{0pt}

\begin{deluxetable}{l@{\extracolsep{0.10in}}ccccc}

\tablecaption{Models for M3 Giants Analyzed from Old Lick Spectra}
\tablehead{
\colhead{Star}                 &
\colhead{T$_{\rm eff}$}        &
\colhead{log $g$}              &
\colhead{$v_t$}                &
\colhead{[Fe/H]}               &
\colhead{[Fe/H]}               \\
\colhead{}                     &
\colhead{K}                    &
\colhead{}                     &
\colhead{km s$^{-1}$}          &
\colhead{I}                    &
\colhead{II}                   
}
\startdata
vZ~1397 (old) & 3925 & 0.10 & 2.00 & --1.53 & --1.53 \\
vZ~1397 (new) & 3925 & 0.10 & 2.00 & --1.53 & --1.42 \\
MB3           & 3925 & 0.20 & 1.90 & --1.57 & --1.48 \\
MB4           & 3950 & 0.40 & 2.00 & --1.55 & --1.42 \\
AA (old)      & 3975 & 0.20 & 2.00 & --1.53 & --1.43 \\
AA (new)      & 3975 & 0.20 & 1.95 & --1.57 & --1.45 \\
II-46         & 4050 & 0.40 & 1.95 & --1.57 & --1.52 \\
vZ~1000       & 4125 & 0.60 & 1.90 & --1.58 & --1.41 \\
vZ~1127       & 4200 & 0.85 & 1.65 & --1.51 & --1.43 \\
\enddata

\end{deluxetable}

\tablenum{6}
\tablecolumns{14}
\tablewidth{0pt}

\begin{deluxetable}{lrrrrrrrrrrrrr}

\rotate

\tablecaption{Revised Abundances from Old Lick Spectra of M3 Giants}
\tablehead{
\colhead{Star}                 &
\colhead{O\tablenotemark{a}}   &
\colhead{Na}                   &
\colhead{Na}                   &
\colhead{Si}                   &
\colhead{Ca}                   &
\colhead{Sc}                   &
\colhead{Sc}                   &
\colhead{Ti}                   &
\colhead{V}                    &
\colhead{Mn}                   &
\colhead{Ni}                   &
\colhead{Ba}                   &
\colhead{Eu}                   \\
\colhead{}                     &
\colhead{[I]}                  &
\colhead{I}                    &
\colhead{I,rev\tablenotemark{b}}&
\colhead{I}                    &
\colhead{I}                    &
\colhead{I}                    &
\colhead{II}                   &
\colhead{I}                    &
\colhead{I}                    &
\colhead{I}                    &
\colhead{I}                    &
\colhead{II}                   &
\colhead{II}                      
}
\startdata
1397 (old) &  +0.31 &   0.00 & +0.25 & +0.45 & +0.22 &  +0.26 & --0.14 & +0.27 &  +0.03 & --0.35 & --0.07 &  +0.01 & +0.61 \\
1397 (new) &  +0.13 & --0.08 & +0.18 & +0.35 & +0.26 &\nodata & --0.02 & +0.31 &   0.00 & --0.30 & --0.07 & --0.02 & +0.43 \\
MB3        &  +0.21 & --0.21 & +0.05 & +0.54 & +0.17 &  +0.26 & --0.08 & +0.21 & --0.01 & --0.28 & --0.02 & --0.11 & +0.66 \\
MB4        &  +0.12 &  +0.27 & +0.52 & +0.42 & +0.24 &  +0.29 & --0.21 & +0.17 & --0.17 & --0.33 & --0.12 &  +0.22 & +0.72 \\
AA (old)   & --0.20 &  +0.30 & +0.04 & +0.21 & +0.20 &   0.00 & --0.15 & +0.14 & --0.08 & --0.44 & --0.14 &  +0.19 & +0.64 \\
AA (new)   & --0.24 &  +0.32 & +0.57 & +0.28 & +0.22 &  +0.09 & --0.12 & +0.19 & --0.06 & --0.35 & --0.06 &  +0.11 & +0.56 \\
II-46      &  +0.32 & --0.20 & +0.03 & +0.21 & +0.15 &  +0.14 & --0.20 & +0.21 & --0.01 & --0.40 & --0.14 & --0.06 & +0.31 \\
1000       & --0.12 &  +0.10 & +0.31 & +0.23 & +0.23 & --0.05 & --0.24 & +0.12 &  +0.04 & --0.30 & --0.03 &  +0.06 & +0.68 \\
1127       &  +0.25 & --0.08 & +0.12 & +0.11 & +0.16 & --0.19 & --0.20 & +0.07 & --0.10 & --0.49 & --0.04 &  +0.16 & +0.58 \\
\multicolumn{14}{c}{Cluster Mean Abundances, Old Data Only} \\
$<>$       &  +0.13 &  +0.03 & +0.19 & +0.31 & +0.20 &  +0.10 & --0.17 & +0.17 & --0.04 & --0.37 & --0.08 &  +0.07 & +0.60 \\
$\pm$      &   0.09 &   0.08 &  0.07 &  0.06 &  0.01 &   0.07 &   0.02 &  0.03 &   0.03 &   0.03 &   0.02 &   0.05 &  0.05 \\
$\sigma$   &   0.21 &   0.21 &  0.18 &  0.16 &  0.04 &   0.19 &   0.05 &  0.07 &   0.08 &   0.08 &   0.05 &   0.13 &  0.14 \\
\enddata

\tablenotetext{a}{All abundance ratios are of the form [X/Fe], where
                  the Fe abundance is that of \ion{Fe}{1} when X is
                  an abundance determined from neutral species lines
                  (except O), or that of \ion{Fe}{2} for O and elements
                  X determined determined from ionized species lines.}

\tablenotetext{b}{Na abundance after correction for the suggested
                  NLTE corrections of Gratton et~al. (2000).}

\end{deluxetable}

\tablenum{7}
\tablecolumns{13}
\tablewidth{0pt}

\begin{deluxetable}{@{\extracolsep{-0.03in}}lrrrrrrrrrrrr}

\tabletypesize{\scriptsize}
\tablecaption{Abundances for M13 Giants}
\tablehead{
\colhead{Star}                             &
\colhead{T$_{\rm eff}$\tablenotemark{a}}   &
\colhead{log~$g$\tablenotemark{a}}         &
\colhead{[Fe/H]\tablenotemark{a}}          &
\colhead{[Fe/H]\tablenotemark{a}}          &
\colhead{[O/Fe]}                           &
\colhead{[Na/Fe]}                          &
\colhead{[Na/Fe],rev\tablenotemark{b}}     &
\colhead{[Mg/Fe]}                          &
\colhead{[Al/Fe]}                          &
\colhead{[Ba/Fe]}                          &
\colhead{[La/Fe]}                          &
\colhead{[Eu/Fe]}                          \\
\colhead{}                                 &
\colhead{}                                 &
\colhead{}                                 &
\colhead{I}                                &
\colhead{II}                               &
\colhead{[I]}                              &
\colhead{I}                                &
\colhead{I}                                &
\colhead{I}                                &
\colhead{I}                                &
\colhead{II}                               &
\colhead{II}                               &
\colhead{II}                               
}
\startdata
\multicolumn{13}{c}{Abundances from HIRES on Keck~I and Upgraded 
Hamilton spectrograph on Lick 3m} \\
L598    & 3900 & 0.30 & --1.62 & --1.47 &  +0.13 & --0.03 &  +0.23 &  +0.28 &     +0.26 &  +0.11 & --0.05 &  +0.31 \\
L629    & 4010 & 0.36 & --1.63 & --1.60 & --0.13 &  +0.27 &  +0.51 &  +0.22 &     +0.70 &  +0.18 &  +0.05 &  +0.49 \\
I-48    & 3950 & 0.34 & --1.57 & --1.56 & --0.78 &  +0.45 &  +0.70 & --0.08 &     +1.17 & --0.07 &  +0.13 &  +0.52 \\
II-67   & 3900 & 0.37 & --1.63 & --1.35 & --1.00 &  +0.42 &  +0.68 &  +0.02 &     +1.16 &  +0.12 & --0.11 &  +0.39 \\
IV-25   & 3975 & 0.38 & --1.64 & --1.36 & --1.08 &  +0.56 &  +0.80 &  +0.08 &     +1.21 &  +0.12 & --0.13 &  +0.34 \\
II-90   & 3960 & 0.38 & --1.63 & --1.59 & --0.47 &  +0.37 &  +0.62 &  +0.06 &     +1.10 &  +0.35 &  +0.01 &  +0.45 \\
L835    & 4035 & 0.45 & --1.60 & --1.47 & --0.70 &  +0.46 &  +0.69 &  +0.01 &     +1.14 &  +0.21 &  +0.04 &  +0.42 \\
J3      & 4520 & 1.31 & --1.63 & --1.62 & --0.10 &  +0.53 &  +0.68 &  +0.15 &     +0.91 &  +0.30 &  +0.22 &  +0.52 \\
A1      & 4550 & 1.34 & --1.72 & --1.67 &  +0 30 &  +0.01 &  +0.15 &  +0.34 &     +0.18 &  +0.28 &  +0.38 &  +0.69 \\
I-12    & 4610 & 1.46 & --1.58 & --1.60 & --0.47 &  +0.29 &  +0.42 & --0.04 &     +1.18 &  +0.28 &  +0.14 &  +0.52 \\
IV-19   & 4610 & 1.52 & --1.62 & --1.56 &  +0.20 & --0.10 &  +0.03 &  +0.25 &     +0.10 &  +0.26 &  +0.08 &  +0.43 \\
IV-22   & 4750 & 2.12 & --1.50 & --1.46 & --0.25 &  +0.34 &  +0.42 & --0.13 &     +1.04 &  +0.32 &  +0.18 &  +0.55 \\
II-9    & 4680 & 1.75 & --1.60 & --1.61 &  +0.11 & --0.21 & --0.12 &  +0.16 &     +0.11 &  +0.18 &  +0.15 &  +0.42 \\
II-41   & 4650 & 1.62 & --1.54 & --1.59 &  +0.25 & --0.17 & --0.06 &  +0.14 &     +0.02 &  +0.08 &  +0.15 &  +0.43 \\
I-72    & 4825 & 1.97 & --1.67 & --1.57 &  +0.26 & --0.21 & --0.15 &  +0.08 & $<$--0.02 &  +0.25 &  +0.34 &  +0.87 \\
II-28   & 4750 & 1.57 & --1.77 & --1.66 &  +0.29 & --0.02 &  +0.05 &  +0.25 &  $<$+0.10 &  +0.24 &  +0.21 &  +0.64 \\
II-1    & 4850 & 2.05 & --1.58 & --1.62 &\nodata &  +0.33 &  +0.38 & --0.20 &     +0.94 &  +0.38 & --0.06 &  +0.46 \\
I-54    & 5050 & 1.74 & --1.64 & --1.62 &  +0.15 &  +0.33 &  +0.36 &  +0.37 &     +0.71 &  +0.64 &  +0.28 &  +0.38 \\
\multicolumn{13}{c}{Abundances from Older Hamilton spectrograph on Lick 3m} \\
L324    & 3925 & 0.35 & --1.60 & --1.25 & --0.04 &  +0.33 &  +0.58 &\nodata &\nodata    &\nodata &\nodata &\nodata \\
III-56  & 4000 & 0.42 & --1.64 & --1.52 & --0.15 &  +0.28 &  +0.52 &\nodata &\nodata    &\nodata &\nodata &\nodata \\
L835    & 4035 & 0.45 & --1.40 & --1.42 & --0.73 &  +0.42 &  +0.65 &\nodata &\nodata    &\nodata &\nodata &\nodata \\
L853    & 4030 & 0.48 & --1.65 & --1.57 & --0.01 &  +0.35 &  +0.58 &\nodata &\nodata    &\nodata &\nodata &\nodata \\
L940    & 4050 & 0.48 & --1.51 & --1.40 & --0.57 &  +0.32 &  +0.55 &\nodata &\nodata    &\nodata &\nodata &\nodata \\
III-63  & 4120 & 0.60 & --1.66 & --1.42 & --0.10 &  +0.39 &  +0.61 &\nodata &\nodata    &\nodata &\nodata &\nodata \\
L261    & 4150 & 0.60 & --1.55 & --1.63 &   0.00 &  +0.22 &  +0.43 &\nodata &\nodata    &\nodata &\nodata &\nodata \\
L262    & 4125 & 0.60 & --1.56 & --1.53 &  +0.07 &  +0.17 &  +0.38 &\nodata &\nodata    &\nodata &\nodata &\nodata \\
II-34   & 4140 & 0.63 & --1.64 & --1.64 & --0.22 &  +0.54 &  +0.75 &\nodata &\nodata    &\nodata &\nodata &\nodata \\
III-73  & 4210 & 0.69 & --1.57 & --1.56 &  +0.25 & --0.05 &  +0.15 &\nodata &\nodata    &\nodata &\nodata &\nodata \\
I-13    & 4230 & 0.80 & --1.55 & --1.49 &  +0.24 & --0.16 &  +0.03 &\nodata &\nodata    &\nodata &\nodata &\nodata \\
II-76   & 4250 & 0.80 & --1.65 & --1.50 &  +0.30 & --0.27 & --0.09 &\nodata &\nodata    &\nodata &\nodata &\nodata \\
III-59  & 4290 & 0.85 & --1.51 & --1.55 & --0.47 &  +0.29 &  +0.47 &\nodata &\nodata    &\nodata &\nodata &\nodata \\
III-52  & 4275 & 0.87 & --1.63 & --1.48 & --0.23 &  +0.32 &  +0.50 &\nodata &\nodata    &\nodata &\nodata &\nodata \\
II-33   & 4300 & 0.90 & --1.53 & --1.55 &  +0.30 &  +0.05 &  +0.23 &\nodata &\nodata    &\nodata &\nodata &\nodata \\
III-18  & 4330 & 0.95 & --1.52 & --1.58 & --0.26 &  +0.44 &  +0.62 &\nodata &\nodata    &\nodata &\nodata &\nodata \\
II-57   & 4350 & 1.00 & --1.53 & --1.48 & --0.09 &  +0.33 &  +0.50 &\nodata &\nodata    &\nodata &\nodata &\nodata \\
\multicolumn{13}{c}{Cluster Mean Abundances} \\
$<>$    &      &      & --1.60 & --1.53 & --0.13 &  +0.21 &  +0.39 &  +0.11 &     +0.75 &  +0.24 &  +0.11 &  +0.49 \\
$\pm$   &      &      &   0.01 &   0.02 &   0.06 &   0.04 &   0.05 &   0.03 &      0.11 &   0.03 &   0.03 &   0.03 \\
$\sigma$&      &      &   0.06 &   0.09 &   0.38 &   0.24 &   0.27 &   0.11 &      0.44 &   0.15 &   0.14 &   0.13 \\
No.     &      &      &     34 &     34 &     33 &     34 &     34 &     18 &        16 &     18 &     18 &     18 \\
\enddata

\tablenotetext{a}{Revised; see text.}

\tablenotetext{b}{Na abundance after correction for the suggested
                  nLTE corrections of Gratton et al. (2000).}

\end{deluxetable}


\clearpage 

\begin{thebibliography}

\bibitem[]{all01}
Allende Prieto, C., Lambert, D. L., \& Asplund, M. 2001, \apj, 556, L63

\bibitem[]{alo99}
Alonso, A., Arribas, S., \& Martínez-Roger, C. 1999, \aaps, 140, 261

\bibitem[]{and89}
Anders, E., \& Grevesse, N. 1989, Geochim. Cosmochim Acta, 53, 197

\bibitem[]{asp01}
Asplund, M., \& García P\'erez, A. E. 2001, \aap, 372, 601

\bibitem[]{bel01}
Bellman, S., Briley, M. M., Smith, G. H., \& Claver, C. F.
2001, \pasp, 113, 326

\bibitem[]{bla80}
Blackwell, D. E., Shallis, M. J., \& Simmons, G. J. 1980, \aap, 81, 340

\bibitem[]{bla90}
Blackwell, D. E., Petford, A. D., Arribas, S., Haddock, D. J., \& Selby, M. J.
1990, \mnras, 232, 396

\bibitem[]{bri94}
Briley, M. M., Bell, R. A., Hesser, J. E. \& Smith, G. H. 1994,
Can. J. Phys., 72, 772

\bibitem[]{bri90}
Briley, M. M., Bell, R. A., Hoban, S., \& Dickens, R. J. 1990
\apj, 359, 307

\bibitem[]{bri96}
Briley, M. M., Smith, V. V., Suntzeff, N. B., Lambert, D. L., 
Bell, R. A., \& Hesser, J. E. 1996, \nat, 383, 604

\bibitem[]{bri03}
Briley, M. M., Cohen, J. G., \& Stetson, P. B. 2003, \apj, 579, L17

\bibitem[]{buo94}
Buonanno, R., Corsi, C. E., Buzzoni, A., Cacciari, C., 
Ferraro, F. R., \& Fusi Pecci, F. 1994, \aap, 290, 69

\bibitem[]{can98}
Cannon, R. D., Croke, B. F. W., Bell, R. A., Hesser, J. E.,
\& Stathakis, R. A. 1998, \mnras, 298, 601

\bibitem[]{cas97}
Castelli, F., Gratton, R. G., \& Kurucz, R. L. 1997, \aap, 318, 841

\bibitem[Cavallo \& Nagar (2000)]{cav00}
Cavallo, R.~M .\& Nagar, N.~M. 2000, \aj, 120, 1364

\bibitem[]{cav96}
Cavallo, R. M., Sweigart, A. V., \& Bell, R. A. 1996, \apj, 464, L79

\bibitem[]{cav98}
Cavallo, R. M., Sweigart, A. V., \& Bell, R. A. 1998, \apj, 492, 575

\bibitem[]{coh99}
Cohen, J. G. 1999, \aj, 117, 2434

\bibitem[]{coh02}
Cohen, J. G., Briley, M. M., \& Stetson, P. B. 2002, \aj, 123, 2525

\bibitem[]{coh78}
Cohen, J. G., Persson, S. E. \& Frogel, J. A. 1978, \apj, 222, 165

\bibitem[]{cot81}
Cottrell, P. L., \& Da Costa, G. S. 1981, \apj, 245, 79

\bibitem[]{cud79a}
Cudworth, K. M. 1979, \aj, 84, 1312

\bibitem[]{cud79b}
Cudworth, K. M., \& Monet, D. G. 1979, \aj, 84, 774

\bibitem[]{dac98}
Da Costa, G. S. 1998, in The Stellar Content of Local Group Galaxies,
IAU Symp. 192, ed. P. Whitelock, R. Cannon San Francisco: Ast. Soc. 
Pac.), p. 13

\bibitem[]{den98}
Denissenkov, P. A., Da Costa, G. S., Norris, J. E.,
\& Weiss, A. 1998, \aap, 333, 926

\bibitem[]{den90}
Denissenkov, P. A. \& Denissenkova, S. N. 1990, 
Sov. Astr. Lett., 16, 275

\bibitem[]{dor98}
Dorman, B. 1998, private communication

\bibitem[]{fit87}
Fitzpatrick, M. J. \& Sneden, C.  1987, \baas, 19, 1129

\bibitem{ful00}
Fulbright, J. P. 2000, \aj, 120, 1841

\bibitem{ful02}
Fulbright, J. P. 2002, \aj, 123, 404

\bibitem[]{gra00}
Gratton, R. G., Sneden, C., Carretta, E. \& Bragaglia, A. 2000,
\aap, 354, 169

\bibitem[]{gra01}
Gratton, R.~G., et al. 2001, \aap, 369, 87

\bibitem[]{gus75}
Gustafsson, B., Bell, R. A., Eriksson, K., \& Nordlund, A. 1975, \aap, 42, 407

\bibitem[]{han98}
Hanson, R. B., Sneden, C., Kraft, R. P., \& Fulbright, J. 1998, \aj, 116, 1286

\bibitem[]{hes77}
Hesser, J. E., Hartwick, F. D. A., \& McClure, R. D. 1977, \apjs, 33, 471

\bibitem[]{hou00}
Houdashelt, M. L., Bell, R. A., \& Sweigart, A. V. 2000, \aj, 119, 1448

\bibitem[]{ibe84}
Iben, I. \& Renzini, A. 1984, Phys. Rev., 105, 329

\bibitem[]{iva01}
Ivans, I. I.,Kraft, R. P., Sneden, C., Smith, G. H., Rich, R. M., \&
Shetrone, M. D. 2001, \aj, 122, 1438

\bibitem[]{iva99}
Ivans, I. I., Sneden, C., Kraft, R. P., Suntzeff, N. B., Smith, V. V.,
Langer, G. E. \& Fulbright, J. P. 1999, \aj, 118, 1273

\bibitem[]{joh98}
Johnson, J. A., \& Bolte, M. 1998, \aj, 115, 693

\bibitem{kar03}
Karakas, A. I., \& Lattanzio, J. C. 2004, \apj, submitted

\bibitem[]{kin98}
King, J. R., Stephens, A., Boesgaard, A. M., Deliyannis, C.
1998, \aj, 115, 666

\bibitem[]{kra94}
Kraft, R. P. 1994, \pasp, 106, 553

\bibitem[]{kra03}
Kraft, R. P., \& Ivans, I. I. 2003, \pasp, 115, 143

\bibitem[]{kra92}
Kraft, R. P., Sneden, C., Langer, G. E., \& Prosser, C. F. 1992,
\aj, 105, 645

\bibitem[]{kra93}
Kraft, R. P., Sneden, C., Langer, G. E., \& Shetrone, M. D.
1993, \aj, 106, 1490

\bibitem[]{kra95}
Kraft, R. P., Sneden, C., Langer, G. E., Shetrone, M. D.m \& Bolte, M. 
1995, \aj, 109, 2586

\bibitem[]{kra98}
Kraft, R. P., Sneden, C., Smith, G. H., Shetrone, M. D., 
\& Fulbright, J. 1998, \aj, 115, 1500

\bibitem[]{kra97}
Kraft, R. P., Sneden, C., Smith, G. H., Shetrone, M. D., Langer, G. E.,
\& Pilachowski, C. A. 1997, \aj, 113, 279

\bibitem[]{kur95}
Kurucz, R. L. 1995, in Workshop on Laboratory and astronomical high
resolution spectra, ASP Conference Ser. \#81 ed. A.J. Sauval, R. Blomme,
and N. Grevesse (San Francisco: Astr. Soc. Pac.), p.583

\bibitem[]{lan86}
Langer, G. E., Kraft, R. P., Carbon, D. F., Friel, E.,
\& Oke, J. B. 1986, \pasp, 98, 473

\bibitem[]{lan95}
Langer, G. E., \& Hoffman, R. D. 1995, \pasp, 107, 1177

\bibitem[]{lan97}
Langer, G. E., Hoffman, R. D., \& Zaidins, C. S. 1997, \pasp, 109, 244

\bibitem[]{lan92}
Langer, G. E., Suntzeff, N. B., \& Kraft, R. P. 1992, \pasp, 104, 523

\bibitem[]{lee94}
Lee, Y.-W., Demarque, P., \& Zinn, R. 1994, \apj, 423, 248

\bibitem[]{lod03}
Lodders, K. 2003, \apj, 591, 1220

\bibitem[]{nor95}
Norris, J., \& Da Costa, G. S. 1995, \apj, 447, 680

\bibitem[]{nor84}
Norris, J., \& Smith, G. H. 1984, \apj, 287, 255

\bibitem[]{osb71}
Osborn, W. 1971, Observatory, 91, 223

\bibitem[]{pet03}
Peterson, R. C., Carney, B. W., Dorman., B., Green, E. M., Landsman, W., 
Liberty, J., O'Connell, R. W., \& Rood, R. T. 2003, \apj, 588, 299

\bibitem[]{pil88}
Pilachowski, C. A. 1988, \apj, 326, L57

\bibitem[]{pil03}
Pilachowski, C., Sneden, C., Freeland, E., \& Casperson, J. 2003, \aj, 125, 794

\bibitem[]{pil96a}
Pilachowski, C. A., Sneden, C. \& Kraft, R. P., 1996, \aj, 111, 1689

\bibitem[]{pil96b}
Pilachowski, C. A., Sneden, C. Kraft, R. P., \& Langer, G. E. 1996, 
\aj, 112, 545

\bibitem[]{pow99}
Powell, D. C., Iliads, C., Champagne, A. E., Grossman, C. A., Hale, S. E., 
Hans per, V. Y., \& McLean, L. K. 1999, Nuclear Phys. A, 660, 349

\bibitem[]{ram03}
Ram\'irez, S. V., \& Cohen, J. G. 2003, \aj, 125, 224

\bibitem[]{ram02}
Ramirez, S. V., \& Cohen, J. G. 2002, \aj, 123, 3277

\bibitem[]{rey01}
Rey, S.-C,, Yoon, S.-J., Lee, Y.-W., Chaboyer, B., 
\& Sarajedini, A. 2001, \aj, 122, 3219

\bibitem[]{san53}
Sandage, A. 1953, \aj, 58, 61

\bibitem[]{san70}
Sandage, A. 1970, \apj, 162, 841

\bibitem[]{sar97}
Sarajedini, A., Chaboyer, B., \& Demarque, P. 1997, \pasp, 109, 1321

\bibitem[]{she96a} 
Shetrone, M. D. 1996a, \aj, 112, 1517

\bibitem[]{she96b} 
Shetrone, M. D. 1996b, \aj, 112, 2639

\bibitem[]{sie02}
Siess, L., Livio, M., \& Lattanzio, J. 2002, \apj, 570, 329

\bibitem[]{smi02a}
Smith, G. H. 2002a, \pasp, 114, 1097

\bibitem[]{smi02b}
Smith, G. H. 2002b, \pasp, 114, 1215

\bibitem[]{smi82}
Smith, G. H., \& Norris, J. 1982, \apj, 254, 594

\bibitem[]{smi96}
Smith, G. H., Shetrone, M. D., Bell, R. A., Churchill, C. W., 
\& Briley, M. M. 1996, \aj, 112, 1511

\bibitem[]{sne73}
Sneden, C. 1973,  \apj, 184, 839

\bibitem[]{sne97}
Sneden, C., Kraft, R. P., Shetrone, M. D., Smith, G. H., Langer, G. E.,
\& Prosser, C. F. 1997, \aj, 114, 1964

\bibitem[]{sne00}
Sneden, C., Johnson, J., Kraft, R. P., Smith, G. H., Cowan, J. J., 
\& Bolte, M. S. 2000, \apj, 536, 85

\bibitem[]{ste96}
Stetson, P. B., Vandenberg, D. A., Bolte, M. 1996, \pasp, 108, 560

\bibitem[]{sun81}
Suntzeff, N. B. 1981, \apjs, 47, 1

\bibitem[]{sun91}
Suntzeff, N. B., \& Smith, V. V. 1991, \apj, 381, 160

\bibitem[]{swe97}
Sweigart, A. V. 1997, \apj, 474, L23

\bibitem[]{the99}
Th\'evenin, F., \& Idiart, T. P. 1999, \apj, 521, 753

\bibitem{tim95}
Timmes, F. X., Woosley, S. E., \& Weaver, T. A. 1995, \apjs, 98, 617

\bibitem[]{tre83}
Trefzger, D. V., Langer, G. E., Carbon, D. F., Suntzeff, N. B.,
\& Kraft, R. P. 1983, \apj, 266, 144

\bibitem[]{ven01}
Ventura, P., D'Antona, F., Mazzitelli, I., \& Gratton, R.
2001, \apj, 550, L65

\bibitem[]{ven02}
Ventura, P., D'Antona, F., \& Mazzitelli, I.  2002, \aap, 393, 215

\bibitem{von08}
von Zeipel, M. H. 1908, Ann. Obs. Paris, 25, F1

\bibitem[]{vog94}
Vogt, S. S., et al. 1994, SPIE, 2198, 362

\bibitem[]{wal97}
Wallerstein, G, et al. 1997, Rev. Mod. Phys., 69, 995

\bibitem[]{wor94}
Worthey, G. 1994, \apjs, 95, 107

\bibitem[]{yon03}
Yong, D., Grundahl, F., Lambert, D. L., Nissen, P. E., \& Shetrone, M. D.
2003, \aap, 402, 985


\end{thebibliography}
\end{document}